\documentclass[useAMS,usenatbib]{mn2e}
\usepackage{epsfig}

\def\mnras{MNRAS}  % Monthly Notices of the RAS
\def\apj{ApJ}      % The Astrophysical Journal
\def\apjl{ApJ}     % The Astrophysical Journal Letters
\def\aap{A\&A}     % Astronomy and Astrophysics
\def\nat{Nature}   % Nature

\title[Shock Heating in Cluster Mergers]{Modelling Shock Heating 
in Cluster Mergers:\\ I. Moving Beyond the Spherical Accretion 
Model}

\author[I. G. McCarthy et al.]{I. G. McCarthy$^{1}$\thanks{E-mail:
i.g.mccarthy@durham.ac.uk (IGM)}, R. G. Bower$^1$, M. L. 
Balogh$^2$, G. M. Voit$^3$, F. R. Pearce$^4$
\newauthor T. Theuns$^{1,5}$, A. Babul$^6$, C. G. Lacey$^1$, and 
C. S. Frenk$^1$
\\
$^{1}$Department of Physics, University of Durham, South Road,
Durham, DH1 3LE\\
$^{2}$Department of Physics and Astronomy, University of Waterloo, 
Waterloo, ON, N2L 3G1, Canada\\
$^{3}$Department of Physics and Astronomy, Michigan State
University, East Lansing, MI 48824, USA\\
$^{4}$Physics and Astronomy Department, University of Nottingham,
Nottingham, NG7 2RD\\
$^{5}$Department of Physics, University of Antwerp, Campus 
Groenenborger, Groenenborgerlaan 171, B-2020 Antwerp, Belgium\\
$^{6}$Department of Physics and Astronomy, University of Victoria,
Victoria, BC V8P 1A1, Canada}

\begin{document}

\date{Accepted XXXX. Received XXXX; in original form XXXX}

\pagerange{\pageref{firstpage}--\pageref{lastpage}} \pubyear{2006}

\maketitle

\label{firstpage}

\begin{abstract}

The thermal history of the intracluster medium (ICM) is complex. 
Heat input from cluster mergers, from AGN, and from winds in 
galaxies offsets and may even prevent the cooling of the ICM. 
Consequently, the processes that set the temperature and density 
structure of the ICM play a key role in determining how galaxies 
form.  In this paper we focus on the heating of the intracluster 
medium during cluster mergers, with the eventual aim of 
incorporating this mechanism into semi-analytic models for galaxy 
formation. 

We generate and examine a suite of non-radiative hydrodynamic 
simulations of mergers in which the initial temperature and 
density structure of the systems are set using realistic scaling 
laws.  Our collisions 
cover a range of mass ratios and impact parameters, and consider 
both systems composed entirely of gas (these reduce the physical 
processes involved), and systems comprising a realistic mixture of 
gas and dark matter.  We find that the heating of the ICM can be 
understood relatively simply by considering evolution of the 
gas entropy during the mergers.  The increase in this quantity in 
our simulations closely corresponds to that predicted from scaling 
relations based on the increase in cluster mass.

We examine the physical processes that succeed in generating the 
entropy in order to understand why previous analytical approaches 
failed.  We find that: (1) The energy that is thermalised during 
the collision greatly exceeds the kinetic energy available when 
the systems first touch.  The smaller system penetrates deep into 
the gravitational potential before it is disrupted. (2) For 
systems with a large mass ratio, most of the energy is thermalised 
in the massive component. The heating of the smaller system is 
minor and its gas sinks to the centre of the final system.  This 
contrasts with spherically symmetric analytical models in which 
accreted material is simply added to the outer radius of the 
system.  (3) The bulk of the 
entropy generation occurs in two distinct episodes.  The first 
episode occurs following the collision of the cores, when a large 
shock wave is generated that propagates outwards from the centre.  
This causes the combined system to expand rapidly and overshoot 
hydrostatic equilibrium.  The second entropy generation episode 
occurs as this material is shock heated as it re-collapses.  Both 
heating processes play an important role, contributing 
approximately equally to the final entropy.  This revised model 
for entropy generation improves our physical understanding of 
cosmological gas simulations.
 
\end{abstract}


\clearpage

\section{Introduction}

The structure of the X-ray emitting gas that is seen in galaxy 
groups and clusters is still not understood. This gas (the 
intracluster medium, hereafter ICM) dominates the baryonic mass 
content of clusters and is material which is left over from 
galaxy formation.  As such its properties present an important 
clue to the galaxy formation process.  If the processes that set 
its temperature and density structure could be understood they 
should provide valuable constraints on galaxy formation models.  
However, linking together models for galaxy formation and 
accurate numerical methods capable of tracing the hydrodynamic 
evolution of the ICM has proved difficult.  If cooling is 
neglected, the emission properties of clusters can be calculated 
in a robust manner (e.g., Evrard et al.\ 1996); however such 
simulations cannot form galaxies and thus omit a fundamental 
component of the physics.  On the other hand, if cooling is 
included in the simulations, the results are not stable 
to changes in numerical resolution unless an effective form of 
heating is introduced to limit the cooling rate in small or early 
halos (e.g., Balogh et al.\ 2001; Borgani et al.\ 2006).  Many 
different ``feedback'' mechanisms have been tried to oppose the 
cooling instability.  Examples include galaxy winds and thermal 
energy injection (e.g., Springel \& Hernquist 2003), delayed 
cooling (e.g., Kauffmann et al.\ 1999), preheating of 
intergalactic gas (e.g., Kaiser 1991; Evrard \& Henry 1991), 
thermal conduction (e.g., Benson et al.\ 2003; Dolag et al.\ 
2004), and heating by AGN (e.g., Dalla Vecchia et al.\ 2004; 
Sijacki \& Springel 2006).  These schemes have all met with 
various levels of success, but have not yet achieved a simultaneous
match to the observed galaxy luminosity function, stellar mass
fraction and temperature and density structure of the ICM.  
One factor slowing the development of these theoretical models
is the lack of understanding of the physical processes at work in 
the simulations. Because it is so difficult to quantify the likely
effect of introducing new processes, the models must be developed
on a largely trial and error basis.

An alternative approach is to develop a semi-analytic model of the 
thermal history of the ICM, in which the complex physics is 
encapsulated in a small number of semi-empirical equations.  This 
approach has been very successful in improving our understanding
of galaxy formation (e.g., Cole et al 2000; Springel et al. 2005;
Bower et al. 2006).
The semi-analytic approach has been tried in several studies 
(e.g., Wu et al.\ 
2000; Bower et al.\ 2001; Benson et al.\ 2003), but the results 
have been limited because the techniques for setting the gas 
distribution in the halos have been ad hoc.  What is needed is 
to take a step back, and to better understand the results of 
simple gas hydrodynamic models.  A better physical understanding 
of how entropy of the ICM is set in these simulations would 
equip us with the techniques needed to attack the much greater 
physical complexity of the problem when cooling and galaxy 
formation are included. This is the subject of this paper.

The starting point for the next generation of semi-analytical 
approaches to the ICM should be to describe the physical state of 
the gas in terms of its entropy distribution function.  Entropy 
is a powerful concept for understanding the density and 
temperature structure of 
clusters, and was initially introduced as a means of quantifying 
the departures of cluster structures from the simple scalings 
expected for their mass and temperature distributions (Evrard \& 
Henry 1991; Kaiser 1991; Bower 1997; Balogh et al.\ 1999; 
McCarthy et al.\ 2003).  By knowing the entropy distribution of 
the gas, we can determine the density and temperature profiles 
of the ICM within a given dark matter gravitational potential 
(e.g., Voit et al. 2002). This process requires that an outer 
boundary condition is set (e.g. by computing the external 
pressure due to infalling gas) but the details of the boundary 
condition have only a weak effect on the profile.  Discussing 
clusters in the language of entropy is extremely powerful, since 
the gas responds adiabatically to slow changes in the 
gravitational potential.  Furthermore, the buoyancy of the ICM 
ensures that a relaxed system will have an ordered structure with 
the lowest entropy gas located in the deepest part of the 
gravitational potential. Only cooling and shock heating or mixing 
events alter the entropy distribution of the gas: cooling lowers 
the gas entropy, while shock heating and mixing can only raise 
it.  Cooling in clusters is already well understood and can be 
modelled with simple semi-analytic techniques (e.g., McCarthy 
et al.\ 2004).  However, we need to understand how entropy is 
generated during cluster growth to self-consistently model the 
cosmological formation of structure in a semi-analytic way.

So what sets the entropy distribution in clusters?  Previous 
papers have looked at how entropy is generated in smoothly 
infalling gas (Cavaliere et al.\ 1998; Abadi et al.\ 2000; Tozzi 
\& Norman 2001; Dos Santos \& Dor{\'e} 2002; Voit et al.\ 2003).  
For a spherically symmetric smooth shell, it is possible to 
determine the infall velocity and density of this material at the 
cluster virial radius.   The bulk infall velocity is converted into 
internal thermal energy at a shock surrounding the cluster.  The 
entropy generated in the shock can be computed from the 
Rankine-Hugoniot shock jump conditions. Modelling the growth of 
clusters in this way produces realistic entropy profiles, however 
the normalisation of the entropy is somewhat too high, so that 
the model predicts clusters with average densities $\rho_{\rm 
gas}$, that are lower than observed (see Voit et al.\ 2003).

However, the smooth accretion approximation is not valid in a 
cold dark matter (CDM) dominated universe since most of the mass 
accreted by clusters has previously collapsed into smaller 
virialised mass concentrations (e.g., Lacey \& Cole 1993; Rowley 
et al.\ 2004; Cohn \& White 2005).  This leads to a problem for 
the picture above since the entropy generated drops rapidly if 
the accreted material is already dense and warm prior to the 
accretion shock.  Importantly, if this deficit is present at 
every step in a system's merger history, the problem becomes 
greatly compounded over time.  This effect is far larger than the 
overestimate of the cluster entropy obtained in the smooth 
accretion case.  Modelling a realistic structure in the accreted 
material leads to systems that are much denser and more luminous 
in X-rays than observed systems.  This issue is discussed by 
Voit et al.\ (2003).  

Recently, considerable progress has been made in simulating the 
growth of clusters using purely numerical techniques (e.g., Borgani 
et al.\ 2004; Kravtsov et al.\ 2005).  The numerical simulations 
clearly show that $\rho_{\rm gas} \approx f_{b} \rho_{\rm tot}$ 
(where $f_{b}$ is the cosmic baryon fraction, and $\rho_{\rm tot}$ 
is total baryon + dark matter density) at large radii, in 
agreement with observations (e.g., Vikhlinin et al.\ 2006; 
McCarthy et al.\ 2007).  This result holds for a wide range of 
simulation parameters and is largely insensitive to numerical 
technique or resolution (Frenk et al.\ 1999).  The agreement 
between simulations and observations indicates that we do not 
yet have an adequate understanding of the lumpy accretion 
process, since analytic attempts to model this process yield 
results in strong discord with the observations.  Our immediate 
aim is to improve our understanding of these simulations.  
Clearly, the approximate methods we wish to develop will not 
replace direct simulations, but they do provide an essential 
tool for understanding their results, for estimating the impact 
of limited numerical resolution, and for exploring the vast 
parameter space of heating mechanisms used in galaxy formation 
models.  In subsequent papers, we will apply this understanding 
to improve the treatment of gas heating in semi-analytic codes 
galaxy formation models (e.g., Cole et al.\ 2000; Bower et al.\ 
2006). By introducing our models for the shock heating of the 
ICM, we will be able to self-consistently model additional 
heating from supernovae and AGN.  At present, the best efforts 
to track the impact of AGN heating in semi-analytic models 
(e.g., Bower et al.\ 2001) cannot follow properly the 
development of entropy and use an ad hoc description based on 
energy  conservation.

In this paper we address the problem of lumpy accretion by focusing 
on idealised simulations of merging systems.  We consider 
simple two-body collisions and explore how the entropy 
of the final system is generated.  In setting up our two-body 
collisions, we remain as faithful as possible to the results of 
cosmological simulations.   We initialise our systems with 
structural properties guided by the results of recent cosmological 
simulations and explore a wide range of representative mass ratios 
and orbits.  These simulations show that the spherical accretion 
model previously considered is a poor representation of the true 
physical process in hierarchical cosmological models.  We show 
that the kinetic energy of the infall lumps is much greater than 
previously estimated.  Furthermore, we show that, if the mass 
ratio of the accretion event is large, only a small fraction of 
the kinetic energy is dissipated in the smaller component.  Most 
of the heating effect is felt by the larger system.  Our 
simulations allow us to present a much improved physical model 
for entropy generation in clusters.

The present paper is outlined as follows.  In \S 2, we present the 
details of our simulation setup, including a discussion of the 
initial structural conditions of our systems, the adopted orbital 
parameters of the mergers, and other characteristics of the 
simulations.  In \S 3, we analyse the entropy history of the 
systems during the merging process and qualitatively compare it 
with the standard spherical accretion model.  In \S 4, we 
present an analytic model that encapsulates the essential 
physics of the merging process in our idealised simulations.  
Finally, in \S 5, we summarise and discuss our findings.

\section{Simulation Setup and Methods}

The baryons in virialised systems formed in non-radiative 
cosmological simulations approximately trace the dark matter 
with $\rho_{\rm gas} \approx f_b \rho_{\rm tot}$.  Any reasonable 
model of shock heating ought to reproduce this basic result.  
However, as outlined in \S 1, it has so far proved quite 
difficult to construct a physical analytic model that can 
successfully pass this test.  Exploration of idealised two-body 
mergers could provide the key insights necessary to achieve this 
goal.  These idealised mergers should be representative of typical 
mergers in cosmological simulations and should themselves reproduce 
the propagation of (near) self-similarity.  However, this is by no 
means a guaranteed result.  First, idealised approaches such as 
those adopted below may be too simplistic to mimic closely enough 
the typical merger event in cosmological simulations.  For example, 
the degree to which the self-similarity of systems formed in 
non-radiative cosmological simulations is sensitive to the 
exact initial conditions of the systems (i.e., structural 
properties) or to the properties of the merger orbits is unclear.  
There are few studies in the literature to date that examine the 
dependence of the properties of the baryons on the detailed merger 
history of a system.  On the other hand, the adopted resolution of 
our idealised mergers is typically better than that of 
cosmological simulations.  So it is possible that we may resolve 
phenomena that are not yet adequately resolved in cosmological 
simulations.  In either case, the reproduction of self-similarity 
in our idealised simulations is not an automatic result.  Below, 
we explore under what conditions self-similarity is achieved in 
our idealised merger simulations.

The following subsections present a description of our simulation 
setup and methods.  This discussion goes into some detail and the 
reader who is mainly interested in the results of the simulations 
may wish to skip ahead to \S 3.

\subsection{Initial conditions}

We make use of the public version of the parallel TreeSPH code 
GADGET-2 (Springel 2005) for our merger simulations.  The 
Lagrangian nature of this code makes it ideal for tracking the 
entropy evolution of specific sets of particles (or in principle 
even on a particle by particle basis) over the course of the 
simulations.  By default the code implements the 
entropy-conserving SPH scheme of Springel \& Hernquist (2002).  
This is ideal for our purposes since the code explicitly 
guarantees that the entropy of a gas particle will be conserved 
during any adiabatic process.  Thus, we are assured that 
any increase in this quantity is rooted in physics.

We perform a series of binary merger simulations involving systems
ranging in mass from $10^{14} M_\odot \leq M_{200} \leq 10^{15}
M_\odot$ with the mass of the primary system (henceforth, we refer 
to the `primary' system as the more massive of the two systems) in 
all collisions set by default to $M_{200}=10^{15} M_\odot$.  
($M_{200}$ is defined in \S 2.1.1.)  Thus, we simulate 
collisions characterised by mass ratios ranging from 10:1 to 1:1.  
Although we 
have experimented with a variety of mass ratios in this range, we 
present the results for simulations with mass ratios of 10:1, 3:1, 
and 1:1 only, but note that these yield representative results.  
Even though we have chosen to focus on the high end of the mass 
spectrum, the results should also be applicable to mergers 
involving lower mass systems (e.g., galaxies) so long as the mass 
ratio is similar.  This is by virtue of the fact that gravity, 
which is scale-free, is completely driving the evolution of the 
systems.  One of the reasons for choosing to focus on clusters 
is that gravitational processes dominate their final properties 
(X-ray observations demonstrate that non-gravitational processes 
such as cooling and AGN feedback influence the very central 
regions only of massive clusters; see, e.g., McCarthy et al.\ 
2007).  
So we expect that our idealised merger simulations will be directly 
applicable to the bulk of the baryons in real clusters, although we 
leave a comparison with X-ray (and/or SZ effect) observations for 
future work.  As outlined in \S 1, our primary goal is to develop a 
physical description of entropy generation in mergers.

\subsubsection{Gas-only models}

We have set up and run collisions of systems composed purely of 
gas.  Of course, observations demonstrate that dark matter 
dominates by mass the baryons in galaxies, groups, and clusters and 
is of fundamental importance in the process of structure formation.  
In this respect, it might be expected that the gas-only merger 
simulations will have limited applicability.  On the other hand, 
we anticipate that these idealised collisions will be considerably 
more straightforward to interpret than the gas + dark matter 
(hereafter, gas+DM) simulations (described below), given that the 
former have only a single phase to consider.  If a physical 
understanding of the gas-only runs can be achieved then this could 
be quite helpful for understanding the role of dark matter in the 
combined runs.  In fact, it is demonstrated in \S 3 that the gas+DM 
collisions exhibit what may be regarded as fairly minor 
deviations from the results of the gas-only runs.  Thus, we find 
the 
gas-only simulations are potentially an excellent tool for 
elucidating the key physical processes in mergers of massive 
systems.

The systems are constructed to initially be structural copies of 
one another.  In particular, the gas is assumed to have a NFW 
density profile (Navarro et al.\ 1997):

\begin{equation}
\rho(r) = \frac{\rho_s}{(r/r_s)(1+r/r_s)^2}
\end{equation}

\noindent where $\rho_s = M_s/(4 \pi r_s^3)$ and

\begin{equation}
M_s=\frac{M_{200}}{\ln(1+r_{200}/r_s)-(r_{200}/r_s)/(1+r_{200}/r_s)} .
\end{equation}

In the above, $r_{200}$ is the radius within which the mean 
density is 200 times the critical density, $\rho_{\rm crit}$, and 
$M_{200} \equiv M(r_{200}) = (4/3) \pi r_{200}^3 \times 200 
\rho_{\rm crit}$.

To define the density profile for a halo of mass $M_{200}$, we must
set the scale radius, $r_s$. The scale radius is often expressed in 
terms of a concentration parameter, $c_{200} \equiv r_{200}/r_s$.  
There are numerous studies that have examined the trend between 
concentration parameter and mass in pure dark matter cosmological 
simulations (e.g., Eke et al.\ 2001).  The 
fact that there is a trend at all implies that the dark 
matter halos in these simulations are not strictly self-similar.  But 
the mass dependence of the concentration parameter is weak, varying 
by only a factor of 2 or so from individual galaxies to galaxy 
clusters (with lower mass halos being more concentrated).  A fixed 
concentration parameter of $c_{200} = 4$, a value typical of 
galaxy clusters simulated in a $\Lambda$CDM concordance cosmology, 
is adopted for all of our systems.

In order to completely specify the properties of the gas we must 
choose an entropy profile and an outer boundary condition.  As 
outlined in \S 1, the gas entropy is probably the most useful 
quantity to track during the simulations.  In order to specify the 
entropy profiles of the systems, it is assumed that the gas is 
initially in hydrostatic equilibrium:

\begin{equation}
\frac{dP(r)}{dr} = - \frac{G M(r)}{r^2} \rho_{\rm gas}(r)  
\end{equation}

The entropy\footnote{We adopt this common re-definition of 
entropy,  which is related to the standard thermodynamic specific 
entropy via a logarithm and an additive term that depends 
only on fundamental constants.}, $K$, is then deduced through the 
equation of state $P = K \rho_{\rm gas}^{5/3}$.  A boundary 
condition 
must be supplied 
before equation (3) can be solved and we specify a value for the 
pressure of the gas at $r_{200}$, where we truncate the 
gas profiles (unless stated otherwise).  There is some freedom in 
our choice of the pressure at the edge of the system.  However, 
for physically reasonable values of $P(r_{200})$ there is in 
fact very little difference between the resulting profiles.  
For the bulk of the gas, we find that the requirement of 
hydrostatic equilibrium forces $K(M_{\rm gas})$ into a near 
power-law distribution with an index of approximately 1.3.  This 
can be understood by noting that at intermediate radii the NFW 
profile is approximated well by an isothermal profile (i.e., $\rho 
\propto r^{-2}$), which has an entropy profile characterised by 
$d\log{K}/d\log{M_{\rm gas}} = 4/3$.  To ensure that our initial 
systems are structural 
copies of one another and to ease the comparison between the 
initial and final systems later on, a value of $P(r_{200})$ that 
establishes this near pure power-law entropy distribution all the 
way out to the system's edge is selected.  In {\it real} systems, 
the hot diffuse baryons are supposedly confined by the ram 
pressure of infalling material.  Experimenting with a physical 
model for this ram pressure, Voit et al.\ (2002) have demonstrated 
that groups and clusters are indeed expected to have near 
power-law entropy distributions out to large radii.

The analytic profiles must be discretised into individual particles 
for input into the GADGET-2 code.  In particular, the initial 
particle positions, velocities, and internal energies (per 
unit mass) must be specified.

For the initial particle positions, we start by generating a glass 
using the ``MAKEGLASS'' compiler option of GADGET-2.  Basically, a 
Poisson distribution of particles is generated and is then 
run with GADGET-2 using $-G$ in place of $G$ for Newton's constant.  
We allow the simulation to run for $\approx$1000 time steps to 
ensure 
the glass achieves a uniform distribution.  
This uniform distribution 
is then morphed into the mass profile that corresponds to the density
profile given in equation (1).  This is done by selecting a point
inside the distribution of particles, which we have taken to be 
the centre of mass, ranking all of the particles according to 
the distance from this point, and then moving each particle 
radially such that the desired mass profile is achieved.

For the particle velocities, the baryons are initially assumed to be
at rest in their hydrostatic configuration.

The specific internal energy, $I$, defined as $I = 
(3/2)P/\rho_{\rm gas}$, of each particle is 
assigned by using the particle's distance from the centre of the 
halo and 
interpolating within the analytic profiles derived above.

Lastly, we surround the systems with a low density medium with a 
pressure set to $P(r_{200})$.  This medium is dynamically-negligible 
and is simply put in place to confine the gas particles near the 
system's edge.  Without a confining medium as much as 20\% of the 
system's mass can leak beyond $r_{200}$.

\begin{figure}
\centering
\includegraphics[width=8.4cm]{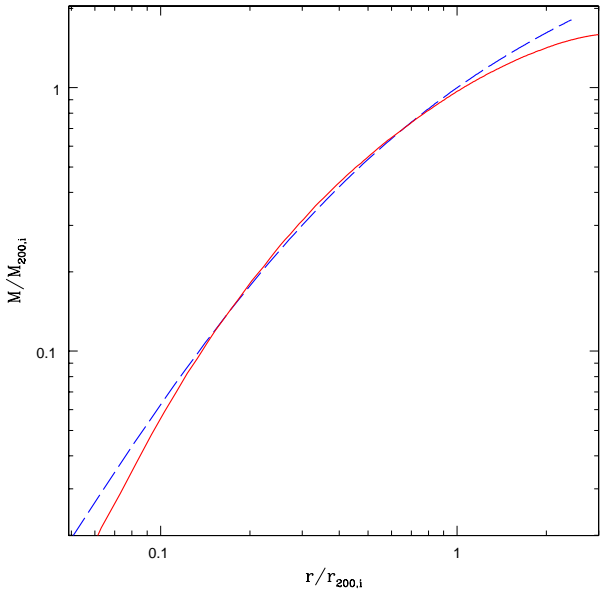}
\caption{Testing the local Maxwellian approximation for the 
primary system.  The dashed
blue curve shows the initial mass profile extended out to
$r_{25}$.  The solid red curve shows the resulting equilibrium 
mass 
profile after running the dark matter halo in isolation for 50 Gyr 
(i.e., many dynamical times).  Over the range the $0.1 \leq 
r/r_{200} \leq 1.0$ the initial and final mass profiles follow 
each other closely.
}
\end{figure}

\subsubsection{Gas + dark matter (gas+DM) models}

For the systems containing both gas and dark matter a slightly 
different procedure is used to construct the systems.  First, we set 
up systems composed entirely of dark matter.  The density 
profiles of the systems are given by equation (1).  However, 
for reasons described below, we extend the dark matter well beyond 
$r_{200}$, to an overdensity of 25 (for our adopted concentration 
$r_{25} \approx 2.44 r_{200}$).  As in the gas-only setup, a glass 
particle distribution is morphed into the desired mass profile.  
Unlike the gas, however, the dark matter particles have no 
thermal pressure and must be assigned appropriate velocities in 
order to maintain this mass profile.  To achieve this, we solve the 
the Jeans equation (see Binney \& Tremaine 1987):

\begin{equation}
\frac{d[\sigma^2_r(r) \rho(r)]}{dr} = - \frac{G M(r)}{r^2} 
\rho(r)
\end{equation}

\noindent for the velocity dispersion profile, $\sigma_r(r)$, of 
the systems.  Equation (4) implicitly assumes that the orbits of the 
dark matter particles are purely isotropic\footnote{Cosmological 
simulations demonstrate that pure isotropy is violated in the outer 
regions of clusters.  However, given that our (hydrostatic) 
gas-only simulations yield results that are remarkably consistent with 
our gas+DM simulations (see \S 3.2), this implies that the 
resulting structural properties of our systems are not sensitive to 
how the initial (pre-merger) mass profiles are maintained.}.

To solve equation (4) it is assumed that virial equilibrium holds 
at the edge of the dark matter halo (see, e.g., Ricker \& Sarazin
2001); i.e.,

\begin{equation}
\sigma_r(r_{25}) = \sqrt{\frac{G M_{25}}{3 r_{25}}}
\end{equation}

We have also experimented with the boundary condition that
$\sigma^2_r(r) \rho(r) \rightarrow 0$ as $r \rightarrow \infty$ 
and obtained very similar results within $r_{25}$.  Henceforth, 
results are presented for the runs that make use of the boundary 
condition given by equation (5).

Each of the three velocity components for a dark matter particle 
are assigned a random value picked from a Gaussian distribution 
with width equal to the 1-D velocity dispersion, $\sigma_r$, at 
that radius.  However, it is known that this local Maxwellian 
approximation does not result in a system that is initially in 
equilibrium (though it is not far removed from one).  Of most 
concern is that a sharp 
truncation of the system at some finite radius\footnote{Note that 
some sort of truncation is necessary since the mass profile 
corresponding to equation (1) diverges at large radii.} results in 
a significant amount of the mass near this radius seeping out to 
much larger distance.  Thus, the resulting equilibrium mass 
profile can be significantly different from that which was 
originally intended.  In fact, this happens at all radii but in 
the system's interior the flux of particles moving to larger 
radii is offset by particles moving inwards (which were 
originally at larger radii).  This obviously cannot happen near 
the system's edge where there are no particles originally at 
larger radii (and moving inwards) to replace those moving away 
from the system's centre.  One way to overcome this problem is to 
apply a smooth tapering of the density profile beyond the 
system's edge and compute the distribution function of the 
particles as opposed to making the Maxwellian assumption (e.g., 
Kazantzidis et al.\ 2004; Poole et al.\ 2006).  However, we adopt 
a more simplistic, but still effective, method for overcoming 
this problem.  In particular, as mentioned above, the initial 
dark matter profile is extended well beyond $r_{200}$, out to 
$r_{25}$.  The aim is to ensure that within $r_{200}$, our region 
of interest, there is a large enough influx of particles to 
maintain the desired mass profile.

The pure dark matter halos are run in isolation for many 
dynamical times.  In Figure 1, we show that by extending the 
dark matter halo out to $r_{25}$, the resulting equilibrium mass 
profile for the primary system traces the initial (intended) 
profile within $r_{200}$ quite well.  A slight deviation 
is apparent for $r < 0.1 r_{200}$, which is due to limited mass 
resolution.  However, we explicitly demonstrate in the Appendix 
that this deviation has negligible consequences for our equal 
mass mergers.  In the case of our unequal mass mergers, the 
secondary systems are resolved with fewer particles and therefore 
the deviation at small radii is slightly larger in these systems.  
However, as we demonstrate in \S 4.2 and discuss further in the 
Appendix, the vast bulk of the energy is thermalised in the more 
massive primary system.  So, as long as the primary system is 
well resolved the properties of the final merged system should 
be robust.  This is what likely accounts for the fact that 
the properties of massive virialised systems formed 
in non-radiative cosmological simulations are robust to 
relatively large changes in numerical resolution (see, e.g., 
Frenk et al.\ 1999).

A drawback of extending the halo out to much larger radii is 
that a significant fraction of the mass is located outside 
the region of interest (thus, we potentially waste a good deal of 
computational effort).  In order to mitigate this issue, we take 
the equilibrium configuration after running the dark halos in 
isolation and simply clip particles that lie beyond $r_{50}$.  We 
have run the clipped halos for a further 10 Gyr and verify that 
the mass profiles within $r_{200}$ is virtually unaffected.
This clipped equilibrium configuration is used to set up our initial
gas+DM systems.  We keep the current positions and velocities
for our initial values for the dark matter particles in the combined 
runs.  For the gas particles we also use the positions of the dark 
matter particles (that is, for those particles that lie within 
$r_{200}$) but reflect them through the centre of mass so that the gas 
particles do not lie directly on top of the dark matter particles.  
The gas particle velocities are set to zero and the internal energy 
densities are determined by placing the gas in hydrostatic 
equilibrium within the total gravitational potential.  As in the 
gas-only models described above, a pressure boundary condition is 
selected such that $K(M_{\rm gas})$ is nearly a pure power-law 
out to $r_{200}$.  

The above procedure implies that we will have equal numbers of gas 
and dark matter particles in our combined runs within $r_{200}$ 
(i.e., double the number of particles used in the isolated dark 
matter run within that radius).  In order to conserve mass, so that 
the equilibrium state for the dark matter is still valid, we 
re-assign the dark matter particle masses to $(1-f_b)$ times that 
used in the isolated run while the gas particles are assigned a 
mass of $f_b$ times the dark matter particle mass in the isolated 
run.  Here $f_b$ is the ratio of gas mass to total mass within 
$r_{200}$, which is set to the universal value of $\Omega_b/\Omega_m 
= 0.02 h^{-2}/0.3 = 0.136$.

The end result of the above procedure is that the gas traces the dark 
matter, both phases are initially in equilibrium, and within 
$r_{200}$ both phases have a form that is very nearly the intended 
NFW distribution.  As in the gas-only models, we surround the 
gas+DM systems with a low density pressure-confining gaseous 
medium.

\subsection{Merger orbits and other simulation characteristics}

We follow the approach of Poole et al.\ (2006) and use the results of 
cosmological simulations to help specify the orbital properties of our 
idealised merger simulations.  In particular, we turn to the study 
of Benson (2005; but see also Tormen 1997; Vitvitska et al.\ 
2002; Wang et al. 2005; Khochfar \& Burkert 2006).  Benson (2005) 
used a large collection of N-body simulations carried out by the 
VIRGO Consortium to study the distribution of orbital parameters of 
substructure falling onto massive (primary) halos.  Particular 
emphasis was given to the 2-D distribution of the relative 
radial and tangential velocities of the substructure as they 
passed through a spherical shell with a radius set to $r = (1 \pm 
0.2) r_{\rm vir}$, where $r_{\rm vir}$ is the virial radius of the 
primary halo.  As anticipated, the peak of the distribution 
corresponds to approximately the circular velocity of the primary 
halo at the virial radius (see his Fig.\ 2).  In particular,  $v 
\equiv (v_r^2 + v_t^2)^{1/2} \approx 1.1 v_c(r_{\rm vir})$, where 
$v_r$ and $v_t$ are the relative radial and tangential 
velocities, respectively.  For the individual components, the 
peak of the distribution is approximately centred on $v_r \approx 
0.9 v_c(r_{\rm vir})$ and $v_t \approx 0.65 v_c(r_{\rm vir})$ and 
it is clear that there is a correlation between $v_r$ and $v_t$.  
However, as pointed out by Benson (2005), the centring of the 
peak also depends on the total mass of systems (see his Fig.\ 4).  
In particular, the orbits of mergers involving massive systems 
are 
more radial and less tangential than those of mergers involving 
low mass halos, presumably because more massive structures tend to 
form at the intersections of large-scale filaments. Unfortunately, 
the sample wasn't large enough to fully quantify this mass 
dependence.

For the purposes of the present study, we adopt the following set 
of orbital parameters, which are representative of the results of 
Benson (2005).  Unless stated otherwise, a fixed relative 
total velocity corresponding to the circular velocity of the 
primary at $r_{200}$, $\approx 1444$ km/s, is adopted initially 
for all of our simulations\footnote{We note that the circular 
velocity at 
$r_{200}$ differs by only a small amount from the 
circular velocity at the virial radius, which is what Benson 
actually compares his velocities to.  This just reflects the 
fact that the NFW mass profile gives rise to a nearly flat circular 
velocity profile at large radii.}.  This is slightly lower than the 
mean value found by Benson (2005), but consistent at the 1-sigma 
level.  For the gas+DM simulations, we examine 
three different orbits for each mass ratio, corresponding to  $v_t 
= 0$, $v_t = v_r/4$, and $v_t = v_r/2$.  We refer to these as the 
``head on'', ``small impact parameter'' and ``large impact 
parameter'' runs, respectively.  In practical terms, this implies 
$v_t \approx 0.243 v_{c,p}(r_{200})$ and $v_t \approx 0.447 
v_{c,p}(r_{200})$ for the latter two.  This choice spans the lower 
half 
of the $v_t/v_r$ distribution for massive halos seen in Fig.\ 4 of 
Benson (2005).  For the gas-only simulations, which we run simply 
to help coax out the key physics during mergers, we simulate 
head on collisions only [i.e., $v_r = v_{c,p}(r_{200})$ and $v_t = 
0$].

\begin{table}
\centering
\begin{minipage}{140mm}
\caption{Merger Orbital Parameters.}
\begin{tabular}{@{}lllll@{}}
\hline
Mass ratio & Sim. type & $d_0$ & $v_r$  & $v_t$\\
 & & (kpc) & (km/s) & (km/s) \\
\hline
1:1 & gas-only & 4126 & 1444 & 0\\
3:1 & gas-only & 3494 & 1444 & 0\\
10:1 & gas-only & 3021 & 1444 & 0\\
\hline
1:1 & gas+DM & 4342 & 1444 & 0\\
1:1 & gas+DM & 4342 & 1400.9 & 350.2\\
1:1 & gas+DM & 4342 & 1291.6 & 645.8\\
3:1 & gas+DM & 3658 & 1444 & 0\\
3:1 & gas+DM & 3658 & 1400.9 & 350.2\\
3:1 & gas+DM & 3658 & 1291.6 & 645.8\\
10:1 & gas+DM & 3170 & 1444 & 0\\
10:1 & gas+DM & 3170 & 1400.9 & 350.2\\
10:1 & gas+DM & 3170 & 1291.6 & 645.8\\
\hline
\end{tabular}
\end{minipage}
\end{table}

Each of the simulations are initialised with the gaseous 
components of the primary and secondary systems just barely 
touching.  Thus, the initial separation, $d_0$, is just the 
summation of $r_{200}$ for the primary and $r_{200}$ for the 
secondary.  For the gas+DM simulations, this implies 
that there is initially some overlap of dark matter halos of the 
primary and secondary systems, which extends beyond the gaseous 
component.  A complete summary of the adopted orbital parameters of 
all of our simulations is presented in Table 1.  Note that for a 
fixed mass ratio $d_0$ differs slightly between the gas+DM and the 
gas-only simulations.  This difference arises owing to the slight 
deviation of the mass profiles in the gas+DM 
simulations from the intended NFW distribution.  Finally, the 
simulations have been set up such that the physical and 
centre-of-mass (of the entire system) coordinates and velocities
are identical, where we have approximated the initial 
configuration as two point masses.

We have carried out a mass resolution study of one of our 
simulations (see the Appendix).  Based on this study we adopt the 
following conditions.  The primary systems in our default runs 
have 50,000 gas particles within $r_{200}$.  This is the case for 
both the
gas-only and gas+DM runs and implies the gas particle mass is
$2\times10^{10} M_\odot$ in the former and $f_b(2\times10^{10}
M_\odot) = 0.272\times10^{10} M_\odot$ in the latter.  In the
gas+DM runs, the primary systems have $\approx$76,500 dark matter
particles within $r_{50}$ and 50,000 within $r_{200}$ with a dark
matter particle mass of $(1-f_b)(2\times10^{10} M_\odot) =
1.728\times10^{10}M_\odot$.  These particle masses are held fixed
for the secondary systems as well (thus, they contain fewer
particles than the primary).

For the gravitational softening length, we adopt a fixed physical 
size of 10 kpc for both the gas and dark matter particles. (We 
have also experimented with softening lengths of 5 kpc and 20 kpc 
and found the differences in the entropy evolution to be 
negligible.)  We use a fairly standard set of SPH 
parameters, including a viscosity parameter, $\alpha_{\rm visc}$, 
of 0.8, a Courant coefficient of 0.1, and the number of SPH 
smoothing neighbours, $N_{\rm sph}$, is set to $50 \pm 2$.

Finally, each simulation is run for a duration of 13 Gyr, i.e.,
approximately a Hubble time.  Particle data is written out in
regularly spaced intervals of 0.1 Gyr.

\begin{figure}
\centering
\includegraphics[width=8.4cm]{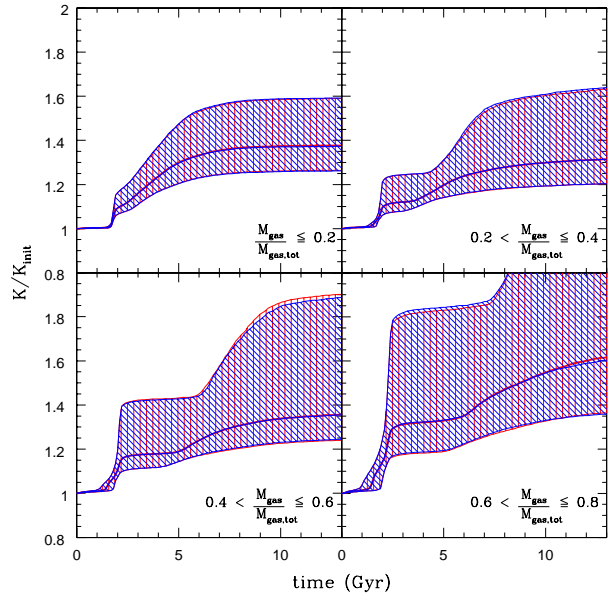}
\caption{Entropy evolution in the 1:1 gas-only merger.  The red
and blue lines/hatched regions represent the primary and
secondary systems, respectively.  The upper and lower bounds of
the hatched regions represent the 75$^{\rm th}$ and 25$^{\rm th}$
percentiles, respectively, while the central (thick) lines
represent the median.  The panels show the evolution of the
entropy of particles separated according to their {\it initial} 
position in the cumulative gas mass profile.  Since the cumulative 
gas mass profile is a monotonic function of radius, the panels also
separate the particles according to initial distance from the 
centre of their respective halos.
 }
\end{figure}

\section{Simulation Results}

Below we present a detailed discussion of the entropy evolution 
of our simulations.  The reader who is mainly interested in a 
general understanding of this progression may wish to skip ahead 
to \S 3.3, which presents a summary of our findings.

\subsection{Gas-only simulations}

\subsubsection{Entropy evolution}

We start by considering the evolution of the entropy in the 
gas-only merger simulations.  Plotted in Figure 2 is the evolution 
for the 1:1 merger subdivided into spherical shells.  Immediately 
apparent is the high degree of symmetry in the 1:1 merger.  In 
all four regions the 25$^{\rm th}$, 50$^{\rm th}$, and 75$^{\rm 
th}$ percentiles follow each other quite closely.  In 
addition, with the exception of the outermost ring, it is evident 
that the particles in both systems have achieved a nearly 
convergent state by the end of the simulation.  Of course, it is 
possible to continue running the simulation to determine exactly 
what the convergent state of the outermost regions will be. 
However, since we have already run the simulation for a Hubble 
time this implies that the outer regions of systems formed from 
similar mergers in cosmological simulations will also not have 
had sufficient time to completely virialise by the present day.  
Since one of our aims is to understand the results of such 
simulations, we limit the duration of our simulations to 13 Gyr.   
Below we will compare this final configuration with a 
scaled up copy of the initial systems.  

Some discussion of how the particles in each system actually get 
their entropy is warranted.  To help aid the discussion we plot in 
Fig.\ 3 a sequence of snapshots of the 1:1 merger with particles 
colour coded according to the fractional change in entropy since 
the previous simulation output.  The general progression of the 
merger can be described as follows.  Since the two systems are 
initially just barely touching and have a relative velocity of 
$v_{c,p}(r_{200})$, they begin interacting as soon 
as the simulation starts.  In particular, two large shock fronts 
are quickly established as the systems approach one another (see 
the panel corresponding to $t = 1$ Gyr in Fig.\ 3).  However, 
there is very little increase in the entropy of the particles in 
any of the four regions plotted in Fig.\ 2 until after the cores 
of the two systems collide at $t \approx 2$ Gyr into the 
simulation. The implication is that the initial shocks are 
actually quite weak (as the colour coding in Fig.\ 3 would also 
indicate).  A plausible explanation for this behaviour comes 
when one compares the sound speed of the gas, $c_s$, defined as

\begin{figure*}
\centering
\includegraphics[width=15cm]{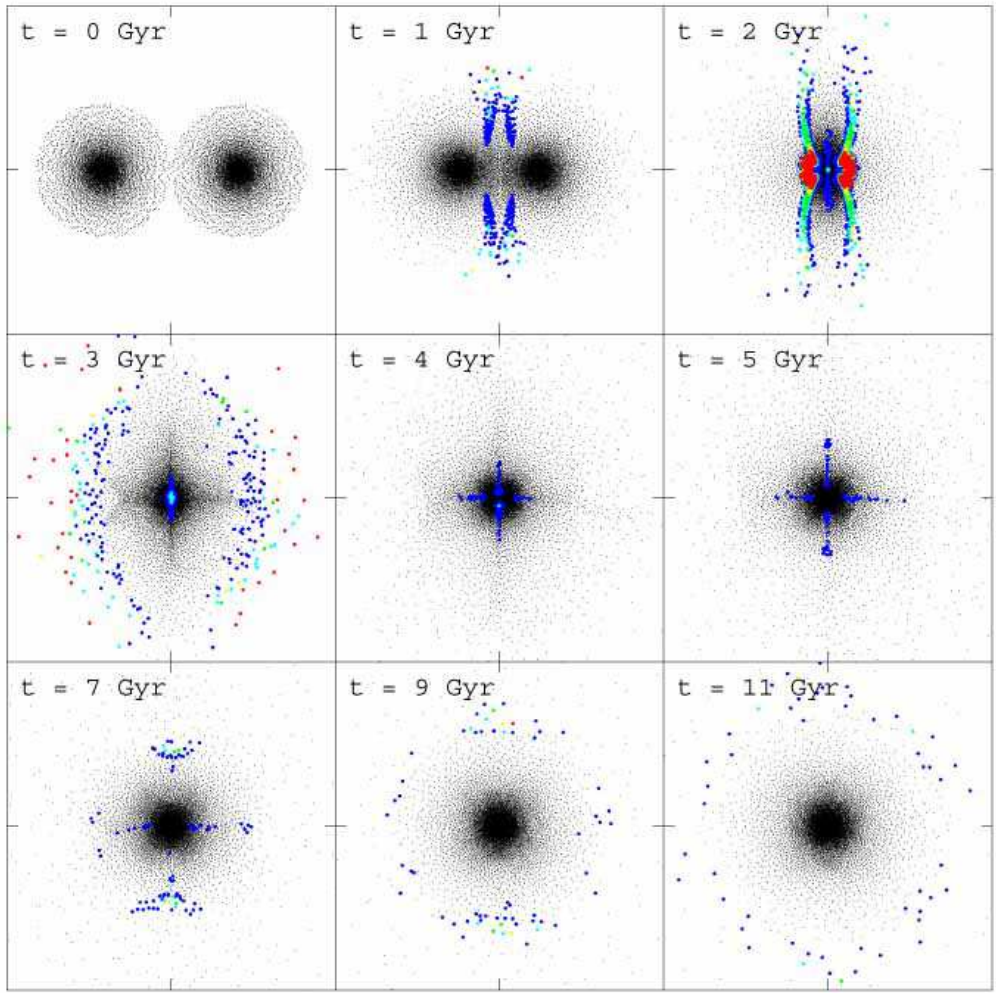}
\caption{Fractional change in particle entropy as a function of
time for the 1:1 gas-only merger.  Shown are particles in a slice
through the centre of thickness 250 kpc (i.e., $|z| < 125$ kpc).
Particles are colour coded according to the fractional change in
entropy since the last simulation output (0.1 Gyr ago).  Black
points are for a fractional change of less than 2\%, blue are for
a 2\%-10\% change, cyan are for a 10\%-20\% change, green are for
a 20\%-30\% change, yellow are for a 30\%-40\% change, and red 
are
for a $>$ 40\% change.  For clarity the surrounding
pressure-confining medium is not displayed.  Each panel is 10 Mpc
on a side.
}
\end{figure*}

\begin{equation}
c_s \equiv \sqrt{\frac{\partial P}{\partial \rho}} = 
\sqrt{\frac{\gamma P}{\rho}} = \sqrt{\frac{5}{3}\frac{k_b T}{\mu m_p}}
\end{equation}

\noindent with the initial relative velocity of the merger.  Since 
the initial relative velocity of the merger is set to the circular 
velocity, we are effectively comparing the sound crossing time of a 
system with its dynamical time.  The condition of hydrostatic 
equilibrium ensures that the gas temperature will be such that 
these two timescales are comparable.  In this case we find the 
mean sound speed of the systems is $\approx$1200 km/s.  Compared 
to 
the initial relative velocity (see Table 1), this implies an 
initial Mach number of only $\cal{M} \approx$ 1.2.  Therefore, we 
should expect only mild shock heating to occur during the early 
stages of the merger (as observed in Fig.\ 2).  However, as we 
illustrate in \S 4, by the time the cores collide the relative 
velocity can approach or exceed nearly 3 times the initial value, 
effectively bringing the merger into the strong shock regime.

Following the collision of the cores, a large shock wave is 
generated.  This shock quickly propagates outwards, heating the 
gas that was initially predominantly on the far side of each 
system and has not yet finished falling in (see the panels 
corresponding to $t = 2$ and 3 Gyr in Fig.\ 3).  In fact, this 
shock wave, combined with the expansion of the shocked high 
pressure/density gas near the core of the merged system, succeeds 
in reversing the infall of the material.  This outflow of gas 
proceeds nearly adiabatically for a period of time that is 
dictated by the initial distance of the particle from its system's 
centre.  For example, for the outermost ring plotted in Fig.\ 2 
this period of adiabatic expansion occurs from $t \approx 2.5$ Gyr 
for a duration of nearly 3.5 Gyr.  Gas particles initially located 
close to the centre of their respective system do not spend such a 
large time in this outflow phase, as they end up being more deeply 
embedded within in the merged system's potential well.

\begin{figure*}
\centering
\leavevmode
\epsfysize=8.4cm \epsfbox{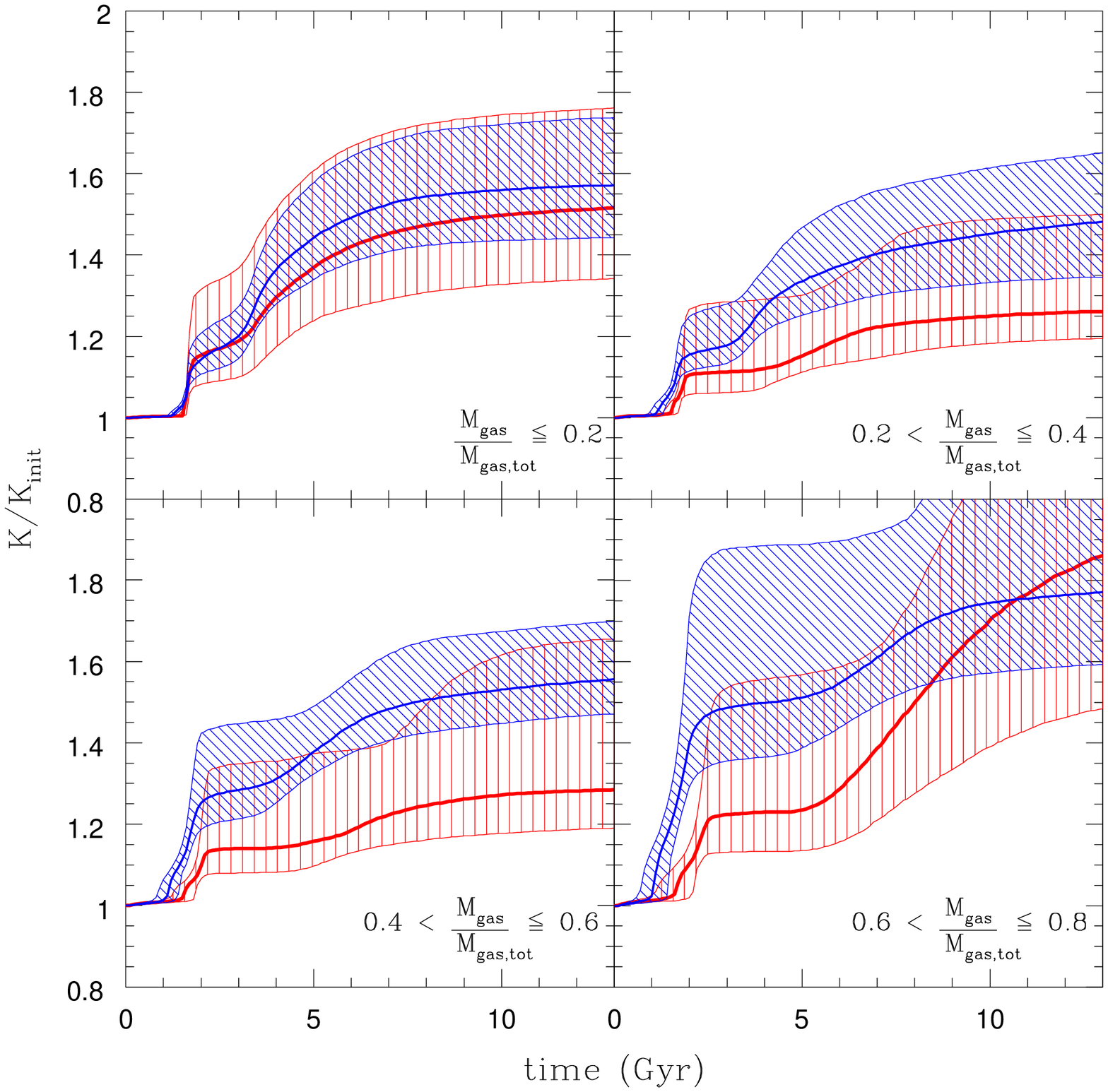}
\epsfysize=8.4cm \epsfbox{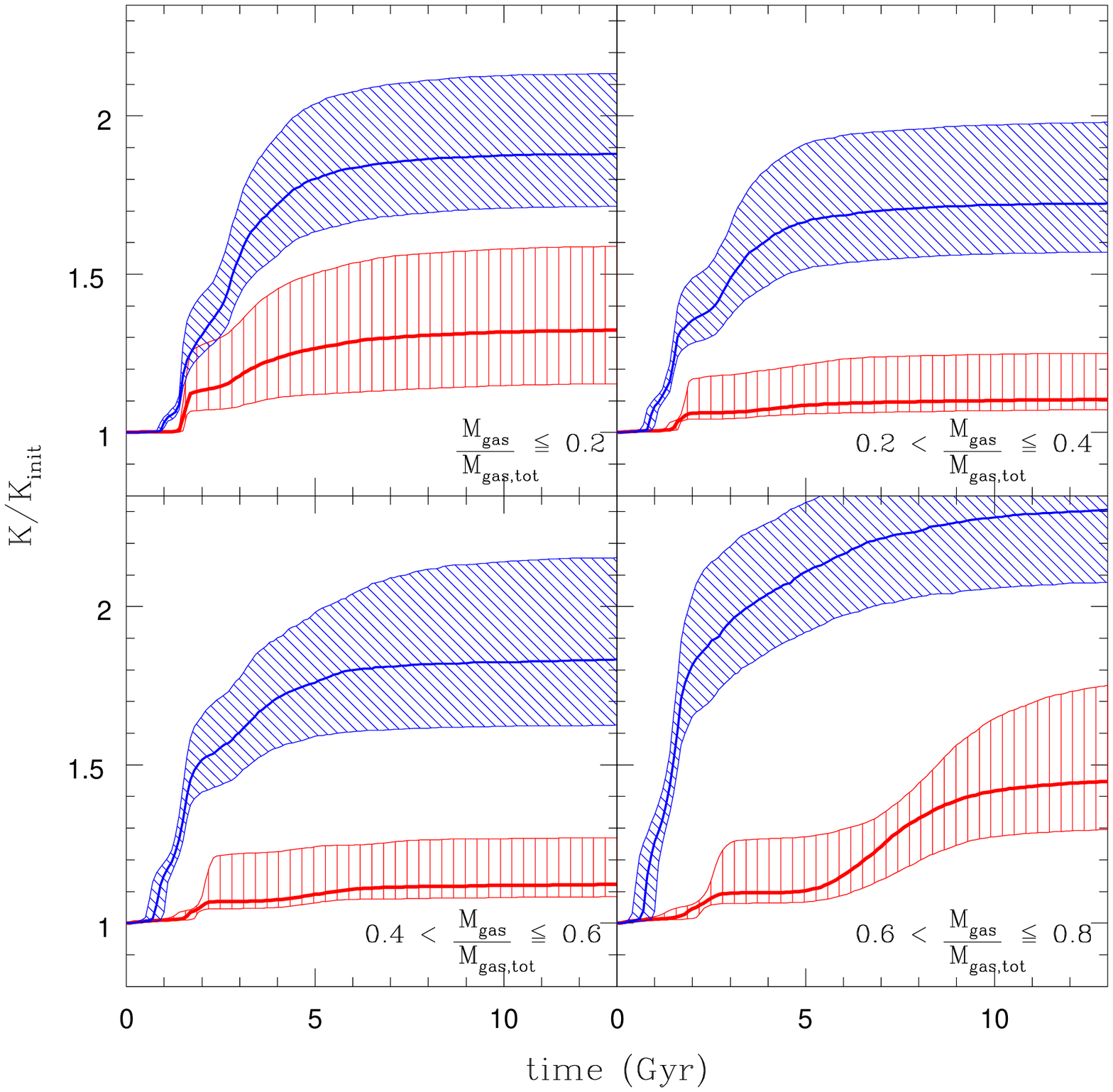}
\caption{Entropy evolution in the 3:1 (left) and 10:1 (right) gas
only mergers.  The red and blue lines/hatched regions represent
the primary and secondary systems, respectively. The upper and
lower bounds of the hatched regions
represent the 75$^{\rm th}$ and 25$^{\rm th}$ percentiles,
respectively, while the central (thick) curves represent the
median.  The panels show the evolution of the entropy of 
particles
separated according to their {\it initial} position in the
cumulative gas mass profile.  Since the cumulative gas mass
profile is a monotonic function of radius, the panels also
separate the particles according to initial distance from the
centre of their respective halos.
}
\end{figure*}

The outflowing gas eventually halts and begins to re-accrete onto 
the core of the merged system.  Although the gas accretes from 
virtually all directions, the symmetry of the 1:1 merger is 
such that the accretion happens preferentially along and 
perpendicular to the original collisional axis.  This occurs for a 
significant period of time, until the merged core has achieved a 
near spherical symmetry.  Following this, gas continues to 
accrete at the outskirts of the system but does so in a more or 
less spherical fashion.  As evidenced from Figs.\ 2 and 3, the 
gas is slowly being shock heated during this period of 
re-accretion.  The character of this episode of entropy 
production therefore differs significantly from the first abrupt 
episode following core collision.  However, as Fig.\ 2 
indicates, both episodes are of comparable importance in terms 
of setting the final state of the gas.  Finally, we note 
that the late time (nearly spherical) accretion of material is 
what gives rise physically to the continued increase of entropy 
in the outermost regions of the system until the end of the 
simulations (e.g., as in the bottom-right panel of Fig.\ 2).

An interesting question is whether or not much mixing occurs 
as a result of the merging process.  We have tested this by 
tracking not 
only the entropy evolution of individual particles but also their 
spatial distributions.  Interestingly, even though the shock 
heating of the systems occurs in an ``inside-out'' fashion, we 
see very little evidence for mixing.  For example, particles 
that were initially located near the centres of the two 
(pre-merger) halos (i.e., particles that initially 
have relatively low entropies) end up being located near the 
centre of the final merged system, whereas the large-radii (high 
entropy) particles end up occupying the outer regions of the 
final system.  A plausible explanation for this behaviour is that 
the initial Mach number distribution of the particles varies 
relatively weakly with radius.  This can be understood by 
considering the initial temperature profiles.  The condition 
of hydrostatic equilibrium results in initial temperature 
profiles that vary by less than a factor of 2 from the cluster 
centre to its periphery.  This then implies the Mach number 
distribution varies by less than a factor of $2^{1/2}$ over the 
entire system.  Therefore, to a large degree, one expects (and 
the simulations bear this out) that most of the particles will 
be shock heated to a similar degree.  As a result, convective 
mixing is minimal.  This agrees well with the idealised cluster 
merger simulations of Poole et al.\ (2006).

The entropy evolution of the 3:1 and 10:1 gas-only mergers
is plotted in Figures 4a and 4b, respectively.  In a qualitative 
sense, both the primary and secondary systems in these mergers 
behave in a similar manner to the systems in the 1:1 case.  For 
example, they both generate two relatively weak shock fronts 
early on, with the secondary driving a bow shock into the 
primary and vice-versa.  As in the 1:1 case, they exhibit two main 
periods of entropy production, the first being associated 
with a strong shock approximately when their cores collide and the 
second being associated with an extended period of re-accretion 
shocks.  Furthermore, we also see little evidence for mixing 
in the 3:1 and 10:1 mergers.  Of course there are some 
differences between the three sets of simulations in detail.  
For example, because the sound speed of the gas in the secondary 
in the 10:1 collision is 
smaller than that in the 1:1 case, the initial Mach number for 
the secondary is higher.  Consequently, the initial bout of shock 
heating before the cores collide is relatively more important 
for the secondary in high mass ratio mergers.

\begin{figure*}
\centering
\leavevmode
\epsfysize=8.4cm \epsfbox{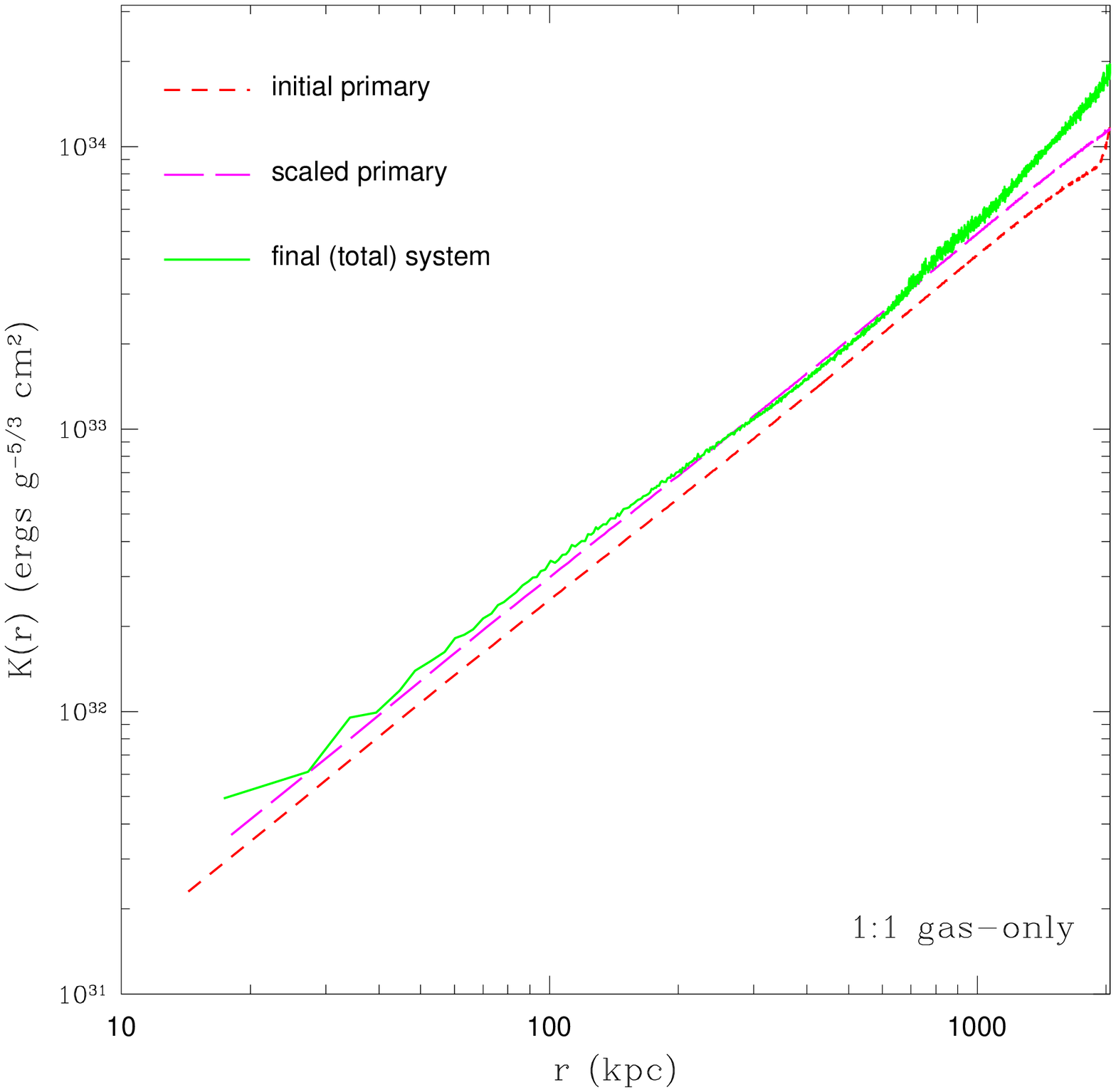}
\epsfysize=8.4cm \epsfbox{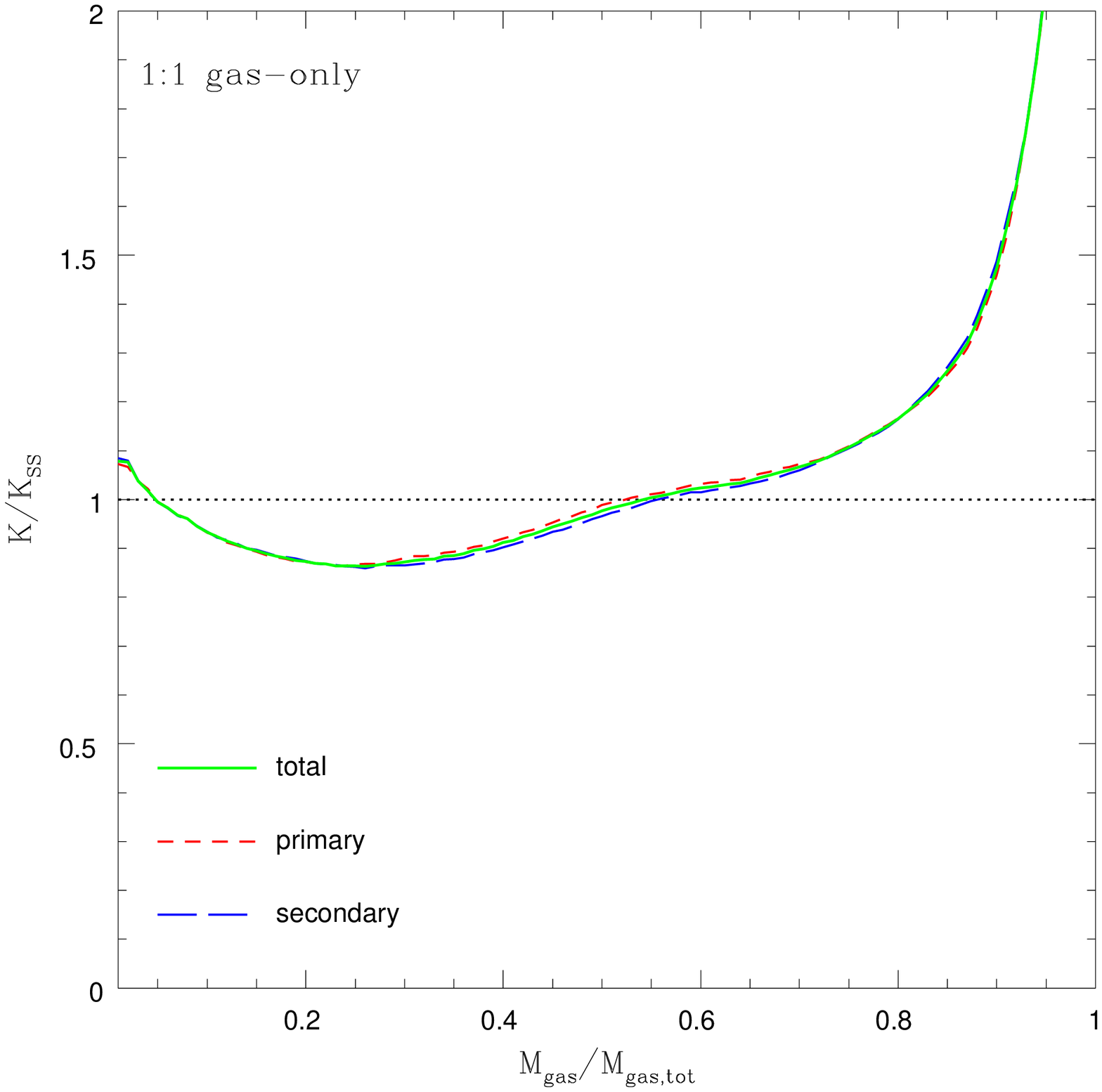}
\caption{Comparison of the resulting entropy distributions
from the 1:1 gas-only merger with self-similar expectations.  
{\it
Left:} Comparison of radial profiles.  The short-dashed red curve
represents the initial entropy profile of the primary, the
long-dashed magenta curve represents a self-similarly scaled up
version of this profile, and the solid green curve represents the
final entropy profile of the total merged system.  {\it Right:}
Comparison of Lagrangian gas mass
profiles.  The short-dashed red, long-dashed blue, and solid 
green
curves represent the final $K(M_{\rm gas})$ distributions of the
primary, secondary, and total merged systems, respectively.  The
horizontal dotted line represents the self-similar scaling.
 }
\end{figure*}

It is clear from a comparison of Figs.\ 2 and 4a,b that the higher 
the mass ratio of the merger the more strongly heated the 
secondary is relative to the primary.  This again may be tied to 
the fact that as one decreases the mass of the secondary its 
characteristic temperature decreases resulting in a decreased 
sound speed and an increased Mach number.  It is also expected if 
the preservation of self-similarity is achieved by heating both 
the secondary and the primary up to same level. To explain, $K = 
P/\rho_{\rm gas}^{5/3} \propto T/\rho_{\rm gas}^{2/3}$, 
but self-similarity means that all systems have the same internal 
mass 
structure and therefore the characteristic entropy depends only 
on the temperature of the system.  The virial theorem relates the 
temperature of a system with its mass via $M \propto T^{3/2}$, 
and therefore the characteristic entropy of a system scales simply 
as

\begin{equation}
K \propto M^{2/3}
\end{equation}

Initially, therefore, the ratio of the secondary's characteristic 
entropy to that of the primary is $K_s/K_p = (M_s/M_p)^{2/3}$.  So 
the secondary requires extra heating to overcome its initial 
deficit compared to the primary if self-similarity is achieved by 
heating both systems to the same level.

\subsubsection{Final configurations}

How do the final entropy distributions of the gas-only mergers 
compare with self-similar expectations?  Figure 5 presents this 
comparison for the 1:1 gas-only merger both in traditional radial 
coordinates (left panel) and in physical gas mass coordinates 
(right panel).  In Figure 5a, we compare the initial entropy 
radial profile of the primary system with the final radial profile 
of the total merged system.  We also show a self-similarly scaled 
up version of the primary's initial entropy profile, achieved by 
scaling the entropy coordinate up by $(M_{\rm tot}/M_p)^{2/3} = 
2^{2/3}$ and the radial coordinate by $(M_{\rm tot}/M_p)^{1/3} = 
2^{1/3}$.  Fig.\ 5a shows that the scaled up version of the 
primary's initial entropy profile follows the final entropy 
profile of the merged system over approximately two decades in 
radius.  This indicates that, by and large, the 1:1 gas-only 
merger has scaled self-similarly through the merger.  However, 
deviations are clearly visible, particularly at large radii.  
These are discussed in more detail below.

\begin{figure*}
\centering
\leavevmode
\epsfysize=8.4cm \epsfbox{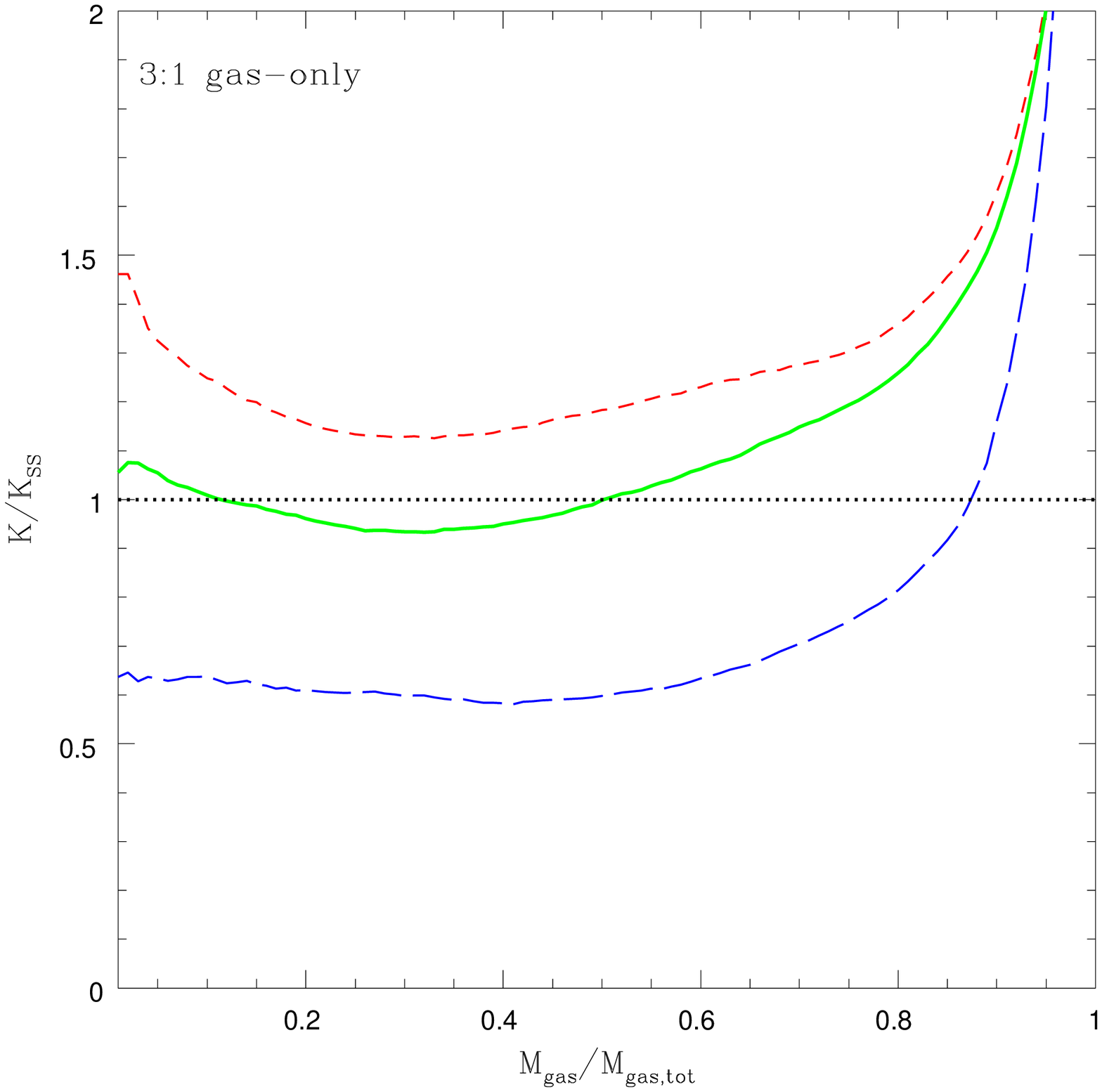}
\epsfysize=8.4cm \epsfbox{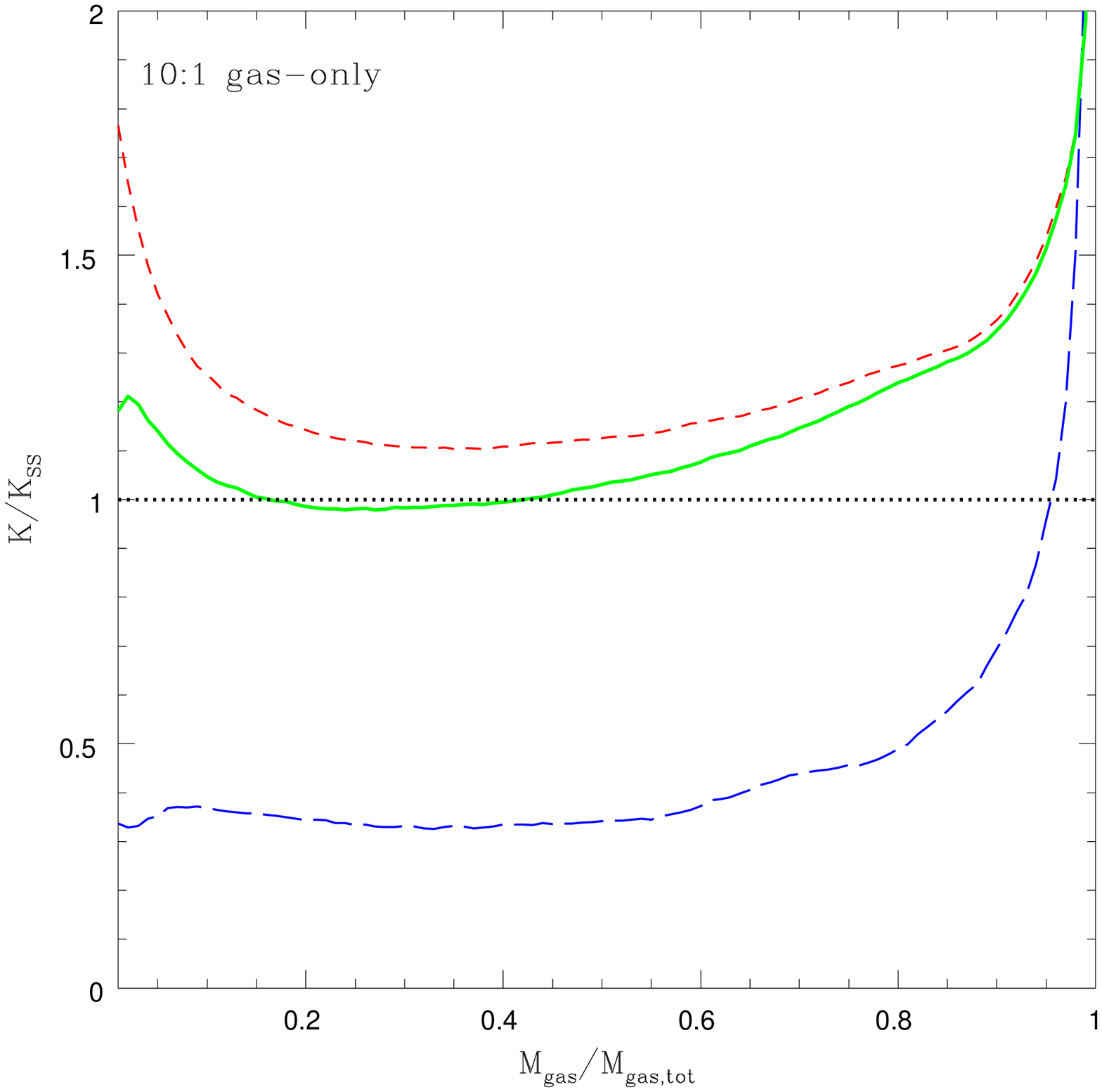}
\caption{Comparison of the resulting $K(M_{\rm gas})$
distributions from the 3:1 (left panel) and 10:1 (right
panel) gas-only merger with self-similar expectations. The
short-dashed red, long-dashed blue, and solid green curves
represent the primary, secondary, and total systems,
respectively.  The horizontal dotted line represents the
self-similar scaling.
 }
\end{figure*}

The entropy distributions of observed systems are most 
commonly presented in radial coordinates (e.g., Piffaretti et 
al.\ 2005; Donahue et al.\ 2006; Pratt et al.\ 2006), and so it is 
useful to present theoretical models in such units if the goal is 
to explain the properties of observed systems.  However, our 
immediate goal in this paper is to develop a physical shock 
heating model. For this purpose, the entropy distributions are 
best presented in Lagrangian (gas mass) coordinates.  In Figure 
5b, therefore, we plot the final $K(M_{\rm gas})$ 
distributions\footnote{The physical meaning of the $K(M_{\rm 
gas})$ distribution is most straightforwardly thought of in terms 
of its inverse, $M_{\rm gas}(K)$, which is the total mass of gas 
with entropy lower than $K$.} for the merged system and for the 
primary and secondary individually for the 1:1 gas-only merger.  
The gas mass coordinate has been normalised by the total gas mass 
in the systems (in this case $10^{15} M_\odot$ for the 
primary and secondary and $2\times10^{15} M_\odot$ for 
the final merged system).  The entropy coordinate has been scaled 
to the self-similar expectations; i.e., the initial $K(M_{\rm 
gas})$ distribution of the primary scaled up by a factor of 
$(M_{\rm tot}/M_p)^{2/3}$.  (Equivalently, we could use the 
initial entropy distribution of the secondary scaled up by a 
factor of $[M_{\rm tot}/M_s]^{2/3}$.)  Perhaps the simplest way 
to achieve this state is by having the primary and secondary 
systems individually obey self-similarity following the merger.  
In this case, we would expect the final distribution of the 
particles belonging to the primary to be $(M_{\rm 
tot}/M_p)^{2/3}$ times the initial distribution, while 
the secondary's final distribution would be a factor of 
$(M_{\rm tot}/M_s)^{2/3}$ larger than its initial distribution. 

If self-similarity is strictly obeyed, then the solid green curve 
in Figure 5b, which represents the final $K(M_{\rm gas})$ 
distribution for the merged system, should lie on the horizontal 
dotted line.  Fig.\ 5b shows that, without any fine tuning of 
the simulations, the central 80\% of the gas mass of the final 
merged system lies within approximately 10\% of the self-similar 
result.  As expected, the symmetry of the 1:1 case is such that 
both the primary and secondary individually obey self-similarity 
as well.

Beyond $M_{\rm gas}/M_{\rm gas,tot} \approx 0.8$ or so there is an 
apparent entropy excess relative to the self-similar expectation.  
Below, in \S 3.1.3, we demonstrate that this effect is 
artificial and its origin can be attributed to the (unrealistic) 
abrupt truncation of the gaseous atmospheres of our idealised 
systems at $r_{200}$.  Fortunately, as demonstrated below, this 
effect is limited to the outermost regions of our systems only.

The $K(M_{\rm gas})$ distributions for the 3:1 and 10:1 
gas-only mergers are plotted in Figures 6a and 6b, respectively.  
In both cases the final merged system is quite close to the 
self-similar result for the central $\sim70$\% of gas mass.  Beyond 
this the edge effect discussed below kicks in.  Thus, the 
3:1 and 10:1 cases both qualitatively and quantitatively resemble 
the 1:1 case.  Interestingly, however, the path through which the 
final merged systems in the 3:1 and 10:1 cases achieve 
self-similarity is by over-heating the primary and under-heating 
the secondary, not by both obeying self-similarity individually.  
This implies that some of the infall energy initially associated 
with the secondary system went into thermalising the gas of the 
primary system.  We return to this point later.

Finally, it is interesting in and of itself that the gas-only 
simulations obey self-similarity.  This implies that the reason the 
baryons trace the dark matter in cosmological hydrodynamic 
simulations is not simply because the baryons are just following 
the orders of the gravitationally dominant dark matter.  In the 
absence of dark matter the baryons would apparently still obey 
self-similarity.  We revisit this result in \S 5.4.

\subsubsection{Excess Entropy at Large Radii}

The resulting $K(M_{\rm gas})$ distributions of all our
idealised merger simulations (including the gas+DM mergers 
discussed below) show evidence for deviations from 
self-similarity beyond $M_{\rm gas}/M_{\rm gas,tot} \approx 0.8$ 
or so.  Is this effect real or artificial?  We hypothesise that 
this effect is artificial and is caused by the (unrealistic) 
abrupt truncation of our idealised systems at some finite 
radius.  In particular, one expects a SPH-based code to 
systematically underestimate the density of the gas particles 
near the edge of the truncated systems.  This, in turn, will have 
the effect of overestimating the entropy boost these particles 
receive in shock heating events.

To test the above, we try a simple modification to the 1:1 
gas-only merger simulation.  In particular, we extend the initial
gas profiles out to an overdensity of 100 (i.e., $r_{100}$) and
re-run the simulation.  The point is to see if initially 
extending the gaseous atmospheres to larger radii will have the 
effect of quenching the (over-efficient) entropy production 
in particles near $r_{200}$.  To make a fair comparison between 
the extended and default runs, in setting up the extended 
gas systems we choose the pressure at $r_{100}$ such that 
$P(r_{200})$ is identical to that in our default run.  This way 
we ensure that within $r_{200}$ the systems in both our default 
and extended runs are identical before being read into GADGET-2. 
An additional consideration is that of the infall velocity.  
Since the systems are more massive when extended (for our 
adopted concentration $M_{100} \approx 1.26 M_{200}$), the 
initial gravitational potential energy between the systems is 
larger (more negative) than that for our default run (despite 
the fact that the centres of the extended systems are initially 
separated by a larger distance; $d_0 = 2 r_{100}$).  Thus, if 
the infall velocity when the centres are separated by $2 
r_{200}$ is to be $v_{c,p}(r_{200})$ (as in the default run), a 
lower initial infall velocity is required.  We calculate this 
velocity by assuming the systems may be  approximated as point 
masses, which should be valid during the early stages of the 
merger.

\begin{figure}
\centering
\includegraphics[width=8.4cm]{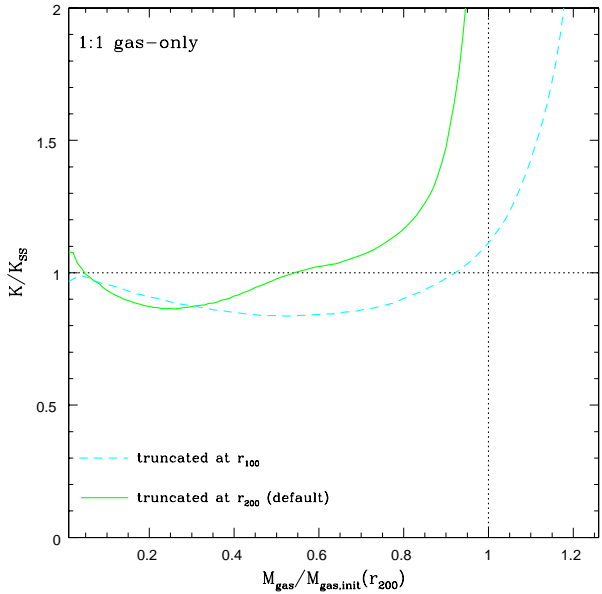}
\caption{Testing the origin of excess entropy at large radii.
Shown are the results of the 1:1 gas-only merger where the 
gaseous atmospheres have initially been truncated at $r_{200}$ 
(i.e., the default setup) and at $r_{100}$.  Extending the 
gaseous atmospheres to larger radii results in a more accurate 
density determination near $r_{200}$ which, in turn, mitigates 
the over-efficient entropy production seen at large radii in the 
default run.
 }
\end{figure}

In Figure 7, we compare the final $K(M_{\rm gas})$ distributions
for the merged systems (normalised to the self-similar result) in
the default and extended runs.  For the central 40\% of gas mass, 
the $K(M_{\rm gas})$ distributions are similar.  However, beyond 
this point differences become readily apparent.  Interestingly, 
for the run where the initial gas profiles were extended to 
$r_{100}$, the final $K(M_{\rm gas})$ distribution remains close 
to the self-similar result all the way out to $M_{200}$.  Beyond 
this point, a deviation from self-similarity is seen, as in the 
default run.  However, since within $M_{200}$ (or, equivalently, 
$r_{200}$) the systems in the extended and default runs are 
initially identical, we must conclude that the deviation from 
self-similarity seen at large in Fig.\ 5 (and similar figures) 
is indeed artificial and its origin is linked to a poor (SPH) 
estimate of the true gas density near the edges of the 
idealised systems.

Fortunately, the central regions of the default run are not 
significantly affected by how we treat the edge and therefore it 
is this region that we focus on when comparing to self-similar 
expectations or our analytic models.  In particular, when 
constructing physical analytic models that describe our 
simulations in \S 4, we restrict our focus to radii corresponding 
to $M_{\rm gas}/M_{\rm gas,tot} < 0.8$.

\begin{figure}
\centering
\includegraphics[width=8.4cm]{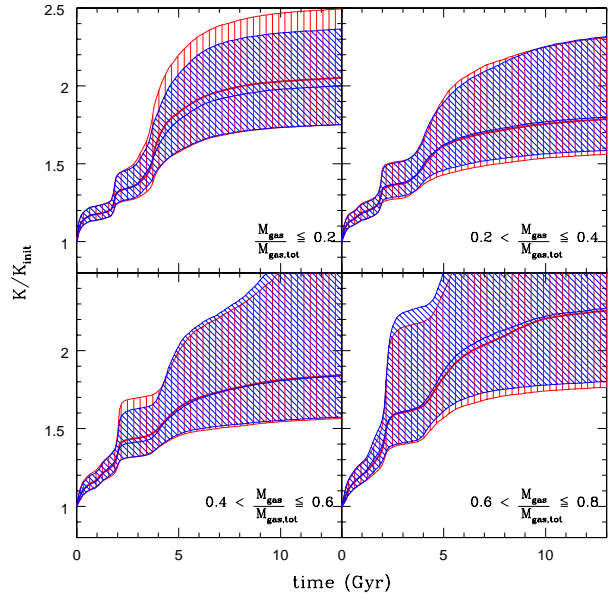}
\caption{Entropy evolution in the head on 1:1 gas+DM merger.
The red
and blue lines/hatched regions represent the primary and
secondary systems, respectively.  The upper and lower bounds of
the hatched regions represent the 75$^{\rm th}$ and 25$^{\rm th}$
percentiles, respectively, while the central (thick) lines
represent the median.  The panels show the evolution of the
entropy of particles separated according to their {\it initial}
position in the cumulative gas mass profile.  Since the
cumulative
gas mass profile is a monotonic function of radius, the panels
also
separate the particles according to initial distance from the
centre of their respective halos.
}
\end{figure}

\subsection{Gas + dark matter (gas+DM) simulations}

\subsubsection{Entropy evolution}

In Figure 8 we plot the evolution of the entropy with time for 
the 
head on 1:1 gas+DM simulation.  Several differences are 
apparent between the evolution in this figure and that of the 1:1 
gas-only simulation plotted in Figure 2.  For example, the gas+DM 
run shows evidence for a small amount of entropy production right 
from the start of the simulation, whereas the gas-only run does 
not (at least not within the central regions).  This difference 
may be attributed to the different methods used for constructing 
the initial conditions.  In an azimuthally averaged sense, both 
the gas-only and gas+DM runs have, for our purposes, nearly 
identical initial conditions.  However, because the initial gas 
particle positions in the gas+DM run were assigned based on the 
positions of dark matter particles (see \S 2.1.2), small local 
inhomogeneities are initially present in these runs.  As a result, 
the gas is not in perfect hydrostatic equilibrium and the systems 
undergo a slight re-adjustment when the simulation starts.  
Fortunately, the amount of entropy generated in this re-adjustment 
is small compared to the entropy produced in the merger 
shock heating and can be safely ignored.  This is demonstrated 
below when we make an explicit comparison of the gas-only and 
gas+DM results.

\begin{figure*}
\centering
\includegraphics[width=15cm]{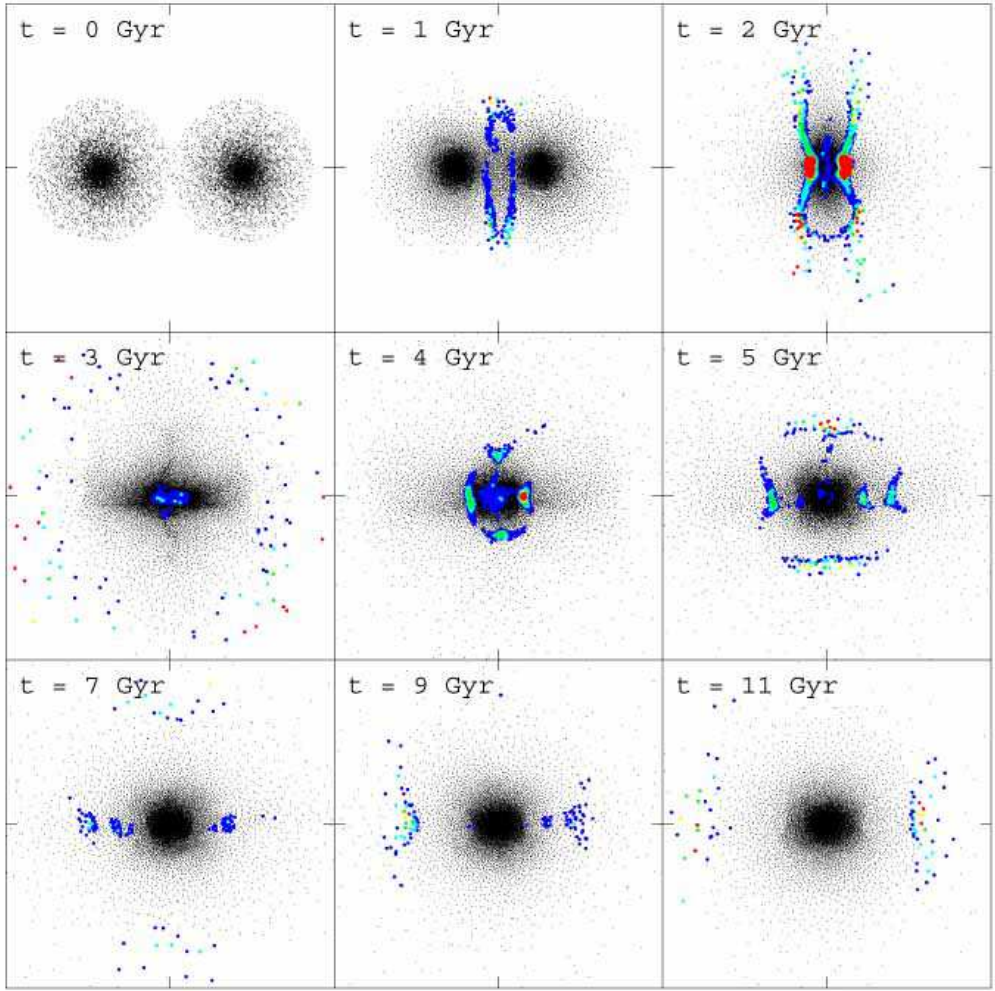}
\caption{
Fractional change in particle entropy as a function of
time for the head on 1:1 gas+DM merger.  Shown are particles in a
slice through the centre of thickness 250 kpc (i.e., $|z| < 125$
kpc).
Particles are colour coded according to the fractional change in
entropy since the last simulation output (0.1 Gyr ago).  Black
points are for a fractional change of less than 2\%, blue are for
a 2\%-10\% change, cyan are for a 10\%-20\% change, green are for
a 20\%-30\% change, yellow are for a 30\%-40\% change, and red
are
for a $>$ 40\% change.  For clarity the surrounding
pressure-confining medium is not displayed.  Each panel is 10 Mpc
on a side.
}
\end{figure*}

The most important changes between Figs.\ 2 and 8 relate 
to the properties of the second (proper) period of entropy 
generation.  In particular, the second phase of heating, which was 
associated with a series of re-accretion shocks in the gas-only 
runs, begins earlier, rises more steeply, and contributes more to 
the final state of the system (particularly in the central 
regions) in the gas+DM run than in the gas-only run.  What is the 
origin of this behaviour?

To help answer this question, we plot in Fig.\ 9 a series of 
snapshots colour coded according to the fractional change in 
entropy since the last simulation output (see Fig.\ 3 for the 
analogous plot for the gas-only run).  The evolution is quite 
similar to that of the gas-only run early on.  However, noticeable 
differences are present between the two at intermediate times ($t 
\sim 4-5$ Gyr).  In particular, in the gas+DM run it is apparent 
that a second fast moving shock wave has been generated.  This 
shock wave propagates outward in a nearly spherical fashion and 
resembles that of the first shock created when the cores 
collided.  The origin of this second shock can be linked to the 
collisionless nature of the dark matter, which allows the dark matter
cores to pass through one another, while the gas cores collide. 
Since the dark matter dominates the gas by mass, and the dark core
is able to drag significant quantities of gas into the other 
hemisphere and away from the overall system's centre of mass.  As 
a result, some of the gas undergoes a second period of collapse, 
collides with gas infalling from the other hemisphere, and produces 
the shock\footnote{In fact, the dark matter cores oscillate back and 
forth a number of times, each time dragging gas away from the 
overall system's centre and consequently generating more shocks.  
However, these additional shocks are extremely weak and generate 
virtually negligible amounts of entropy.}.  It is demonstrated 
later that the extra energy required for this second shock is 
derived from tapping the dark matter component (see, e.g., Fig. 
21).  
Following this second shock, there is an extended period of 
re-accretion that proceeds in a similar fashion to that in the 
gas-only run.  It is the combination of this second shock and 
the re-accretion phase that is responsible for the increased 
importance of the second major phase of entropy production in 
Fig.\ 8.

The head on 3:1 and 10:1 gas+DM mergers also show the same 
qualitative trends.  But the head on case is special and, 
therefore, the final state of such collisions may not be 
representative of typical mergers in cosmological simulations.  
On the other hand, cosmological simulations suggest that 
groups and clusters acquire the bulk of their mass by substructure 
falling in along filaments.  So the head on scenario will not be 
as far removed from the typical merger event as in the case of 
collisions between galaxies.  However, it is still important to 
quantify the differences between the head on and non-zero impact 
parameter cases.

In Figures 10-15 we compare snapshots showing the spatial
distributions of the primary and secondary particles and also 
the fractional change in entropy since the last simulation output 
for the all of the gas+DM merger simulations.  We omit the panel 
showing the initial particle positions at $t = 0$, since they 
are the same for the three different orbital cases (thus, the 
energy to be thermalised is the same for all three cases).  
In these figures, the centre of each panel corresponds to the 
overall centre of mass, with the secondary system initially 
approaching from the left and the primary from the right.  In 
the non-zero impact parameter runs, the secondary's initial 
motion is towards the upper right, while the primary is 
initially moving towards the lower left.

\begin{figure*}
\centering
\includegraphics[width=13.5cm]{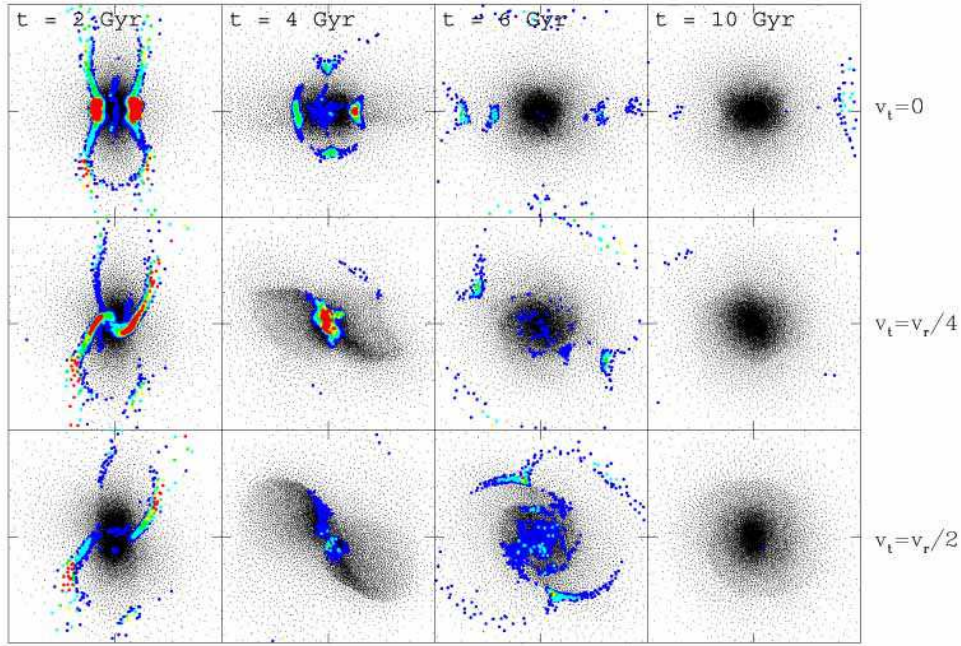}
\caption{
Fractional change in particle entropy as a function of
time for the 1:1 gas+DM mergers.  {\it Top:} Head on case.  {\it 
Middle:} Small impact parameter case.  {\it Bottom:} Large impact 
parameter case.  Particle colour 
coding is the same as in Figure 9.  Each panel is 6 Mpc on a 
side.}
\end{figure*}

\begin{figure*}
\centering
\includegraphics[width=13.5cm]{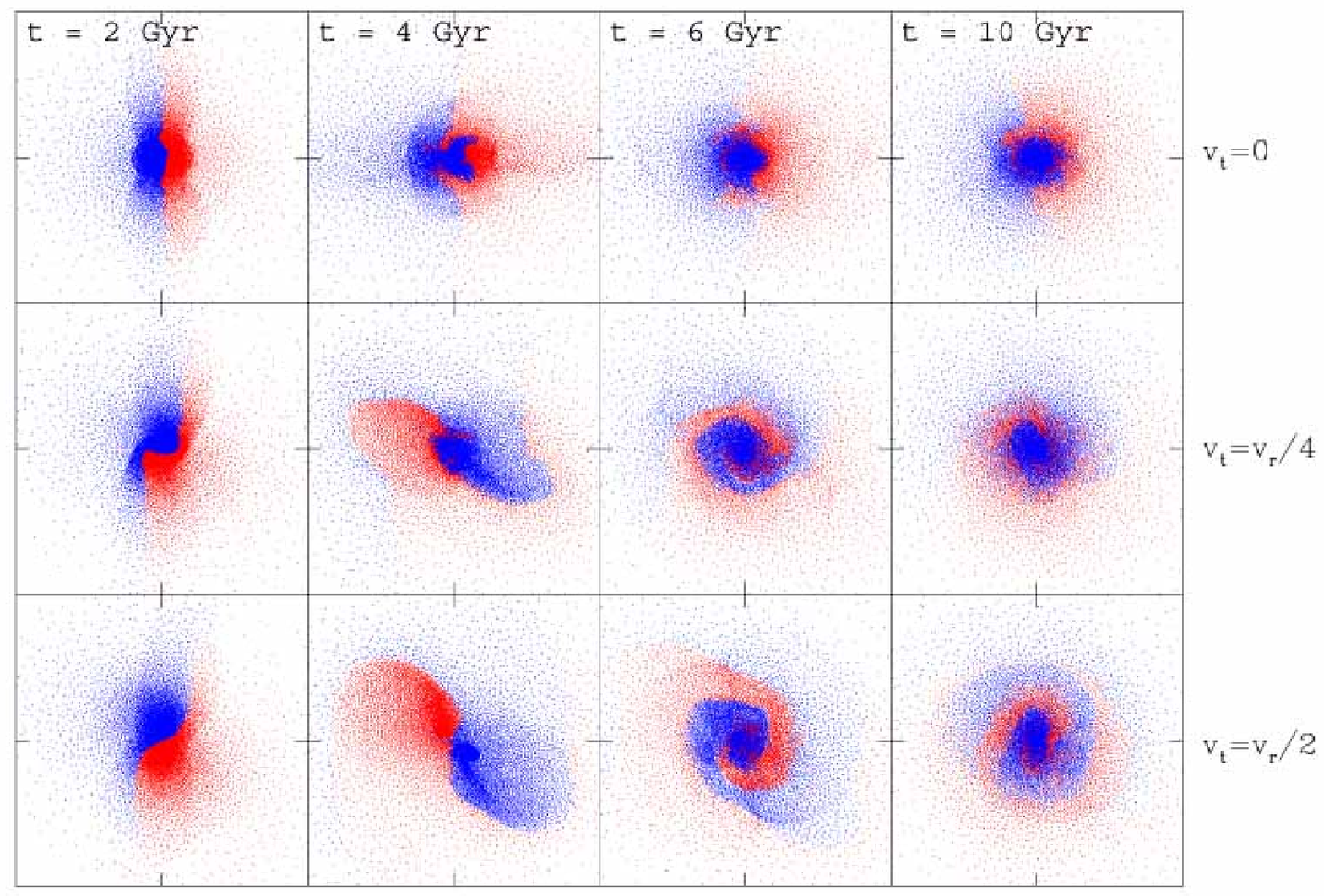}
\caption{Spatial distributions of particles originally belonging 
to the primary (red) and secondary (blue) systems for the 1:1 
gas+DM mergers.  {\it Top:} Head on case.  {\it Middle:} Small 
impact parameter case.  {\it Bottom:} Large impact parameter case. 
For clarity, the secondary particles are plotted on top of the 
primary particles.  Each panel is 6 Mpc on a side.
 }
\end{figure*}

\begin{figure*}
\centering
\includegraphics[width=13.5cm]{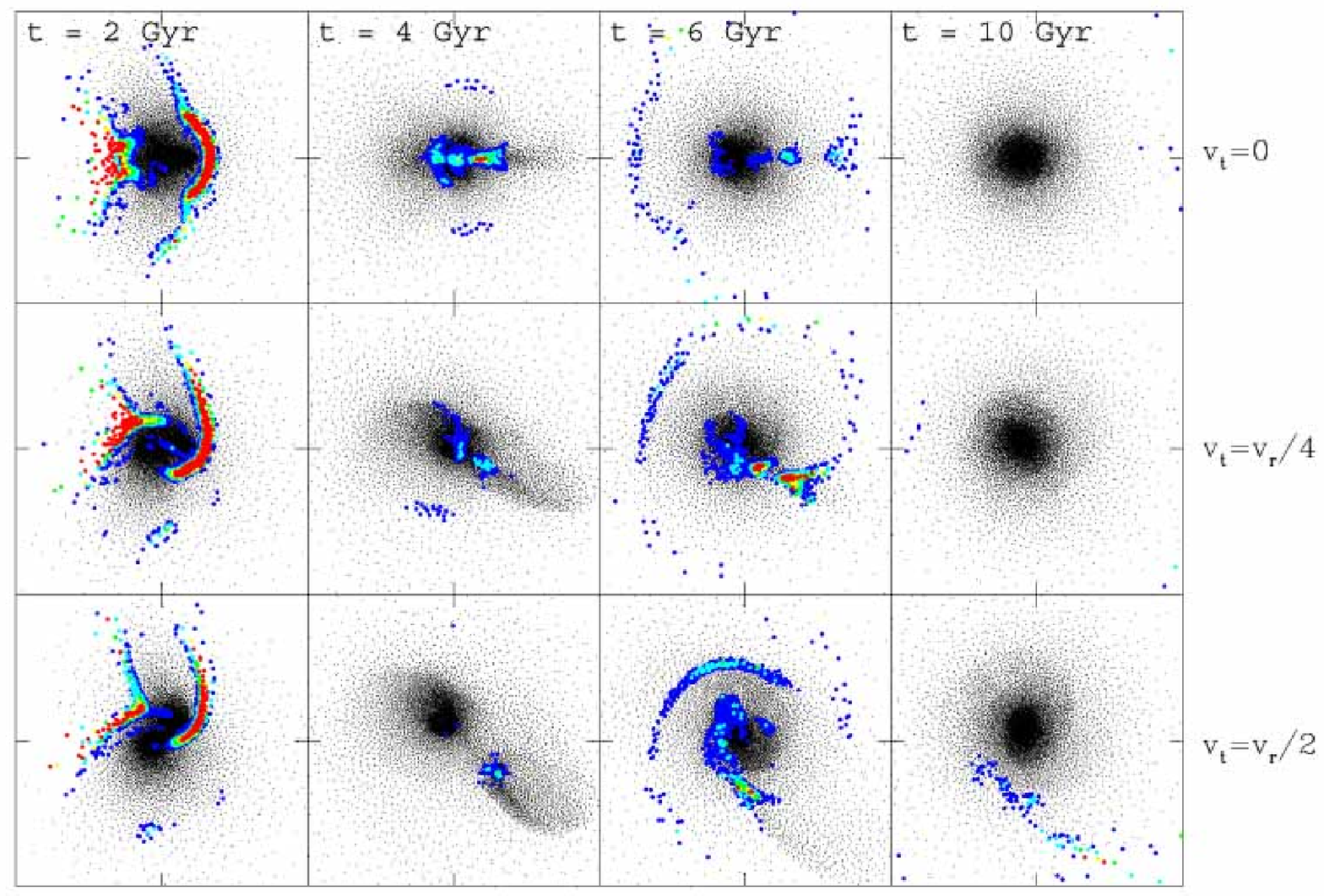}
\caption{
Fractional change in particle entropy as a function of
time for the 3:1 gas+DM mergers.  {\it Top:} Head on case.  {\it
Middle:} Small impact parameter case.  {\it Bottom:} Large impact
parameter case.  Particle colour
coding is the same as in Figure 9.  Each panel is 6 Mpc on a
side.}
\end{figure*}

\begin{figure*}
\centering
\includegraphics[width=13.5cm]{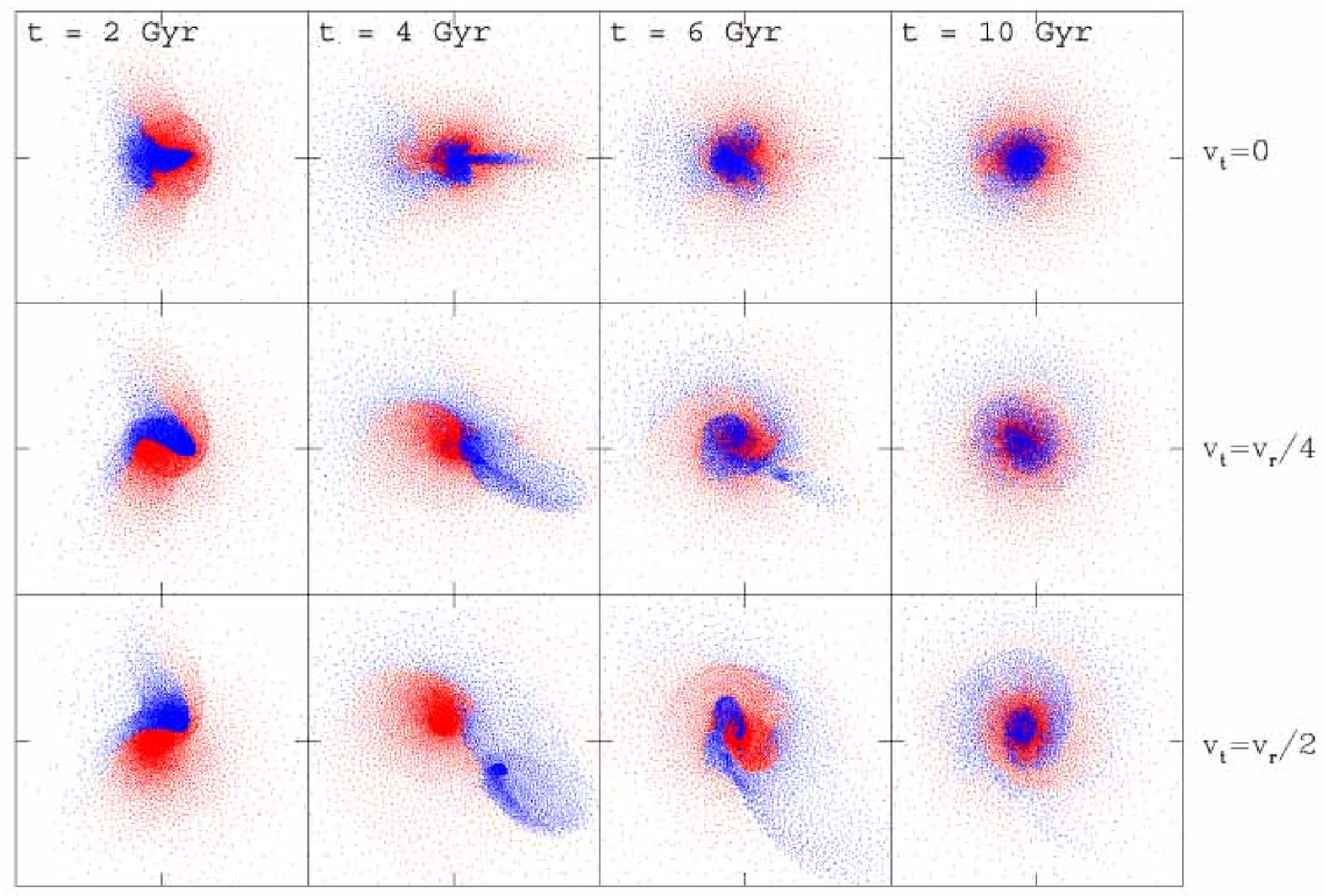}
\caption{Spatial distributions of particles originally belonging
to the primary (red) and secondary (blue) systems for the 3:1
gas+DM mergers.  {\it Top:} Head on case.  {\it Middle:} Small
impact parameter case.  {\it Bottom:} Large impact parameter case.
For clarity, the secondary particles are plotted on top of the
primary particles.  Each panel is 6 Mpc on a side.}
\end{figure*}

\begin{figure*}
\centering
\includegraphics[width=13.5cm]{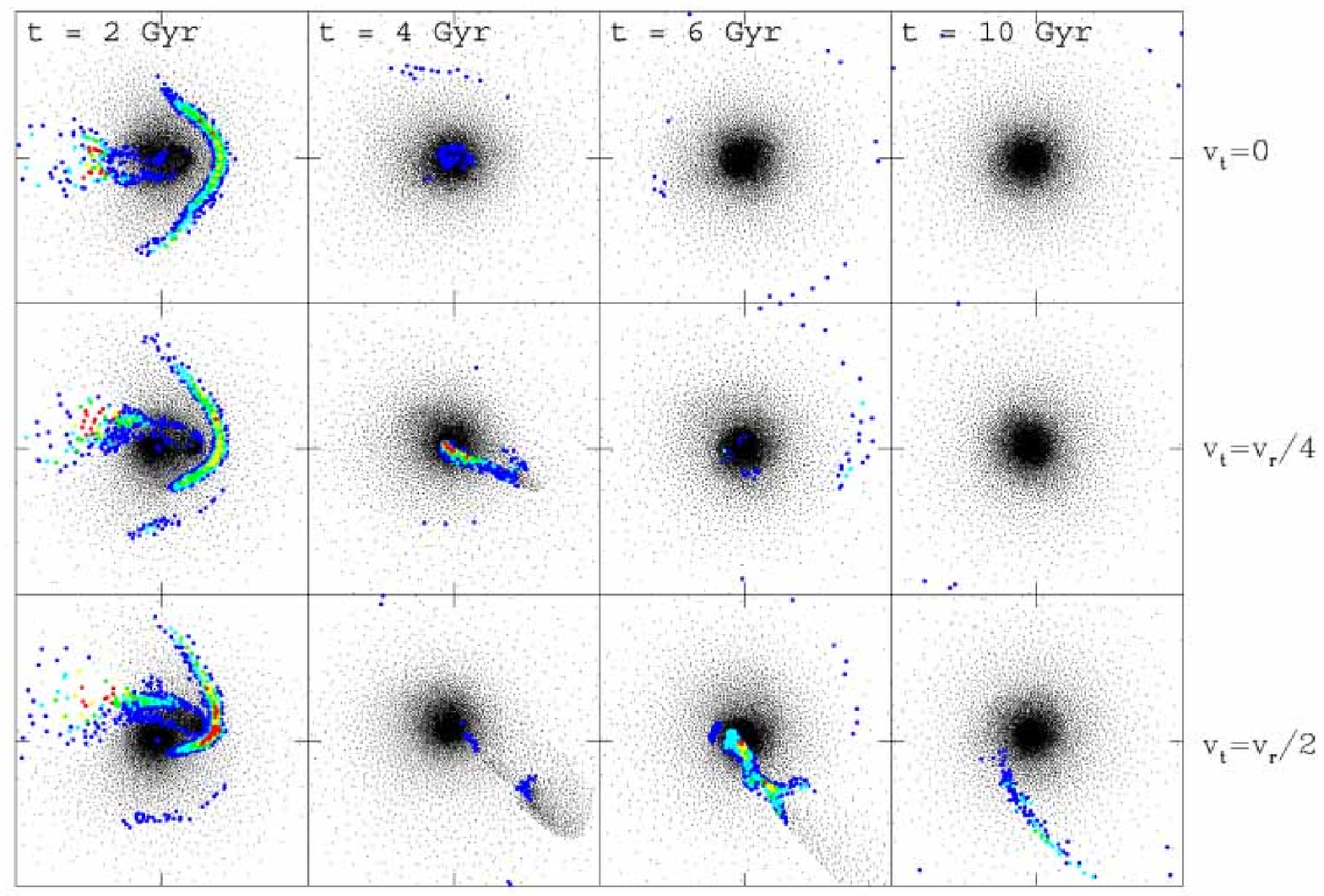}
\caption{Fractional change in particle entropy as a function of
time for the 10:1 gas+DM mergers.  {\it Top:} Head on case.  {\it
Middle:} Small impact parameter case.  {\it Bottom:} Large impact
parameter case.  Particle colour
coding is the same as in Figure 9.  Each panel is 6 Mpc on a
side.}
\end{figure*}

\begin{figure*}
\centering
\includegraphics[width=13.5cm]{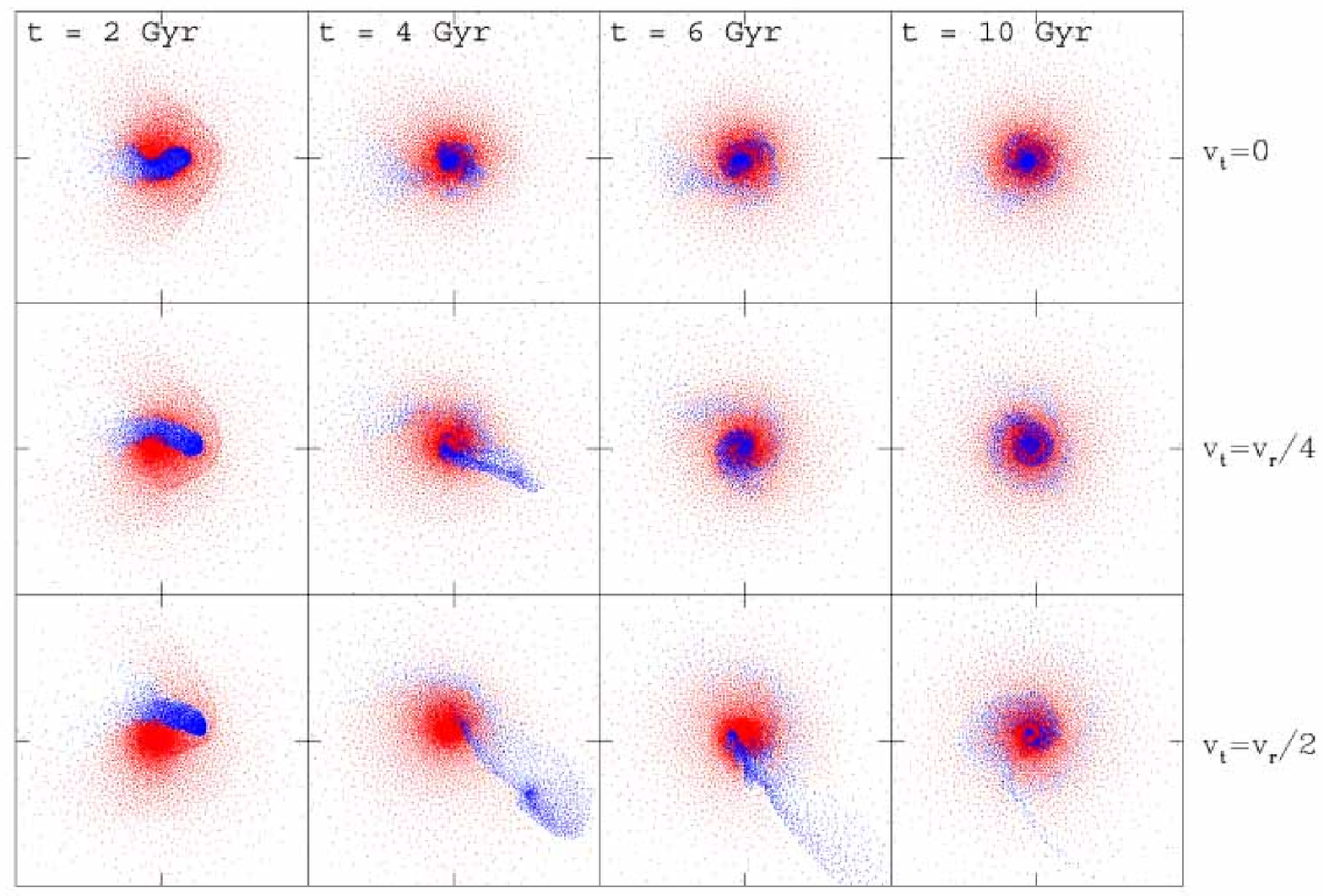}
\caption{
Spatial distributions of particles originally belonging
to the primary (red) and secondary (blue) systems for the 10:1
gas+DM mergers.  {\it Top:} Head on case.  {\it Middle:} Small
impact parameter case.  {\it Bottom:} Large impact parameter case.
For clarity, the secondary particles are plotted on top of the
primary particles.  Each panel is 6 Mpc on a side.}
\end{figure*}

The general progressions of the 1:1 non-zero impact parameter 
cases, presented in Figs.\ 10 and 11, are as follows.  In the 
small impact parameter case (where $v_t = 
v_r/4$ initially) the cores graze each other after $t \approx 2$ 
Gyr 
of infall.  As a result, the cores are temporarily spared from 
significant shock heating.  However, the systems are 
gravitationally bound and cannot avoid a major collision for long.
In short order the cores cease moving apart and begin falling 
inward nearly radially, setting up an almost head on secondary 
collision (at $t \approx 4$ Gyr) that greatly heats the core gas.  
Additional small shocks are generated by the back and forth 
sloshing of the dark matter cores, but in general their 
contribution to the entropy production is minor.  
As in the cases examined above, an extended period of 
re-accretion then ensues.  In this case, the bulk of the 
accretion first happens preferentially along and perpendicular 
to the axis of the secondary (near head on) collision. 
Qualitatively, the large impact parameter case proceeds much the 
same way.  The main differences are as follows.  The larger impact 
parameter means that the gas cores go virtually unheated during 
the first pericentric passage, as the cores of the two systems 
completely miss one another.  The larger impact parameter also 
implies the amount of time spent between the first and second 
pericentric passages will be longer.  Furthermore, the secondary 
collision is not directly head on in this case, meaning that some 
of the gas actually undergoes a third pericentric passage before 
settling.  In both the small and large impact parameter cases, 
significant angular momentum is imparted to the gas and large, 
nearly 
circular, bulk motions remain present until the end of the 
simulations.

Figs.\ 12-15 are completely analogous plots for the 3:1 and 
10:1 cases.  In both head on collisions, the secondary essentially 
acts like a bullet, driving a large shock into the gas of the 
primary system and is able to easily penetrate all the way to 
the core of the primary.  In fact, unlike the head on 1:1 case, 
the core of the secondary in the 3:1 and 10:1 cases is able to 
remain somewhat intact even after a collision with the core of 
the primary.  As a result, there is actually a period where the 
cores pass through one 
another.  However, the tidal forces exerted on the secondary's 
core significantly stretch it along the collision axis.  A second 
head on collision between the cores of the two systems then occurs.  
At the same time, the sloshing back and forth of the dark matter 
cores is generating small shocks.  The small and large impact 
parameter 3:1 and 10:1 cases show evidence for even more 
complicated behaviour, with multiple collisions occurring and 
shock waves propagating in various directions simultaneously.

\begin{figure*}
\centering
\leavevmode
\epsfysize=8.4cm
\epsfbox{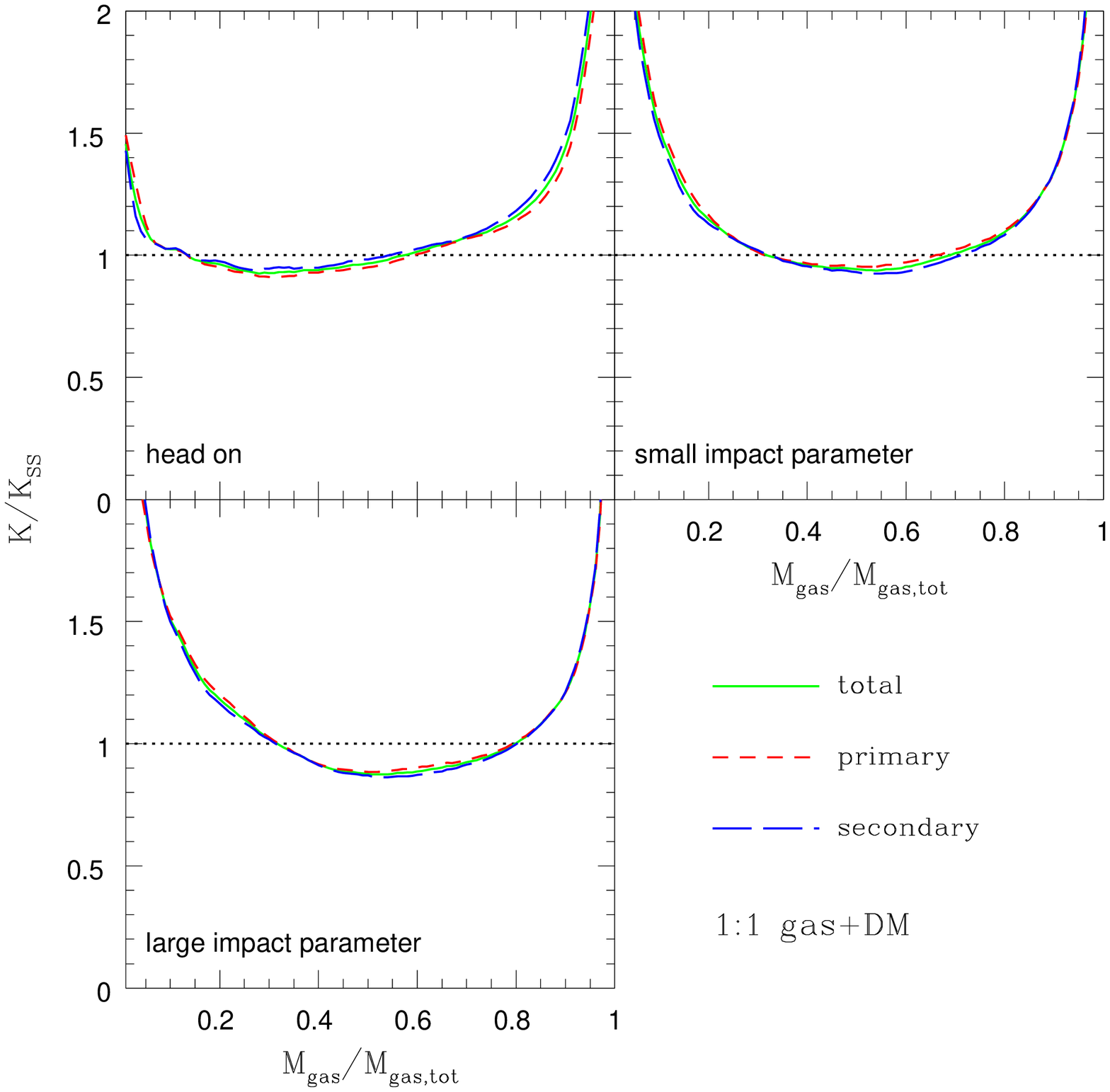}
\epsfysize=8.4cm \epsfbox{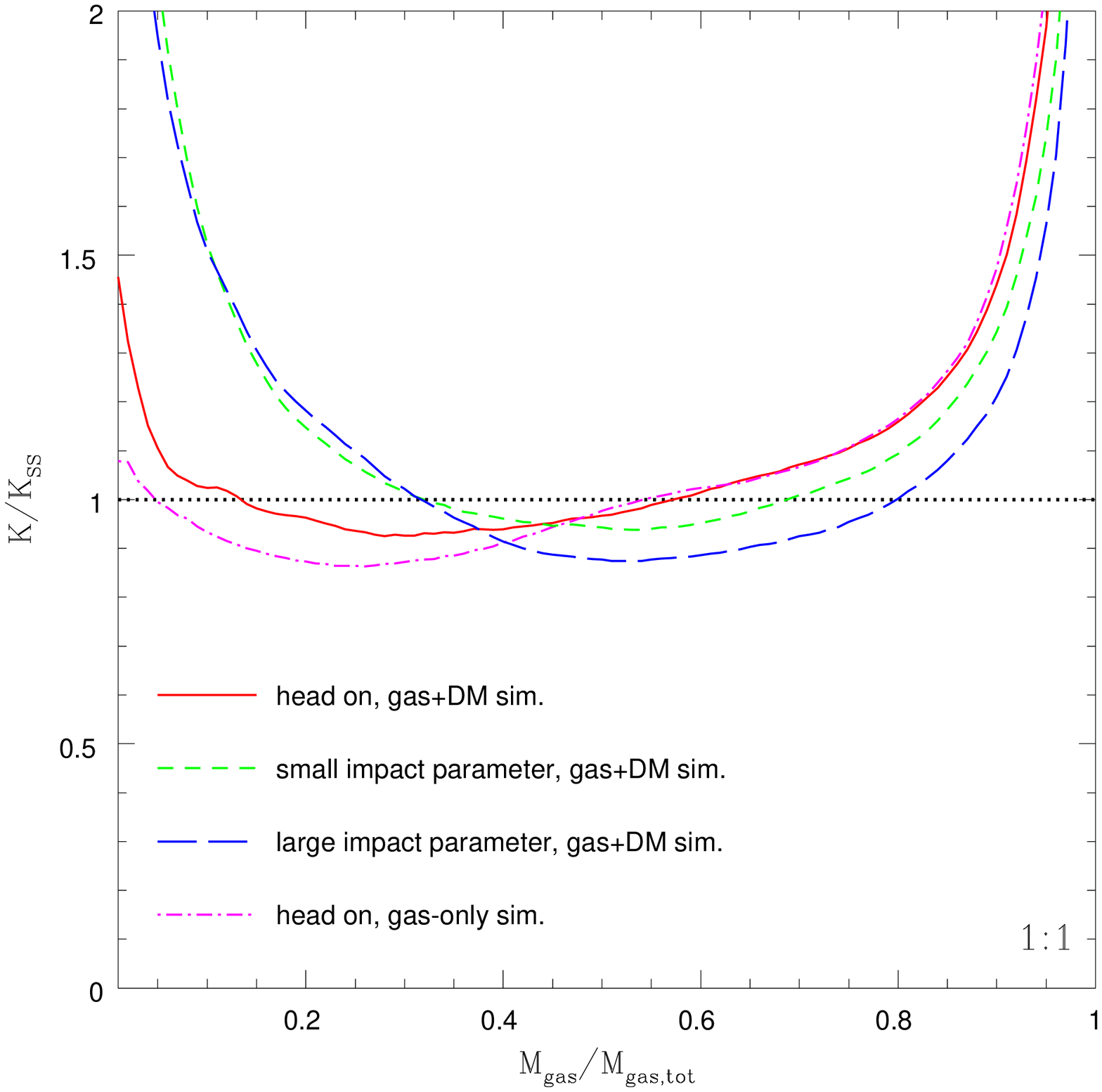}
\caption{The resulting $K(M_{\rm gas})$ distributions for the 1:1
gas+DM mergers.  {\it Left:} Comparison of the 
final entropy distributions for the primary (short-dashed red), 
secondary (long-dashed blue), and total (solid green) systems for 
the three different orbital cases.  {\it Right:}  Comparison of 
the final entropy distributions of the total systems in the gas+DM 
mergers for the head on (solid red), small impact parameter 
(short-dashed green), and large impact parameter (long-dashed 
blue) systems and the gas-only head on merger (dot-dashed 
magenta).  The horizontal dotted line represents the self-similar 
scaling.
 }
\end{figure*}

\subsubsection{Final configurations}

Given the wide variety of behaviours seen in Figs.\ 10-15, one 
might reasonably expect that the resulting gas properties of the 
various mergers to be significantly different from one another.  
But an analysis of the final entropy distributions indicates 
otherwise.  Plotted in Figures 16a,b are the 
final $K(M_{\rm gas})$ distributions for the various 1:1 mergers.  
In Fig.\ 16a, it is demonstrated that the bulk of the gas in the 
primary, secondary, and total systems nearly maintain
self-similarity.  However, it is apparent that the small and 
large impact parameter cases show evidence for excess heating 
of the core gas relative to the head on case.  This is illustrated 
most clearly in Fig.\ 16b, which compares the $K(M_{\rm gas})$ 
distributions for the total systems of the three 
different gas+DM simulations.  This plot shows that the cases with 
non-zero impact parameter heat the core at the expense of the 
outer regions of the systems.  However, it is remarkable that only 
the central $\sim$20\% of the gas mass show significant deviations 
from the head on case given the large modifications to the initial 
orbital parameters.  It is worth noting that non-radiative 
cosmological simulations also show evidence that the gas departs 
from a pure hydrostatic NFW profile in cores of systems (e.g., 
Frenk et al.\ 1999; Voit et al.\ 2005).  So some deviation from 
self-similarity in our simulations might be expected.

Also plotted in Fig.\ 16b is the $K(M_{\rm gas})$ distribution 
for 
the 1:1 gas-only merger.  Outside the central regions, it traces 
the distribution of the head on gas+DM simulation remarkably well.  
The two distributions begin to deviate from one another for 
$M_{\rm gas}/M_{\rm gas,tot} < 0.4$ or so.  The excess heating in 
the gas+DM can plausibly be attributed to energy exchange between 
the gas and the dark matter (see, e.g., Lin et al.\ 2006).  In \S 4, 
we show that typically 5-10\% of the dark matter's energy is 
transferred to gas.  This energy exchange occurs during the period 
when the dark matter cores of the primary and secondary are 
oscillating back and forth.

In Figures 17a,b we compare the final $K(M_{\rm gas})$ 
distributions for the various 3:1 mergers.  Fig.\ 17a shows that 
the bulk of the gas for the total systems in all three orbit cases 
nearly obey self-similarity.  As in the gas-only mergers discussed 
in \S 3.1, self-similarity is achieved by over-heating the primary 
and under-heating the secondary.  Similar to the 1:1 gas+DM 
mergers discussed immediately above, the 3:1 mergers also show 
evidence for departures from self-similarity for the central 20\% 
or so of the total gas mass.  A comparison of the final 
distributions for the total systems in Fig.\ 17b illustrates this 
more clearly.  Interestingly, 
even the head on 3:1 case shows evidence for strong central 
heating.  In fact, the head on case appears to be even slightly 
more efficient at heating the core than the case characterised by 
a large impact parameter. (But we note that in the large impact 
parameter case there is still some residual heating occurring at 
the end of the simulation, so in the long run it may be the more 
efficient of the two.)  A comparison of the entropy distributions 
of the head on gas+DM and gas-only mergers again highlights the 
fact that the two differ from each other only in the very central 
regions.

Finally, in Figures 18a,b we show the same set of plots for the 
10:1 gas+DM mergers.  The trends in these plots follow those 
discussed above for the 1:1 and 3:1 mergers.  The only difference 
that we would mention is that the 10:1 cases exhibit a much lesser 
degree of central heating than the simulations discussed 
previously.

\begin{figure*}
\centering
\leavevmode
\epsfysize=8.4cm
\epsfbox{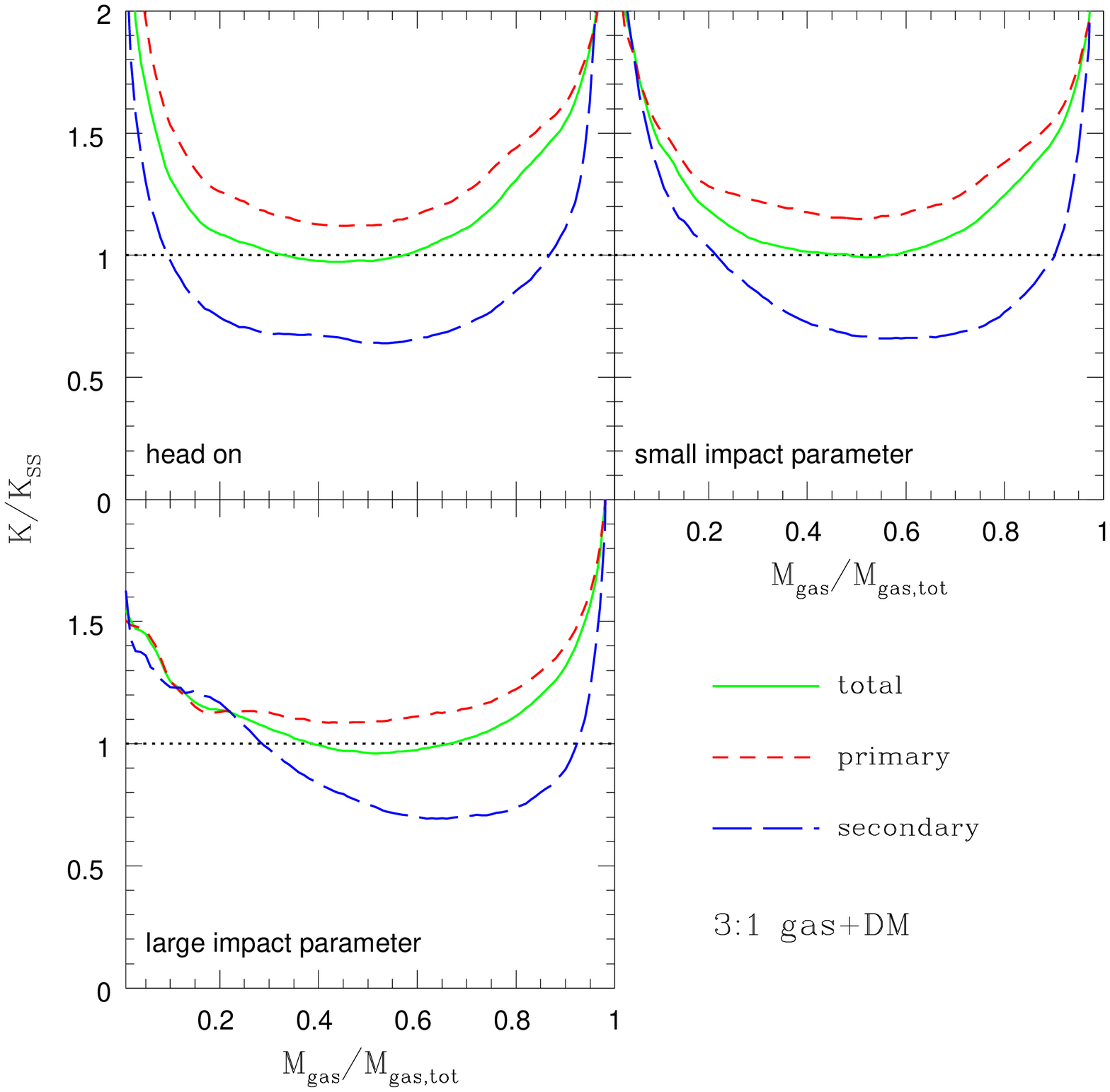}
\epsfysize=8.4cm \epsfbox{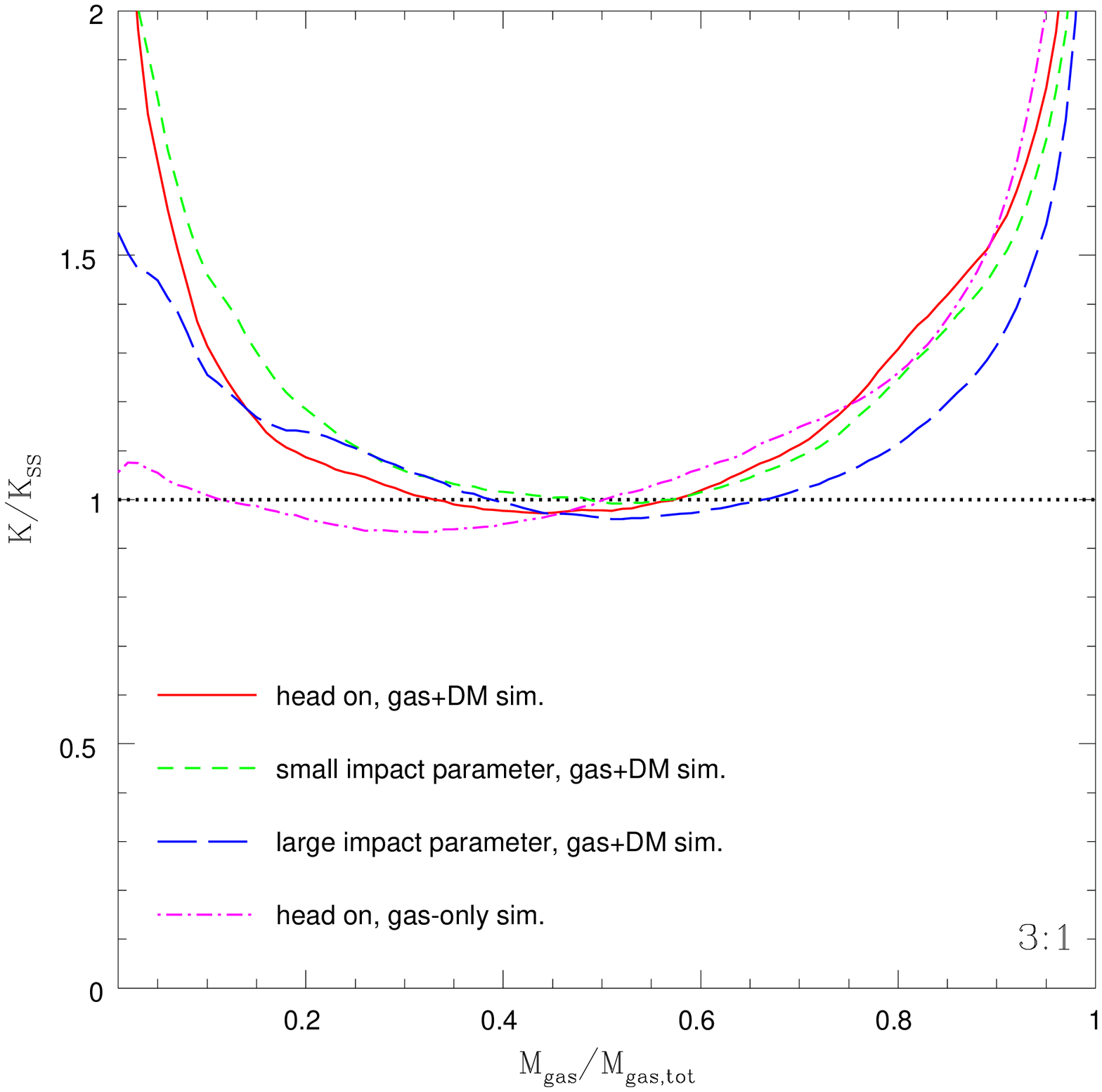}
\caption{
The resulting $K(M_{\rm gas})$ distributions for the 3:1
gas+DM mergers.  {\it Left:} Comparison of the
final entropy distributions for the primary (short-dashed red),
secondary (long-dashed blue), and total (solid green) systems for
the three different orbital cases.  {\it Right:}  Comparison of
the final entropy distributions of the total systems in the gas+DM
mergers for the head on (solid red), small impact parameter
(short-dashed green), and large impact parameter (long-dashed
blue) systems and the gas-only head on merger (dot-dashed
magenta).  The horizontal dotted line represents the self-similar 
scaling.
}
\end{figure*}

\begin{figure*}
\centering
\leavevmode
\epsfysize=8.4cm
\epsfbox{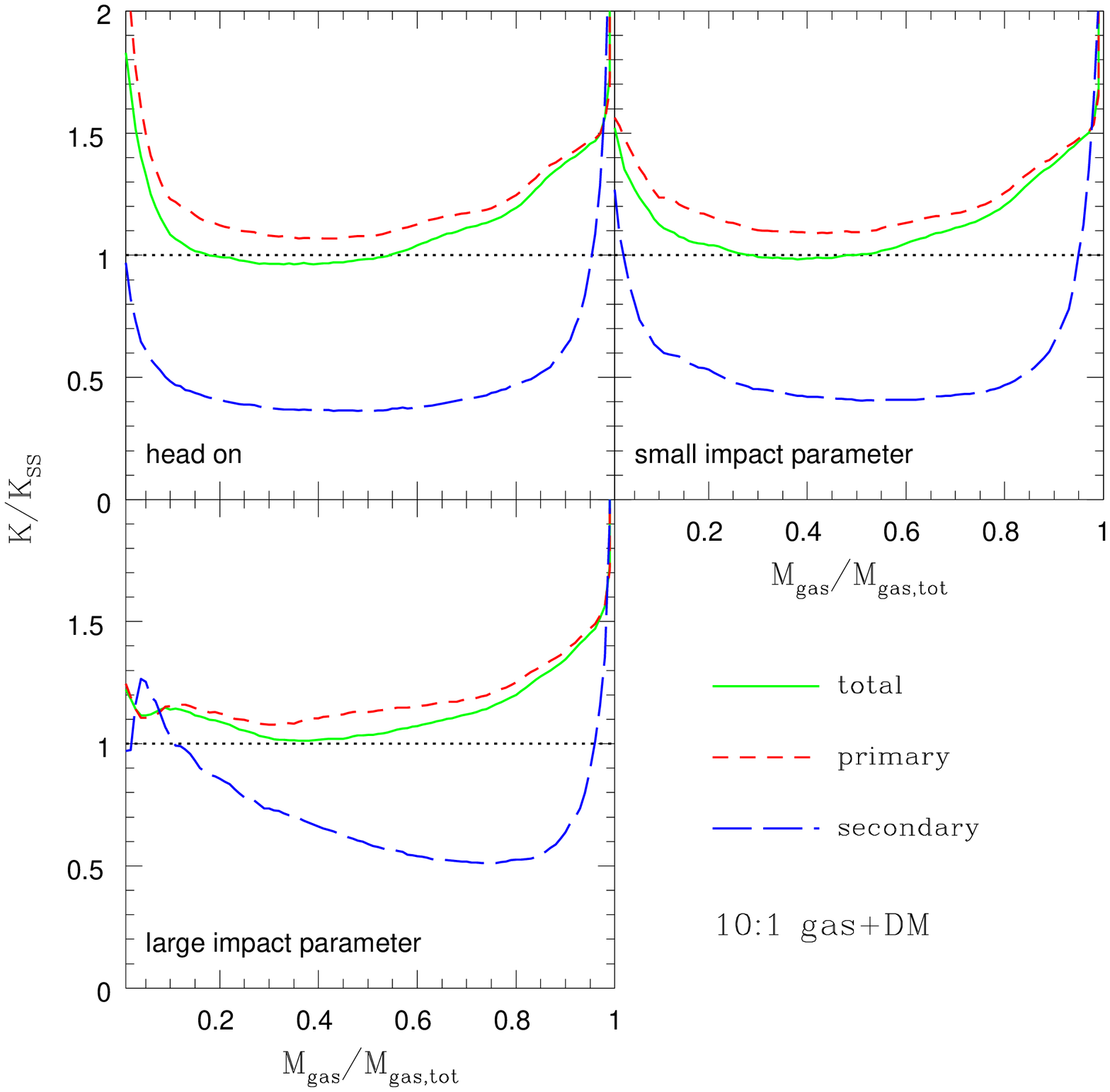}
\epsfysize=8.4cm \epsfbox{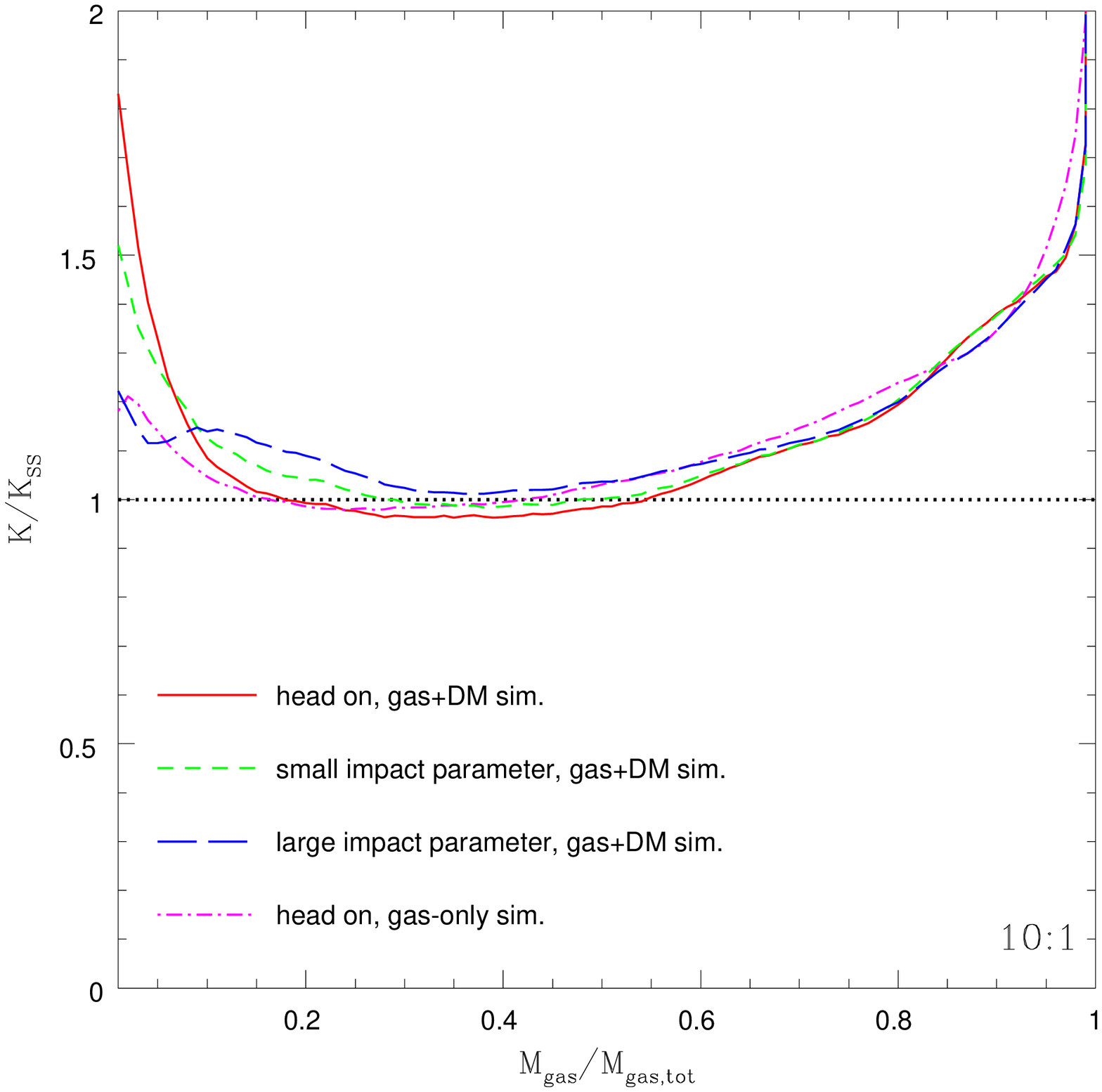}
\caption{
The resulting $K(M_{\rm gas})$ distributions for the 10:1
gas+DM mergers.  {\it Left:} Comparison of the
final entropy distributions for the primary (short-dashed red),
secondary (long-dashed blue), and total (solid green) systems for
the three different orbital cases.  {\it Right:}  Comparison of
the final entropy distributions of the total systems in the gas+DM
mergers for the head on (solid red), small impact parameter
(short-dashed green), and large impact parameter (long-dashed
blue) systems and the gas-only head on merger (dot-dashed
magenta).  The horizontal dotted line represents the self-similar 
scaling.
}
\end{figure*}

\subsection{Summary of Simulation Results}

In the next section we develop a physically-motivated analytic 
model that attempts to encapsulate the key physics of the merging 
process just described.  However, before presenting this model we 
summarise the results of our simulations:

\begin{itemize}
\item{Both the primary and secondary systems gain the bulk of their 
final entropy through two distinct episodes.  The first episode is
associated with a strong, quickly propagating shock wave that is
generated approximately when the cores of the two systems collide.
The second episode of heating occurs over an extended period of
time.  In the gas-only simulations, this second phase is driven 
entirely by a series of re-accretion shocks.  In the gas+DM 
simulations, it is driven by a combination of re-accretion shocks 
and energy transfer from the dark matter to the gas, with the
re-accretion shocks playing the dominant role.}  
\item{In all of our simulations, the contributions of the first and 
second episodes of entropy production to the final distribution 
are comparable.}
\item{We find that the bulk of the gas in our simulations matches 
the self-similar result to within 10\%.  Deviations from 
self-similarity are seen at large radii in both the gas-only and 
gas+DM simulations.  This is almost certainly due to a truncation 
effect in our idealised setup (see \S 3.1.3).  Deviations from 
self-similarity are also seen at small radii in the gas+DM 
simulations (which are real) and are at least partially due to 
energy exchange between the gas and dark matter.  Some deviation 
might be expected at small radii, as the gas in systems formed in 
non-radiative cosmological simulations also show departures there 
from a pure hydrostatic NFW distribution.
 }
\item{With the obvious exception of the symmetric 1:1 case, 
self-similarity of the final merged system is achieved by 
over-heating the gas in the primary while under-heating the gas in 
the secondary.  This implies that some of the infall energy 
initially associated with the secondary has gone into heating the 
primary.
}
\end{itemize}

Our simulations therefore differ markedly from the 
standard spherical smooth accretion model where the ICM is built 
up one shell at a time from the inside out and where each shell is 
shocked a single time as it enters the virial radius.  It seems 
remarkable that these two very physically different scenarios yield 
structural properties that are as similar as they are.  However, as 
discussed in detail by Voit et al.\ (2003), the spherical accretion 
model fails to match the results of cosmological 
simulations when it is modified to account for more realistic 
``lumpy'' accretion.  The reason for this failure is that 
increasing the density but keeping the total energy to 
be thermalised fixed results in decreased entropy production in the 
shock.  However, our simulations show that significant 
shock heating does not occur until core collision.  As a result, 
there is significantly more infall energy available for 
thermalisation.  To a large extent, this boost offsets the 
otherwise reduced level of entropy owing to the increased density 
of infalling gas in the lumpy model relative to the smooth 
accretion model.

\section{Discussion: Understanding the Results}

\subsection{A Simple Single-Shock Model}

We have demonstrated that our idealised merger simulations 
approximately preserve self-similarity, as seen in non-radiative 
cosmological 
simulations.  As a result, we are now in a position to try to 
develop a physical analytic model for the entropy 
evolution of the primary and secondary systems in our idealised 
simulations.  Taking our cue from the study of Voit et al. (2003), 
we consider a simple model whereby gas with some initial density 
$\rho_1$ and entropy $K_1$ is moving with a velocity 
$v_{\rm in}$ in the system's centre-of-mass frame.  If this velocity 
is supersonic it will generate a shock front.  Assuming that the 
post-shock gas is at rest in the centre-of-mass frame 
(i.e., that all of the energy associated with $v_{\rm in}$ 
is thermalised), then the shock propagates with a velocity 
$v_{\rm shock}$ into the gas in the centre-of-mass frame.  In the 
rest-frame of the shock, the gas therefore has a velocity $v_1 = 
v_{\rm in} + v_{\rm shock}$, and the post-shock gas has a 
velocity $v_2 = v_{\rm shock}$.  The 
pre-shock (upstream) conditions are related to the post-shock 
(downstream) conditions via the well-known Rankine-Hugoniot jump 
conditions (e.g., Shu 1992):

\begin{equation}
\frac{\rho_2}{\rho_1} = \frac{v1}{v2} =  
\frac{(\gamma+1){\cal M}_1^2}{(\gamma-1){\cal M}_1^2+2},
\end{equation}
\begin{equation}
\frac{P_2}{P_1} = 1+\frac{2\gamma({\cal M}_1^2-1)}{\gamma+1}
\end{equation}
\begin{equation}
\frac{T_2}{T_1} = 
\frac{[1+\gamma(2 {\cal M}_1^2-1)][2+(\gamma-1){\cal 
M}_1^2]}{(\gamma+1)^2{\cal M}_1^2}
\end{equation}

\noindent where ${\cal M}_1 \equiv v_1/c_s$ is the Mach number.  
Equations (8) and (9) can be used to yield the jump condition 
relating pre-shock and post-shock entropy:

\begin{equation}
\frac{K_2}{K_1} = \biggl[\frac{5{\cal M}_1^2-1}{4} \biggr]
\biggl[\frac{4{\cal M}_1^2}{{\cal M}_1^2+3} \biggr]^{-5/3}
\end{equation}

\noindent where we have used $\gamma = 5/3$.  Therefore, one can 
calculate the final entropy distribution if the Mach number of the 
shock is known.  Unfortunately, it is non-trivial to measure the 
Mach number of a shock directly from the simulations.  Shock 
heating is implemented in SPH simulations via an artificial 
viscosity term which significantly broadens the shocks both 
spatially and temporally.  This prevents one from easily applying 
equations (8)-(11) to individual SPH particles (see Pfrommer et 
al. 2006).  Another difficulty is that the Mach number is expressed 
in terms of pre-shock velocity in the rest-frame of the shock.  So 
application of these equations to individual particles would 
require one to carefully track the evolution of the shock itself 
during the simulation.  To avoid these difficulties, we use 
equation (8) to instead express the Mach number in terms of the 
difference between the pre-shock and post-shock velocities:

\begin{equation}
v_1 - v_2 = v_{\rm in} = \frac{3}{4} \biggl[1-\frac{1}{{\cal 
M}^2_1} \biggr] v_1
\end{equation}

Rearranging and replacing $v_1$ by ${\cal M}_1 c_s$, we obtain

\begin{equation}
v_{\rm in} = \frac{3}{4} \biggl[\frac{{\cal 
M}_1^2-1}{{\cal M}_1} \biggr] c_s  = \frac{3}{4} 
\biggl[\frac{{\cal M}_1^2-1}{{\cal M}_1} \biggr] 
\biggl[\frac{5}{3} K_1 \rho_1^{2/3} \biggr]^{1/2}
\end{equation}

\begin{figure*}
\centering
\leavevmode
\epsfysize=8.4cm \epsfbox{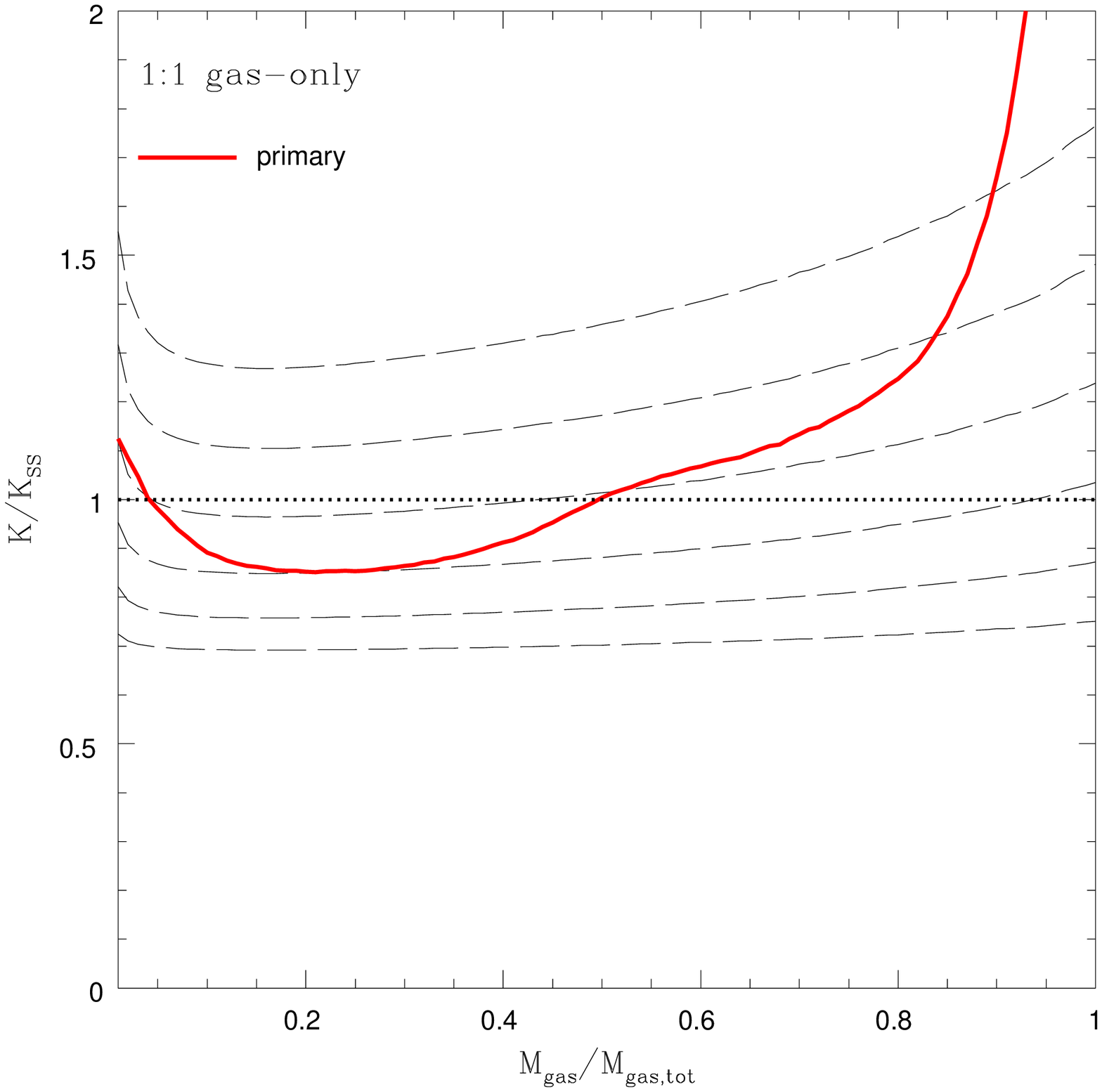}
\epsfysize=8.4cm \epsfbox{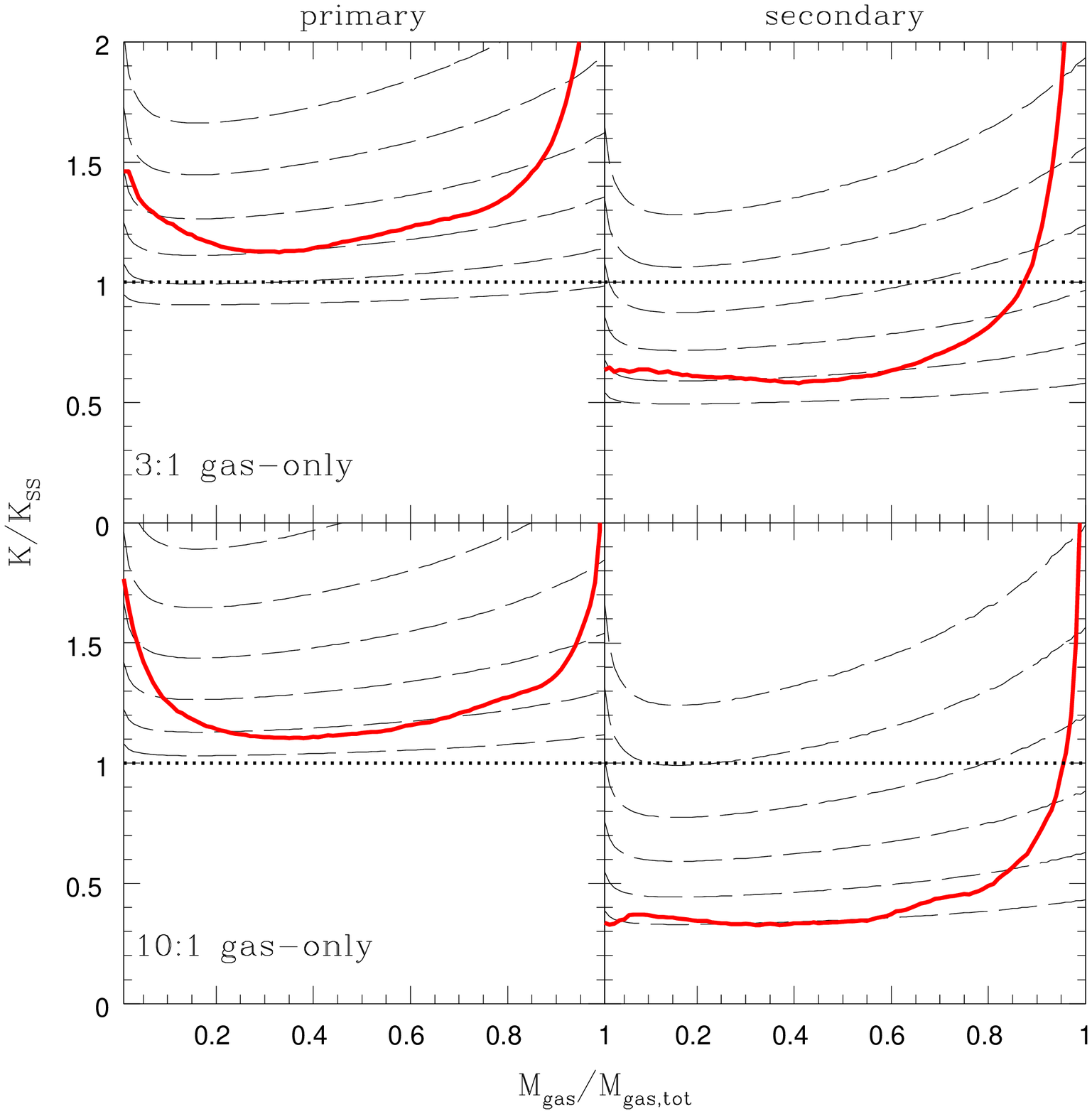}
\caption{
Comparison of the final entropy distribution of the
primary in the 1:1 (left) and 3:1 and 10:1 (right) gas-only
simulation with our simple analytic shock heating model.  The
thick red curve represents the simulation result.  The dashed
curves represent the simple analytic model (equations 11 \& 13)
for choices of $v_{in}/v_{c,p}(r_{200})$ ranging from $0.75 -
2.00$ in steps of $0.25$ (bottom to top).
}
\end{figure*}

Therefore, given the initial entropy and density distributions of 
the gas and the velocity of the gas in the centre-of-mass frame, it 
is possible to solve the quadratic equation (13) for the Mach 
number of each particle.  Deriving the final entropy distribution 
is then 
simply a matter of plugging these Mach numbers into equation (11).  
Of course in our idealised 
merger simulations we know precisely what the initial entropy 
and density profiles of the systems are, but what velocity 
do we pick for $v_{\rm in}$?  Various possibilities exist, but 
$v_{\rm in}$ should not exceed the relative velocity of the cores as 
they are about to collide (i.e., when the gravitational energy 
between the two systems has been maximally converted to infall 
energy).  In fact, the velocity will have to be quite a 
bit lower than this since, for example, in the 1:1 case using the 
maximum relative velocity for the gas in both systems would require 
twice as much energy as there is available to be thermalised.  
Unfortunately, given the complicated nature of the simulations, the 
correct value of $v_{\rm in}$ could fall any where between zero and 
this upper bound.  
Previous analytic studies (e.g., Voit et al.\ 2003) adopted the 
infall velocity at the virial radius.  
However, it is evident from \S 3 that significant shock heating 
does 
not occur in our simulations until the cores of the two systems 
nearly collide.  As a result, the infall velocity will be much 
larger than assumed in those analytic studies.  Furthermore, it is 
also clear that energy is being exchanged between the primary and 
the secondary systems (for example, even in the 10:1 case the 
secondary is capable of driving a shock that significantly heats 
virtually all of the gas in the primary).  This makes it even more 
difficult to assess {\it a priori} what is the appropriate value of 
$v_{\rm in}$ to assign for the primary and secondary systems.

We sidestep this problem by inverting the question: i.e., given the 
initial entropy and density profiles, what velocity is required to 
explain the final entropy distribution?  To answer this question we 
simply try a range of different values for $v_{\rm in}$ and assess 
which gives the best match to the final entropy distribution.  
This velocity can be cast in terms of a requirement for the total 
amount of energy that must have been thermalised in order to 
explain the final entropy distribution.  In \S 4.2, we examine 
whether or not there is enough bulk energy available to explain our 
findings.

Although the simulated systems undergo two periods of entropy 
production, we start by trying to use the single-shock model outlined
above to explain the observations. In principle this model, which 
just uses continuity and conservation equations to link upstream
and downstream conditions, can effectively describe physical 
situations that are more complex than a single shock.  Therefore 
it is the natural starting point for our investigation.

We start by first examining the gas-only simulations, which make a 
useful benchmark for the more realistic gas+DM runs.  In Figures 
19a and 19b, the final entropy distributions of the primary 
and secondary systems in the 1:1, 3:1, and 10:1 gas-only mergers 
are compared with the simple analytic shock heating model proposed 
above.  For the analytic model, the upstream values of the entropy 
and density are taken from the initial conditions of the simulations 
(see \S 2).  We try several different values of $v_{\rm in}$ for 
both the primary and secondary systems, each corresponding to a 
unique prediction for the final entropy distribution of both 
systems.  All curves plotted in Figs.\ 19a \& 19b have been 
normalised to the self-similar expectation.
In the discussion that follows, we exclude the high entropy tail at 
large values of $M_{\rm gas}/M_{\rm gas,tot}$ from consideration.
The \S 3.1.3 shows that this tail results from the truncation of 
our idealised halos. 

Figs.\ 19a \& 19b show that a 
centre-of-mass velocity ranging approximately from $0.9 < v_{\rm 
in}/v_{c,p}(r_{200}) < 1.25$ is required to explain the final 
entropy distributions of the primary systems in the three runs.  
The secondary systems require slightly lower velocities ranging 
from $0.75 < v_{\rm in}/v_{c,p}(r_{200}) < 1.25$.  For both the 
primary and secondary systems there is a trend with the mass 
ratio of the merger, in the sense that the higher the mass ratio 
the lower the required velocity is to explain their final entropy 
distributions.  It is worth noting, however, that no single choice 
of $v_{\rm in}$ can explain the entire $K(M_{\rm gas})$ profiles 
for the primary and secondary systems.  In particular, if the 
analytic model is normalised to explain the intermediate regions 
of the entropy profiles (say $0.3 < M_{\rm gas}/M_{\rm gas,tot} < 
0.6$), it systematically under-predicts the level of the lowest 
entropy gas ($M_{\rm gas}/M_{\rm gas,tot} < 0.2$) compared to the 
simulations.  Nevertheless, the bulk of the gas in the primary 
and secondary systems can be adequately modelled by a fairly 
small range of velocities.

\begin{figure*}
\centering
\leavevmode
\epsfysize=8.4cm \epsfbox{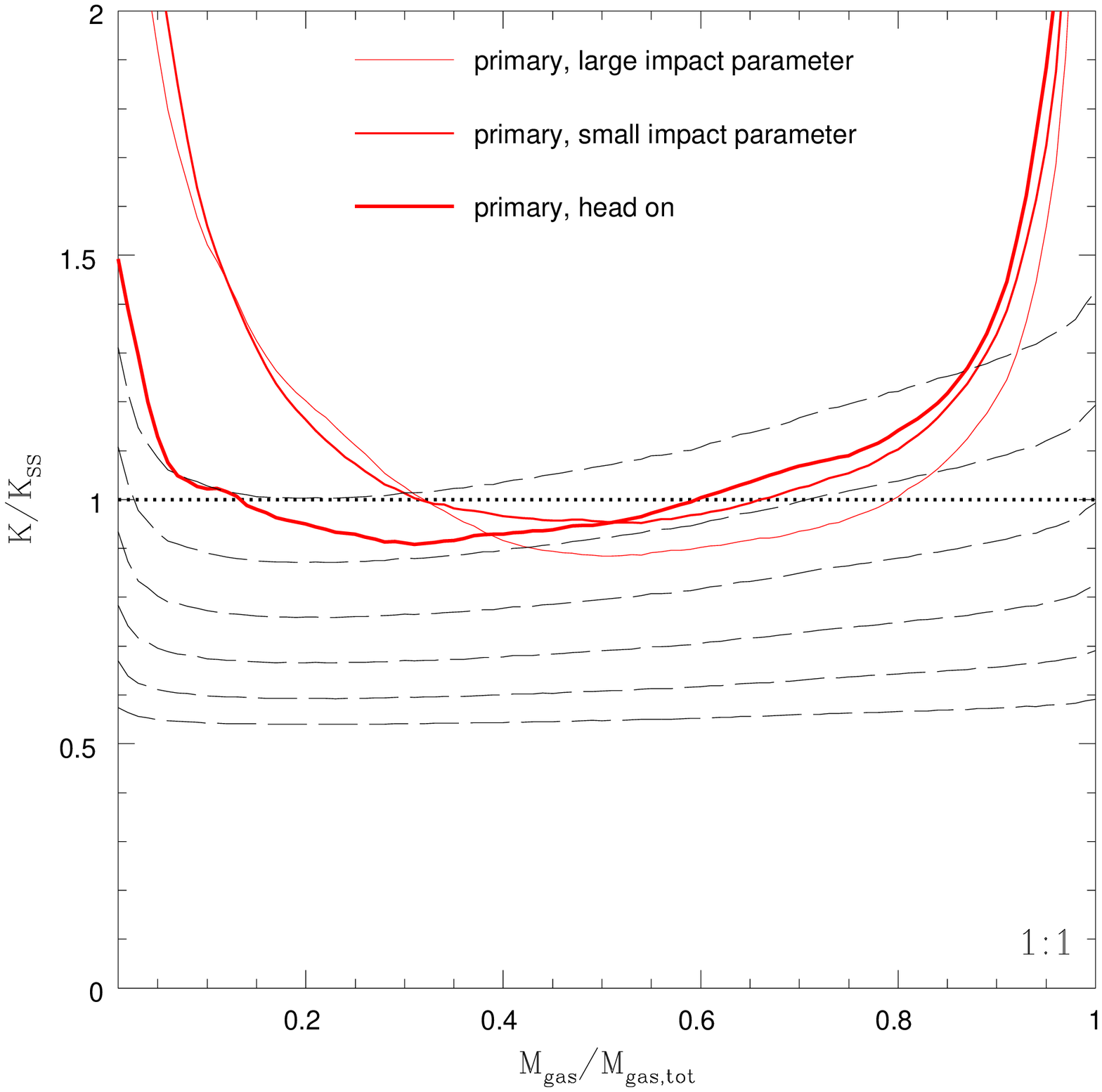}
\epsfysize=8.4cm \epsfbox{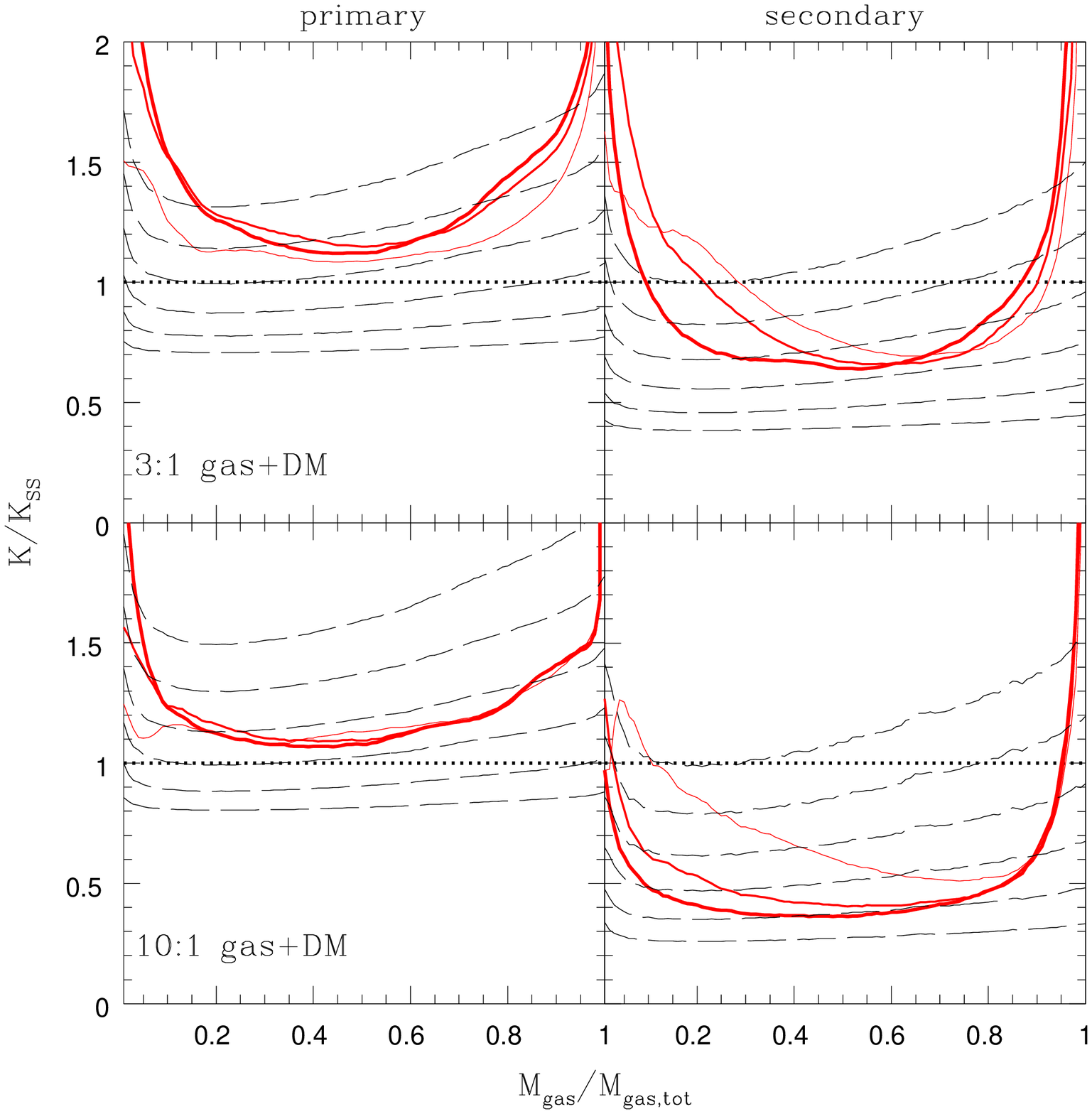}
\caption{
Comparison of the final entropy distribution of the
primary in the 1:1 (left) and 3:1 and 10:1 (right) gas+DM
simulation with our simple analytic shock heating model.  The
thick red curve represents the
simulation result.  The dashed curves represent the simple
analytic model (equations 11 \& 13) for choices of
$v_{in}/v_{c,p}(r_{200})$ ranging from $0.75 - 2.00$ in steps of
$0.25$ (bottom to top).
}
\end{figure*}

An identical set of plots is presented in Figures 20a,b for 
the 
gas+DM simulations.  Note that significantly higher 
velocities are required to explain the final entropy distributions 
of the primary and secondary systems in the gas+DM simulations.  In 
particular, the primary systems typically require velocities 
ranging from $1.35 < v_{\rm in}/v_{c,p}(r_{200}) < 1.80$, while the 
secondaries typically require $1.00 < v_{\rm in}/v_{c,p}(r_{200}) < 
1.80$ (see Table 2).  Even though the systems in the gas+DM 
simulations require significantly higher velocities than those in 
the gas-only simulations this does not necessarily imply that the 
energetic requirements of gas+DM mergers exceed those of the 
gas-only mergers.  It should be kept in mind that for a system of 
mass
$M_{200}$ there is simply much more gas that requires heating in
the gas-only simulations (the baryon to total mass ratio of the gas-only
simulations is $\approx$7 times larger than the gas+DM 
simulations).
Below we compare the energy required by the simple shock heating 
model to match level of entropy production seen in the idealised 
simulations with our best estimates of the amount of energy 
available.

\subsection{Energy Considerations}

If we assume that the simple shock heating model provides a 
reasonable description of what is taking place in the simulations, 
the velocity $v_{\rm in}$ can be used to estimate the total amount 
of energy that was thermalised in producing the final entropy 
distributions.  Conveniently, $v_{\rm in}$ is defined such that the 
post-shock gas is at rest in the centre-of-mass frame, so the 
total thermalised energy is just:

\begin{equation}
E_{\rm T,tot} = E_{\rm T,p} + E_{\rm T,s} = \frac{1}{2}M_{\rm 
gas,p} v_{\rm in,p}^2 + \frac{1}{2}M_{\rm gas,s} v_{\rm in,s}^2
\end{equation}

Typically, this results in values ranging from approximately $2-6 
\times 10^{64}$ ergs for the gas-only mergers and $0.5-2 \times 
10^{64}$ ergs for the gas+DM mergers.  So even though 
the gas+DM mergers require 
higher velocities to preserve self-similarity, their energy 
requirements are lower than those of the gas-only simulations.  
As mentioned above, this is simply because there is more gas that 
requires heating in the gas-only simulations.

Interestingly, even though the energetic requirements are 
different between the two types of simulations, the way the energy 
is distributed does not appear to be.  In particular, in both 
types of simulations the thermalisation of the primary's gas 
dominates the thermalisation energy budget with 50\%, 
$\approx$80\%, and $\approx$95\% of the total energy in the 1:1, 
3:1, and 10:1 mergers, respectively (see Table 2).  The following 
analytic relationship yields a remarkably good fit to our 
simulated mergers (including those involving mass ratios not 
presented in this paper):

\begin{equation}
\frac{E_{\rm T,p}}{E_{\rm T,s}} \approx 
\biggl(\frac{M_{\rm p}}{M_{\rm s}}\biggr)^{5/4}
\end{equation}

Interestingly, this is quite close to the case where the 
primary and secondary thermalise each other's infall energy, 
i.e., $E_{\rm T,p}/E_{\rm T,s} = M_p/M_s$.  We also note that 
this type of scaling is naturally achieved if each gas particle 
thermalises the same fraction of the {\it total} available 
energy.  This interesting result deserves further investigation, 
which we take up in the next paper of this series.

\begin{table}
\centering
\begin{minipage}{140mm}
\caption{Single-Shock Model Velocity/Energy Requirements }
\begin{tabular}{@{}lllll@{}}
\hline
$M_p/M_s$ & Sim. type & $v_{\rm in,p}$ & $v_{\rm in,s}$ &
$E_{T,p}/E_{T,tot}$\\
 &  & $[v_{c,p}(r_{200})]$ & $[v_{c,p}(r_{200})]$ & \\
\hline
1:1  & gas-only & 1.25 & 1.25 & 0.50 \\
3:1  & gas-only & 1.25 & 1.00 & 0.82 \\
10:1 & gas-only & 0.9  & 0.75 & 0.94 \\
1:1  & gas+DM   & 1.8  & 1.8  & 0.50 \\
3:1  & gas+DM   & 1.65 & 1.35 & 0.82 \\
10:1 & gas+DM   & 1.35 & 1.00 & 0.95 \\
\hline
\end{tabular}
\end{minipage}
\end{table}

An important consistency check of the simple shock heating model is 
to test whether or not there is enough energy available to meet 
the requirements of the model.  Having too much energy available 
isn't necessarily a problem, since there are numerous ways the 
available energy could be tapped (e.g., some could go into bulk 
kinetic circular motions of the gas or into the dark matter in 
the case of the gas+DM simulations).  
However, if there isn't enough energy available, the only 
possibility is that the model is incorrect or, at best, 
incomplete.

With this in mind, we calculate the energy available 
to be thermalised in our idealised mergers.  We do so via two 
different methods, with both yielding similar results.  The first 
method, which we refer to as the ``simulation method'', takes 
advantage of the excellent energy conservation of the GADGET-2 
simulations.  The total energy of the gas is

\begin{equation}
E_{\rm tot,gas}(t) = E_{\rm K_{\rm gas}}(t) + E_{\rm U_{\rm gas}}(t) 
+ E_{\rm I_{\rm gas}}(t)
\end{equation}

\noindent where $E_{\rm K_{\rm gas}}(t)$, $E_{\rm U_{\rm gas}}(t)$, 
and $E_{\rm I_{\rm gas}}(t)$ are the kinetic, potential, and 
internal (thermal) energies of the gas at time $t$.  $E_{\rm 
tot,gas}(t)$ is conserved in the gas-only simulations to better 
than 1\%.  Therefore, we can write

\begin{equation}
E_{\rm K_{\rm gas}}(t) + E_{\rm U_{\rm gas}}(t) + E_{\rm I_{\rm 
gas}}(t) = E_{\rm K_{\rm gas}0} + E_{\rm U_{\rm gas}0} + E_{\rm 
I_{\rm gas},0}
\end{equation}

\noindent where $E_{\rm K_{\rm gas},0}$, $E_{\rm U_{\rm gas},0}$, 
and $E_{\rm I_{\rm gas},0}$ are the values of the three different 
energies at the start of the simulation.  The total energy that 
is available to be thermalised at time $t$ is just the initial 
(centre-of-mass) kinetic energy plus the 
change in the gravitational potential energy:

\begin{equation}
E_{\rm T,tot}(t) = E_{\rm K_{\rm gas},0} - [E_{\rm U_{\rm 
gas}}(t) - E_{\rm U_{\rm gas},0}]
\end{equation}

\noindent which is relatively trivial to measure in the 
simulations.  This energy estimate can be compared directly with 
the simple shock heating model estimate in equation (14).  

However, the above treatment is strictly only valid for the 
gas-only simulations.  In the gas+DM simulations, energy can be 
exchanged between the gas and the dark matter.  But we can 
take advantage of the fact that the summation of the total energy 
of the gas and the total energy of the dark matter is conserved 
in these simulations.  Thus, for the gas+DM simulations we modify 
equation (18) to read

\begin{equation}
E_{\rm T,tot} = E_{\rm K_{\rm gas},0} + [E_{\rm U_{\rm 
gas}}(t) - E_{\rm U_{\rm gas},0}] - E_{\rm DM \leftrightarrow 
gas}(t)
\end{equation}

\noindent where 

\begin{eqnarray}
E_{\rm DM \leftrightarrow gas}(t) & \equiv 
 & [E_{\rm K_{DM}}(t)+E_{\rm U_{DM}}(t)]\\
 & & -[E_{\rm K_{DM},0}+E_{\rm U_{DM},0}] \nonumber
\end{eqnarray}

\noindent is the energy exchanged between the gas and dark matter.

In principle, this energy exchange can go either way, but in 
general we find that the dark matter loses energy to the gas.  In 
Figure 21, we plot the evolution of the energy associated with 
the dark matter in the gas+DM simulations (see caption).  By the 
end of the simulations, we find that the dark matter in the 1:1, 
3:1, and 10:1 mergers has lost approximately 7-10\%, 7-9\%, and 
2-3\%, 
respectively, of its energy to the gas.  Interestingly, this 
estimate doesn't depend much on the adopted orbital parameters, 
nor the mass resolution of the simulation.  We will return to 
this issue of energy exchange below.

The second method that we use to calculate the energy available 
to 
be thermalised, which we refer to as the ``analytic method'', is as 
follows.  If we assume that the primary and secondary remain 
completely intact from their initial setup until the point when 
the centres of the two systems coincide, it is straightforward to 
calculate the total energy to be thermalised.  Neglecting the 
thermal energy of the systems (which remains fixed with time by 
construction), the total energy of the gas in the centre-of-mass 
frame is

\begin{equation}
E_{\rm tot,gas} = \frac{1}{2}\mu v_{rel}^2 + U_{\rm gas,ps}(r_{\rm 
ps})\end{equation}

\noindent where $\mu \equiv M_{\rm gas,p}M_{\rm gas,s}/(M_{\rm 
gas,p} + M_{\rm gas,s})$ is the reduced mass, $v_{rel}$ is the 
initial relative velocity between the primary and secondary 
systems, $r_{\rm ps}$ is the separation of their centres, and 
$U_{\rm gas,ps}$ is the gravitational potential energy.  Assuming 
that the primary and secondary systems are rigid and ignoring the 
potential energy of the systems due to themselves (which, again, 
does not change by construction), the potential energy of 
interaction {\it between} the two systems is

\begin{eqnarray}
U_{\rm gas,ps}(r_{\rm ps}) & = &\frac{1}{2} \int \phi_{\rm p} 
dM_{\rm gas,s} + \frac{1}{2} \int \phi_{\rm s} dM_{\rm gas,p} \nonumber\\
 & = & \frac{1}{2} \int \rho_{\rm gas,s} \phi_{\rm p} d^3V + 
\frac{1}{2} \int \rho_{\rm gas,p} \phi_{\rm s} d^3V
\end{eqnarray}

\noindent where $\phi$, the gravitational potential, is defined as

\begin{equation}
\phi(x) \equiv - G \int \frac{\rho(x')}{|x'-x|} d^3V' 
\end{equation}

\begin{figure}
\centering
\includegraphics[width=8.4cm]{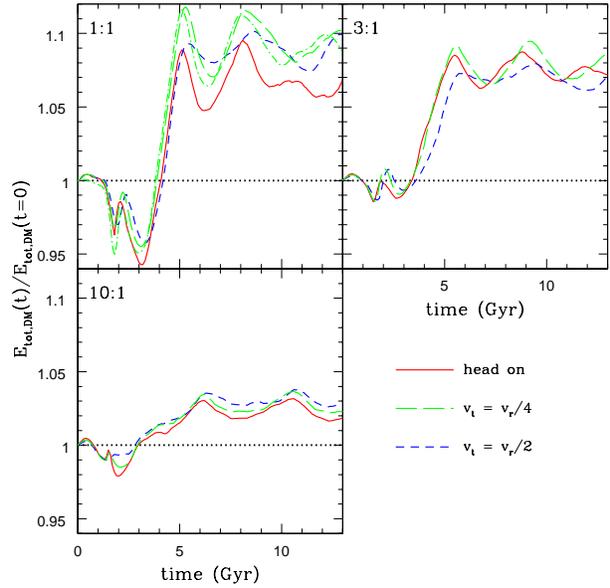}
\caption{Evolution of the total energy of the dark matter in the
gas+DM simulations.  The energy has been normalised to its initial
value.  Since the systems are bound the total energies are
negative.  Thus, a value of $E_{\rm tot,DM}(t)/E_{\rm
tot,DM}(t=0)$
greater than unity implies that energy has been {\it lost} to the
gas, while a value less than unity means energy has been extracted
from the gas.  The green dot-dashed curve in the 1:1 panel is 
for the highest resolution simulation in the mass resolution 
study in the Appendix.
  }
\end{figure}

Approximating the systems as point masses initially, when there is 
little or no overlap between them, the potential energy of the gas 
is just

\begin{figure*}
\centering
\leavevmode
\epsfysize=8.4cm \epsfbox{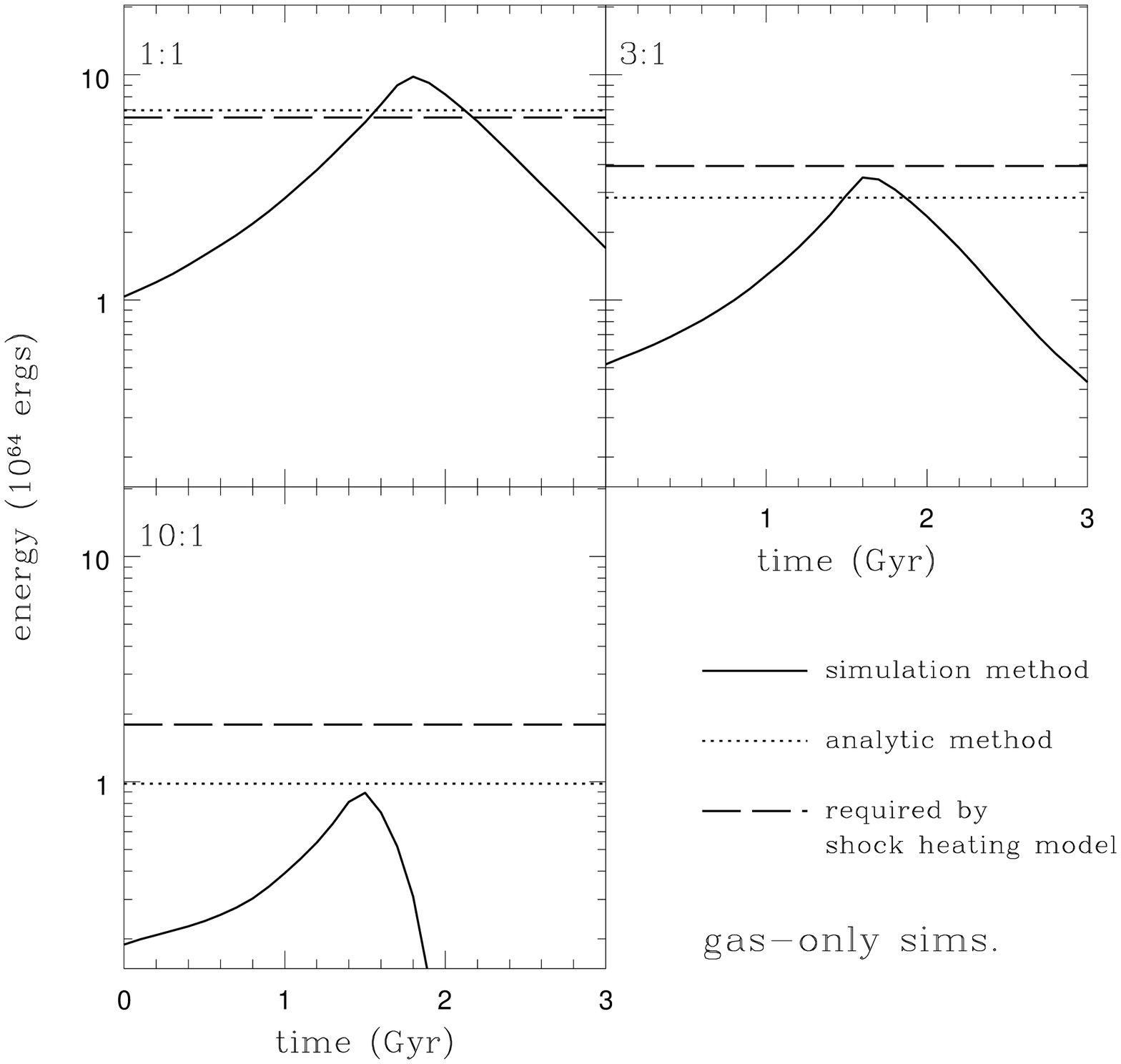}
\epsfysize=8.4cm \epsfbox{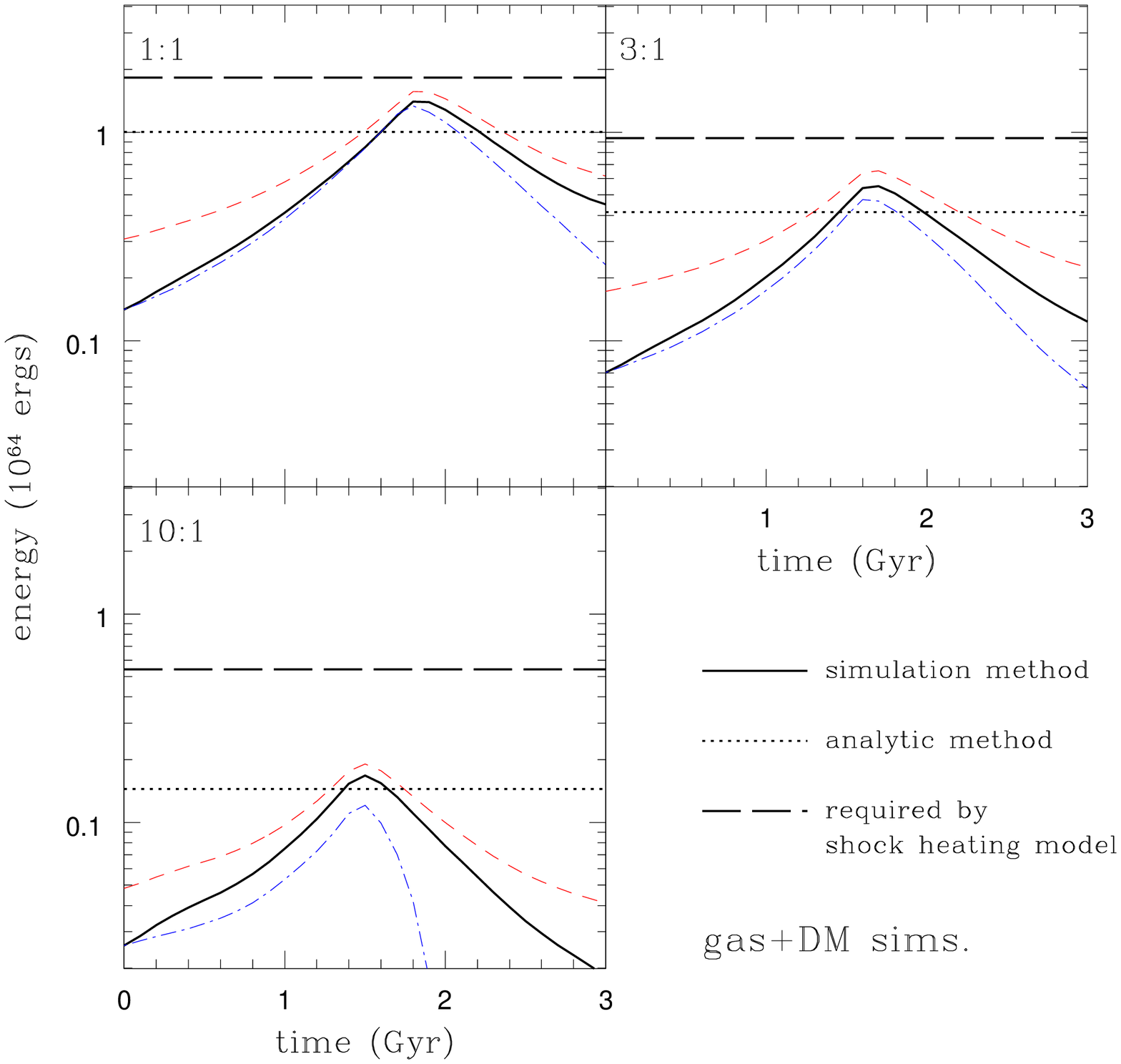}
\caption{
Comparison of the shock heating model's energy requirements
with the maximum energy available to be thermalised in the
gas-only (left) and gas+DM (right) simulations.  In the panels on
the right, the short-dashed red curves represent the
results of the simulation method shifted up by $E_{\rm
tot,DM}(t=13 {\rm Gyr})-E_{\rm tot,DM}(t=0)$, i.e., by
the total amount of energy lost by the dark mark to gas by the
end of the simulation.  For comparison, the dot-dashed blue lines
represent the gas-only simulations, which have been renormalised
by multiplying the energy by the baryon fraction of the gas+DM
systems.
}
\end{figure*}

\begin{eqnarray}
U_{\rm gas,ps}(r_{\rm ps}=d_0) & = & -\frac{1}{2} \frac{G}{d_0} 
\biggl[M_{\rm p} M_{\rm gas,s} + M_{\rm p,gas} M_{\rm s} \biggr]\\
 & = & - \frac{G f_b M_p M_s}{d_0} \nonumber
\end{eqnarray}

\noindent where the second line is true only if both the 
primary and secondary systems have the same baryon fraction.

Therefore, the total energy of the gas is

\begin{equation}
E_{\rm tot,gas} = \frac{1}{2}\mu [v_{\rm c,p}(r_{200})]^2 - 
\frac{G f_b M_p M_s}{d_0}
\end{equation}

The maximum energy available to be thermalised occurs when 
the centres of the primary and secondary coincide.  We can 
therefore calculate the maximum energy available to be thermalised 
by subtracting the potential energy when the systems coincide 
(calculated by evaluating equation 22) from the total energy given 
in equation (25).

In Figures 22a,b, we compare the simple shock heating model's 
energy requirements (see Table 2) with the energy available to be 
thermalised as estimated by both the simulation and analytic 
methods described above.  We focus first on the gas-only results 
plotted in Fig.\ 22a.  The solid curves represent the amount of 
energy available to be thermalised as estimated with the 
simulation method.  The peak of the curves are reached at $t 
\approx 1.5$-$1.8$ Gyr, i.e., just 
slightly before the cores of the primary and secondary collide.  
The amplitude of the peak is within a few tens of percent of the 
maximum energy estimated via the analytic method (dotted line).  
This agreement indicates that our estimate of the maximum energy 
available for thermalisation is robust and also demonstrates that 
one can estimate this energy reasonably well using simple analytic 
modelling.  A comparison to the dashed line, which represents the 
energy requirement of the simple shock heating model described in 
\S 4.1, yields interesting results.  Apparently, the 1:1 gas-only 
merger has sufficient energy available to accommodate the shock 
heating model's requirements.  So the simple model provides a 
viable explanation for this collision.  For the 3:1 gas-only 
merger there is a very small deficit of energy.  However, in the 
case of the 10:1 gas-only merger, both the simulation method and 
analytic method estimates of the maximum amount of energy there 
is to be thermalised fall short of the required amount by roughly 
a factor of two.

Moving on to the gas+DM simulations in Fig.\ 22b, we find that 
there 
is insufficient energy available in any of the simulations to 
accommodate the requirements of the simple shock heating model 
(although the 1:1 is close).  This is the case even when we 
account for the energy exchange between the gas and the 
dark matter.  Thus, even though both the gas-only 
and gas+DM simulations 
preserve self-similarity, they do not give a consistent answer 
when compared to the simple analytic model.  However, there are 
some similarities between the two in terms of their comparison 
with the analytic model.  For example, both show a similar trend 
with mass ratio, in the sense that agreement gets worse for 
higher mass ratios.  Furthermore, the energy shortfall is less 
than about a factor of $3$ for all of the simulations we have 
performed.  Thus, while the simple analytic model fails 
to explain the results, it does not fail by a huge margin.  This 
motivates us to consider modifications of the model.

\subsection{A Double-Shock Model}

In \S 3 we found that there are actually two major periods of 
entropy production experienced by the primary and secondary 
systems in our merger simulations.  We now seek to modify the 
simple shock heating model presented in \S 4.1 to account for 
this behaviour.  The relevant question is, for a fixed amount of 
energy to be thermalised, is a double-shock model capable of 
generating more entropy than a single-shock model?  In the case 
where the post-shock conditions, as dictated by the jump 
conditions, simply become the pre-shock conditions for the second 
shock, we find that the answer is `no'.  In general, we find that 
as one increases the number of shocks over which the energy is to 
be thermalised, the resulting final entropy decreases.

\begin{figure}
\centering
\includegraphics[width=8.4cm]{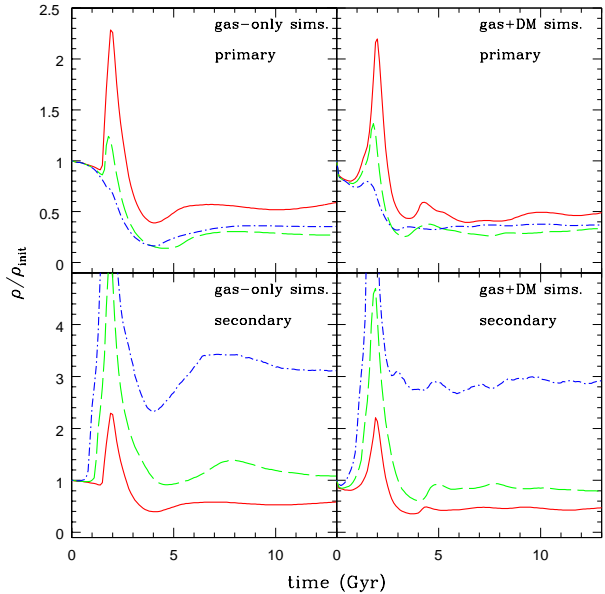}
\caption{
Evolution of the median density for particles that were
located within the spherical shell $0.25 \leq M_{\rm gas}/M_{\rm
gas,tot} \leq 0.75$ initially.  The solid red, dashed green, and
dot-dashed blue curves represent the 1:1, 3:1, and 10:1 mergers,
respectively.  For the gas+DM simulations only the head on results
are plotted.  Note for the primary systems that
the density drops below its initial (pre-merger) value between the
end of the first shock and the onset of the second shock at
$t \sim 4$ Gyr.  For the secondary systems the density increases
(except for the 1:1 case), as expected.  However, only a small
fraction of the total energy is thermalised in the secondary.
  }
\end{figure}

However, it quickly becomes apparent from an examination 
of the 
simulations that the properties of the gas evolve significantly 
between the end of the first shock and the onset of the second.  In 
particular, there is a period of adiabatic expansion between the 
two shocks which likely arises as a result of the fact that not all 
the kinetic energy was thermalised in the first shock\footnote{The 
expansion could also be partially due to a re-adjustment of the 
gas towards a new hydrostatic configuration.  Note, however, that 
this cannot be the whole story since no further shock heating 
would be expected in this case.  A period of re-accretion is 
required.}.  The net result is that the typical density of the 
gas is significantly reduced between the shocks and can even drop 
below its pre-merger value (see Fig.\ 23).  Furthermore, the drop 
is largest for the highest mass ratio mergers, precisely where we 
find the largest energy deficits between the simulations and the 
single-shock model.  Dropping the density between the shocks will 
have the effect of 
increasing the amount of entropy generated in the second shock 
relative to the case where there is no adiabatic expansion between 
the shocks.  More importantly, is the decrease in density between 
shocks large enough to generate more entropy than the single-shock 
model?

To answer this question, we have compared the single- and 
double-shock models head to head for an idealised parcel of gas 
with an initial density $\rho_{\rm init}$ and an initial entropy 
$K_{\rm init}$. In particular, in Fig.\ 24 we plot three sets of 
curves representing three different comparisons, each 
characterised by a different total amount of energy to be 
thermalised.  The total energies have been chosen such that, in 
the single-shock model, the (shock frame) Mach numbers are 1.5, 
2.5, and 3.5 (bottom to top).  For the single-shock model, the 
Mach number is all that is required to predict the ratio of final 
to initial entropy (see equation 11).  The three horizontal 
dotted lines in Fig.\ 24 represent the predictions of the 
single-shock model.

\begin{figure}
\centering
\includegraphics[width=8.4cm]{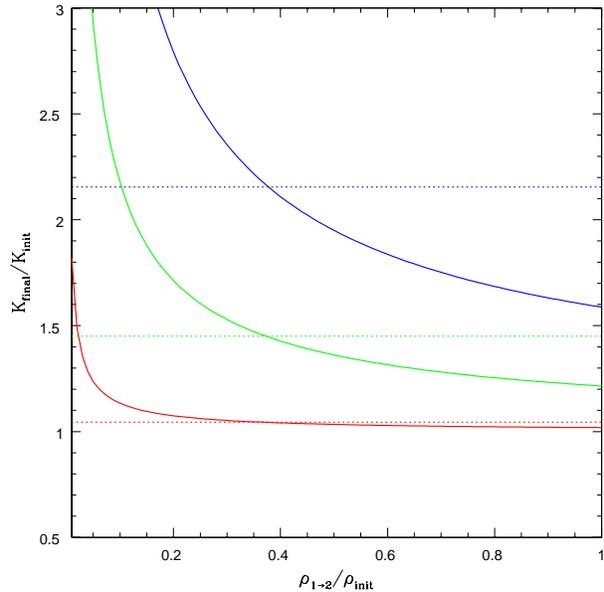}
\caption{A comparison of the entropy generated by the single- and
double-shock heating models.  Horizontal dotted lines show the
predictions of the single-shock heating model for (shock frame)
Mach numbers of 1.5, 2.5., and 3.5 (bottom to top).  The solid
curves show the predictions of the double-shock model as a
function of the density between the two shocks.  Note that if the
density drops below $\approx 40$\% of its pre-merger value then
the double-shock model generates more entropy than the
single-shock model for a fixed amount of energy to be thermalised.
  }
\end{figure}

\begin{figure*}
\centering
\leavevmode
\epsfysize=8.4cm \epsfbox{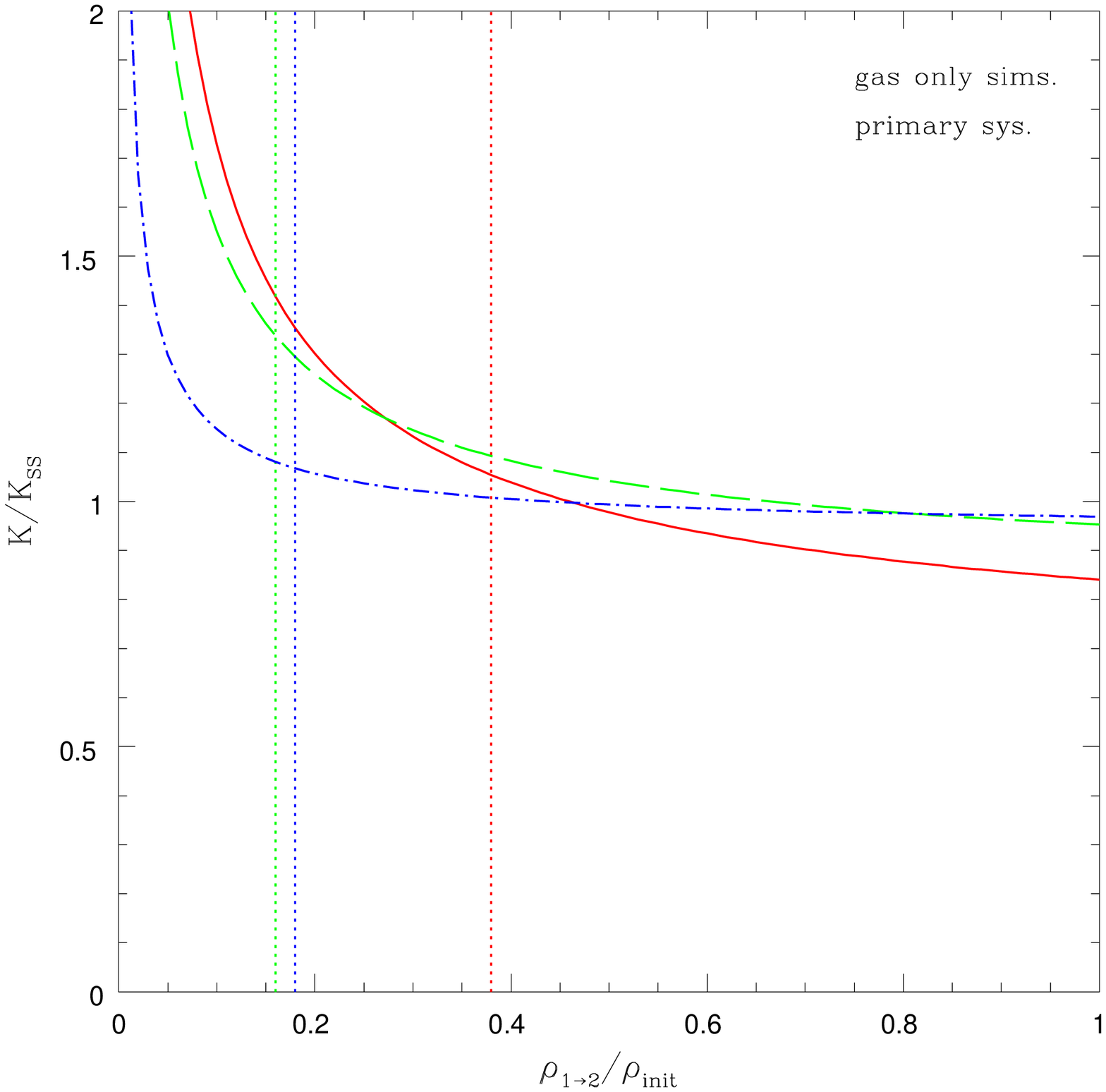}
\epsfysize=8.4cm \epsfbox{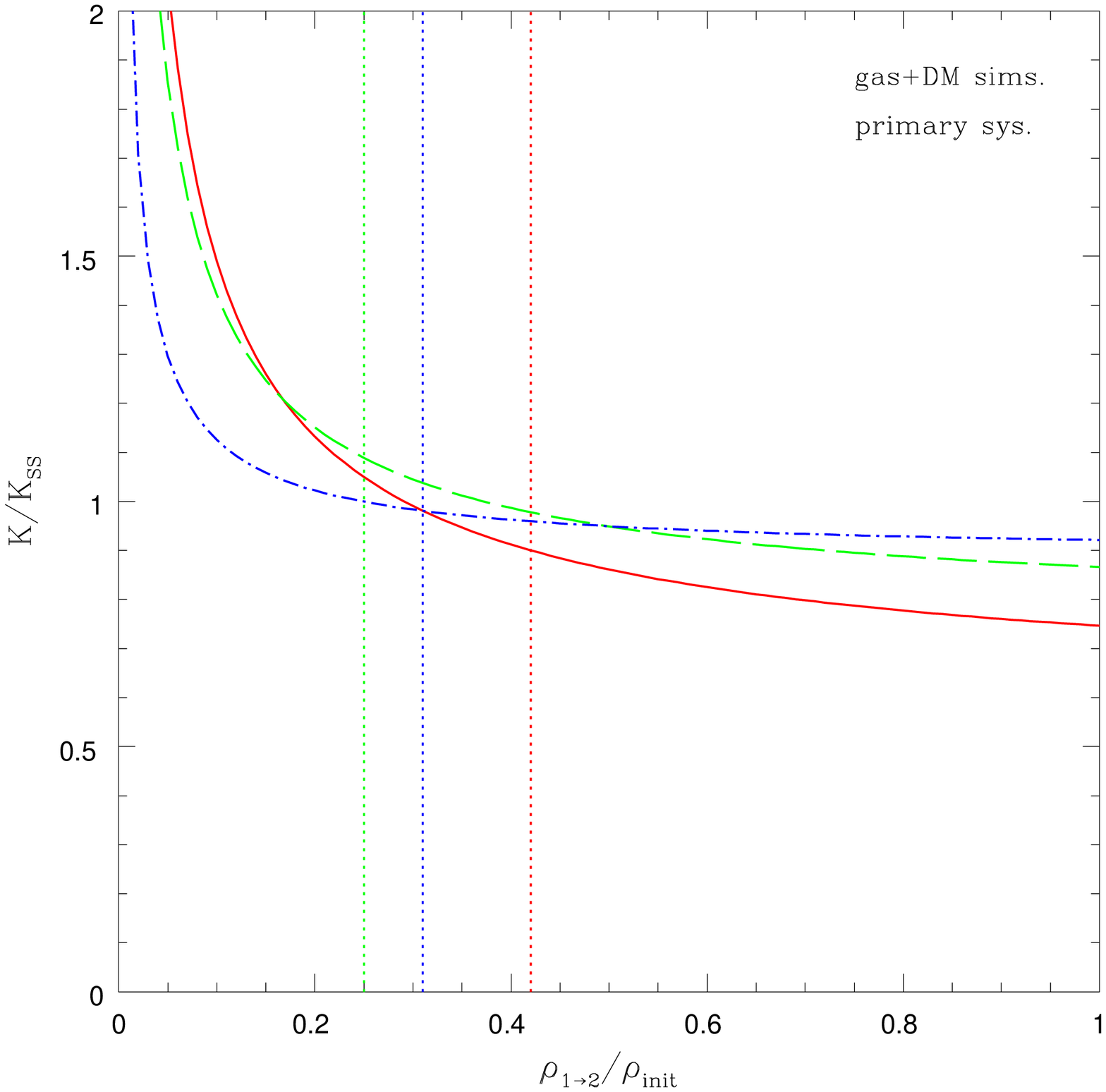}
\caption{
Final entropy predicted by the double-shock model as
a function of the density $\rho_{1\rightarrow2}$ between the two
shocks for the gas-only (left) and gas+DM (right) simulations.
The solid red, dashed green,
and dot-dashed blue curves represent the 1:1, 3:1, and 10:1
mergers, respectively.  The entropy has been normalised to the
expected self-similar result, while the density has been
normalised to its initial pre-merger value.  The vertical dotted
lines show the minimum density estimated from Fig.\ 23 at
$t \approx 4$ Gyr.
}
\end{figure*}

For the double-shock model, we use the entropy and density 
evolution plots of the simulations as guides.  Examination of the 
evolution of the entropy of our mergers (e.g., see Figs.\ 2, 4, 
and 
8) indicates that the first and second shocks contribute 
comparable amounts of entropy to the final state.  To first order, 
therefore, we surmise that the first and second shocks thermalise 
comparable amounts of energy (i.e., for the purposes of this toy 
model, we assume each shock thermalises half of the total energy.)
Between the first and second shocks, there is a period of 
adiabatic expansion during which the density drops to some value 
$\rho_{1\rightarrow2}$.  Using the new density and entropy, we 
can solve for thermal energy of the gas (or, equivalently the 
sound speed).  The ratio of the thermal energy to the remaining 
energy to be thermalised (i.e., half the initial energy) sets the
Mach number of the second shock which, in turn, allows us to
compute the final entropy.

In Fig.\ 24, the solid curves show the predicted trend between 
final entropy and density between the two shocks for double-shock 
model.  A comparison between the single- and double-shock models 
demonstrates that if the density drops below about 40\% of its 
initial pre-merger value then the double-shock model does indeed 
generate more entropy than the single-shock model.  This is quite 
promising since Fig.\ 23 demonstrates that the primary systems, 
which dominate the thermalisation budget, have their densities 
reduced to at least this level (and lower for the 3:1 and 10:1 
cases).  

Given these results, we test the model further by tailoring 
the total energy to be thermalised (and, therefore, the Mach 
numbers) to match our merger simulations more closely.  In 
particular, we assume the total energy is taken to be the peak of 
the solid curves plotted in Figs.\ 22a,b corresponding to 
the 
``simulation method'' estimate.  For simplicity, we further 
assume that each gas particle thermalises the same amount of 
energy.  So, for example, the primary system thermalises 
$M_p/[M_p+M_s]$ times the total energy.  As above, the Mach 
numbers are set by computing the ratios of thermal energy 
to energy to be thermalised for the first and second shocks.

In Figures 25a,b we plot the final entropy predicted by the 
double-shock model for the primary systems in our simulations.
The entropy has been scaled to the self-similar result while 
$\rho_{1\rightarrow2}$, the density between the two shocks, has 
been scaled to the initial pre-merger density.  We use the 
vertical dotted lines, which are meant to represent the density 
minimum between the shocks seen in Fig.\ 23 (at $t \approx 4$ 
Gyr), to select the appropriate predicted entropy.  

A comparison of the predicted entropy in Figs. 25a,b
with that of the primary in our simulations in Figs.\ 5-6 
(gas-only) and Figs.\ 16a, 17a, and 18a (gas+DM) demonstrates 
that in all cases, including the 10:1 mergers, the double-shock 
model can reproduce the simulation results to within 10\%.  
Furthermore, we have found that this result is not very 
sensitive to the way in which we've distributed the energy over 
the two shocks.  For example, we obtain very similar results if 
the second shock thermalises anywhere between 20-80\% of the 
total energy, as opposed to half.

The double-shock heating model therefore appears to provide a 
simple framework for understanding the typical level of entropy 
generated in our mergers.  In \S 5.3, we present a point by point 
algorithm that the reader can use to apply the double-shock 
model.

\section{Conclusions}

We have presented a series of simulations aimed at exploring the 
generation of entropy during cluster mergers.  The results show that 
the entropy generated is remarkably robust.  We show that the 
generation of entropy is largely independent of the impact parameter 
of the collision, and that similar results are obtained for 
simulations only involving gas and for mergers of systems that 
contain a cosmological mixture of gas and dark matter.  The 
resulting entropy profiles also depend little on the resolution of 
the simulation, once more than $\sim10^4$ particles are placed in 
each halo.  These results hint that a general principle is at work,
and that the generation of entropy can be understood as a general
process that converts the gravitational potential energy released
during the collapse of the system into the thermal energy
of the ICM.  Based on this reasoning, it should be possible to 
develop a simple model for the evolution of the entropy distribution 
as clusters grow in mass through mergers by examining the potential 
energy released in the collapse and the efficiency with which this is
converted into thermal energy.

\subsection{Equal mass mergers}

We explore mergers of systems with a variety of mass ratios. In 
each case, we find that the entropy generated is 
approximately sufficient to place 
the final system on the same entropy scaling relation that was used 
to generate the original system.  During equal mass mergers 
symmetry ensures that both systems are heated to the same degree. 
For the bulk of the gas, the powerlaw slope of the resulting entropy 
profile changes little compared to the original (see, e.g., Fig.\ 
5a) and the degree of heating raises the normalisation 
of the entropy profile by $\approx 2^{2/3}$.  As a result the final 
entropy distribution is scaled so that the final system is a 
self-similar copy of the original.

The exact process by which this entropy is generated is far from 
simple. The two clusters are in contact with each other at the 
start of the simulation, however, their infall velocity is not 
sufficiently high compared to the sound speed to generate a strong 
shock. The system 
becomes highly compact along the infalling axis with compressed 
material tending to flow out along the orthogonal plane.  A strong 
shock is not generated until the cores of the two systems are 
superposed.  At this point, a strong shock is generated, propagating 
rapidly out from the central regions.  However, not all of the 
available infall energy is thermalised in this first shock.  Some of 
it remains in kinetic form and succeeds in driving a period of 
adiabatic expansion.  Eventually, however, the remaining energy is 
thermalised in a series of shocks as material is re-accreted by the 
merger remnant.

The generation of entropy thus shows two distinct peaks. The first 
corresponds to the strong shock generated as the cores collide, the
second corresponding to the re-accretion of material that tried 
initially to escape from the system.

The delay to the initial shock in this system plays an important
role in determining the entropy generated.  Because the system
collapses prior to the first shock, considerably greater binding
energy is available to be thermalised. Superposing the initial 
mass distributions provides a good estimate of the available binding
energy, and modelling this energy as being thermalised in a single
strong shock provides a reasonable approximation to the entropy
generated in the equal mass mergers.
  
\subsection{Unequal mass mergers}

For mergers between unequal mass halos, the generation of entropy is 
distributed unequally.  Visually, we see that the smaller component 
remains essentially intact as it plunges into the centre of the 
main system (see, e.g., Figs.\ 13 and 15).  We find that, 
although the 
kinetic energy of the collapsing system is primarily localised in 
the smaller mass system, this energy is largely thermalised in 
the more massive progenitor.  As a result, the heating of the 
more massive progenitor exceeds what is predicted by the 
self-similar scaling relation

\begin{eqnarray}
K \propto M^{2/3} \nonumber ,
\end{eqnarray}

\noindent while the heating of the less massive component falls 
short of that needed for self-similarity. Despite this, the over- 
and under-heating of the components combine is such a way that the
final system comes close to following the self-similar relation.  
We provide an analytic fit to the ratio of the energy dissipated 
in the two components in equation (15).  This is close to assuming 
that the energy is exchanged between the primary and secondary 
components.

Thus we find that the mergers tend to produce scaled up copies of the
original systems.  This is good news since we started with systems
having properties close to those of observed clusters, and we chose
cosmologically likely values for the infall velocities. Our simulations
show that the normalisation of this entropy profile is not a 
coincidence. Given the infall velocity distribution expected in a CDM
universe this profile is a stable configuration. 

As with the 1:1 merger case, we can estimate the energy that is 
available to be thermalised by tracking the evolution of the 
potential energy.  This energy significantly exceeds the initial 
kinetic energy of initial system, showing that the long survival 
time of the secondary is responsible for much of the entropy 
generation.  However, we find that a single shock model for 
the thermalisation of this energy under-predicts the entropy 
generated, particularly in the case of the 10:1 mass ratio. In 
order to match the entropy generation we find that it is necessary 
to model the two shock process that is seen in the simulations. A 
key 
ingredient of the entropy generation is the drop in the gas density 
as the system responds to the first shock.  As the remaining binding 
energy is thermalised in this more diffuse medium, the entropy 
generation is more efficient. 

\subsection{An algorithm for computing shock heating}

A good way to summarise the findings of this paper is to sketch out
an algorithm for computing the entropy generated during the merger.
Future papers will present the results from implementing this
within cosmological merger trees. We summarise the algorithm as follows.
\\
\\
\noindent 1.) Calculate the energy available for thermalisation.  
Cosmological simulations suggest that the secondary (less massive) 
system crosses the virial radius of the primary system with a 
total (relative) velocity of approximately 
the circular velocity of the primary system at that radius.  The 
total energy is therefore given by equation (25).  (Note, a 
similar energy estimate can be derived by calculating the potential 
energy between the systems at turnaround, when the relative 
kinetic energy is zero.)  The energy available for thermalisation 
can then be obtained by subtracting from this the potential energy 
between the systems when their centres coincide (see \S 4.2).
\\
\\
\noindent 2.) Distribute this energy in appropriate proportions to 
the primary and secondary systems.  Our simulations indicate that 
the bulk of the energy is thermalised in the more massive primary 
system.  Equation (15) provides a fit to how the energy should be 
divided up as a function of mass ratio.  
\\
\\
\noindent 3.) If the merger is 1:1 (or very nearly so), calculation 
of the post-shock properties is now straightforward.  The energy 
estimated above can be converted into an estimate for the 
(centre-of-mass) velocity, $v_{in}$, for both systems.  This, in 
turn, may be converted into an estimate of the (shock frame) Mach 
number (see equation 13).  Calculation of the post-shock 
properties is then simply a matter of 
evaluating the Rankine-Hugoniot jump conditions (equations 8-11).
\\
\\
\noindent 4.) If the mass ratio is different from unity, 
distribute the energy over two shocks.  Our simulations suggest 
that the two shocks contribute comparably to the final entropy 
(see Figs.\ 2, 4, \& 8; note also that as the mass ratio 
approaches
unity, the double shock model converges to the single shock result).  
To first order, therefore, one can assume the shocks each thermalise 
half of the total energy estimated in steps (1) and (2) above 
(note, however, that the results are not very sensitive to exactly 
how the energy is distributed over the two shocks).  For the first 
shock, one can calculate the post-shock conditions as in step (3).  
Next, assume the systems adiabatically expand and the density 
drops to approximately 20\% of its pre-merger value for the primary 
system and to some appropriate value for the secondary system (see 
Fig.\ 23).  Using the post-shock entropy from the first shock, 
this new 
density, and 
the appropriate value for $v_{in}$ (i.e., which corresponds to 
the remaining half of the total thermalisation energy), one can 
calculate the Mach number of the second shock (equation 13).  The 
final post-shock conditions are then determined as usual via the 
jump conditions.

\subsection{What next?}

These simulations have allowed us to develop a good understanding 
of how entropy is generated during cluster mergers.  We have 
provided an algorithm that encapsulates this physical process. 
The practical application of this is to be able to predict the 
evolution of the entropy profiles of groups and clusters as they 
grow in mass in a CDM universe.  In future papers, we will 
consider this in detail.  In particular, we will focus on the 
problem of explaining the self-similar growth of clusters seen in 
hydrodynamical simulations.

The key application of our results is to understand how 
perturbations in the entropy distribution of clusters propagate 
through the merging hierarchy. The entropy profile is modified both 
by cooling, which lowers the entropy of the system, and 
non-gravitational heating (e.g., from supernovae, AGN outflows, 
thermal conduction), which raises it.  In addition to the 
inclusion of merger rates derived from numerical simulations, 
this aim requires us to validate the heating model developed here 
using simulations of merging clusters with ``perturbed'' initial 
entropy profiles (i.e., where gas does not trace dark matter).  
Among others, we will explore common physically-motivated examples 
of entropy modification include shifting and truncating the 
distributions (e.g., Babul et al.\ 2002; Voit et al.\ 2002).  
This study is currently underway (McCarthy et al.\ , in 
prep.).

Aside from such practical applications, there 
is remaining academic work to be done as well.  In the current 
study we 
have presented a detailed exploration of {\it how} entropy is 
generated in merger shock heating events.  However, {\it why} the 
entropy is generated in this fashion needs further clarification.  
For example, an interesting result of our study is that 
the gas-only mergers preserve self-similarity.  It is well-known 
that dark matter-only simulations also approximately preserve 
self-similarity through the hierarchy (e.g., NFW).  Why it should 
be that the gas, which exchanges energy with itself through shock 
heating, and the dark matter, which exchanges energy with 
itself through phase mixing and violent relaxation, both give rise 
to the same equilibrium state is not immediately obvious (see, 
e.g., Faltenbacher et al.\ 2006).  Presumably, this is the result 
of both the gas and dark matter adhering to the virial theorem, 
but demonstrating this explicitly is non-trivial.  Another 
interesting result is that the simplest of shock heating models, 
a single-shock model, does not provide an adequate description of 
mergers characterised by large mass ratios.  What fundamental 
factor determines how many shocks are necessary to completely 
thermalise the available energy?  One possibility is that the 
system is following the course of {\it maximum entropy} 
generation.  For example, we have found that if the density drop 
between shocks is determined by the amount of kinetic energy 
remaining to be thermalised (i.e., the leftover kinetic energy 
fixes the degree of adiabatic expansion between shocks), the 
maximum amount of entropy generated almost corresponds to the 
case where the total energy is thermalised equally over two 
shocks.  This also appears to be roughly the route the simulated 
systems are following.  These and other basic matters require 
further attention.

\section*{Acknowledgments}

IGM thanks Christoph Pfrommer and Gregory Poole for helpful 
discussions and acknowledges support from a NSERC Postdoctoral 
Fellowship and a PPARC rolling grant for extragalactic astronomy 
and cosmology at the University of Durham. RGB acknowledges the 
support of a PPARC senior fellowship.  MLB and AB acknowledge 
support from NSERC Discovery Grants.  GMV acknowledges support 
from NASA grant NNG04GI89G.  TT thanks PPARC for the award of an 
Advanced Fellowship.

\clearpage

\section*{Appendix}

\subsection*{I. MASS RESOLUTION STUDY}

Given that our simulations are non-radiative, the number of
particles required to accurately capture the evolution of the
systems should not be particularly stringent.  However, to be sure
we have carried out a mass resolution study for one of our
simulations.  In particular, we have simulated the 1:1 gas+DM
small impact parameter merger at several different mass
resolutions (see Table 3 for a summary).

In Figure 26 we show the final entropy profiles of the merged
system for the five different mass resolution runs.  Instead of
radius, we use integrated gas mass along the abscissa.  Both
coordinates have been scaled to the anticipated self-similar
result (see \S 3.1).

Figure 26 shows that resulting profile is remarkably insensitive
to the adopted mass resolution.  For example, there is only a 20\%
shift between the resulting entropy profiles of the lowest and
highest resolution simulations, even though the resolution differs
by a factor of 100 between the two.  To strike a balance
between speed and accuracy, we adopt the characteristics of the
medium resolution run for all of our other simulations.  As
indicated by Fig.\ 26, the medium resolution run yields a final
entropy profile that differs only by a few percent from our
highest resolution run.

We point out that the increased entropy in the lowest resolution 
run is likely due to an underestimate in the gas density which, 
in turn, results in more efficient entropy generation in the 
shocks.  However, as one increases the resolution (i.e., particle 
number), one obtains a more accurate density determination and, 
therefore, a more accurate entropy jump.  In the case of high 
mass ratio mergers, our simulations indicate that most of the 
energy is thermalised in the more massive primary system.  
Therefore, so long as the primary system is well-resolved 
the results should be robust.  This likely accounts for the fact 
that the distribution of gas in massive virialised systems formed 
in non-radiative cosmological simulations does not depend much on 
resolution (e.g., Frenk et al.\ 1999), even though the properties 
of the small systems that merge to form the massive system change 
significantly with resolution.

\begin{table}
\centering
\begin{minipage}{140mm}
\caption{Mass Resolution Study.}
\begin{tabular}{@{}lllll@{}}
\hline
Sim. label & $N_{\rm gas}$ & $N_{\rm dm}$
& $m_{\rm gas}$ & $m_{\rm dm}$\\

&  &  & ($M_\odot$) & ($M_\odot$)\\
\hline
lowest res. & $10^4$ & $1.5\times10^4$ & $2.7\times10^{10}$ &
$1.7\times10^{11}$ \\
low res. & $3.3\times10^4$ & $5.1\times10^4$ & $8.2\times10^9$ &
$5.2\times10^{10}$\\
medium res. & $10^5$ & $1.5\times10^5$ & $2.7\times10^9$ &
$1.7\times10^{10}$\\
high res. & $3\times10^5$ & $4.6\times10^5$ & $9.1\times10^{8}$ &
$5.8\times10^9$\\
highest res. & $10^6$ & $1.5\times10^6$ & $2.7\times10^{8}$ &
$1.7\times10^9$\\

\hline
\end{tabular}
\end{minipage}
\end{table}

\begin{figure}
\centering
\includegraphics[width=8.4cm]{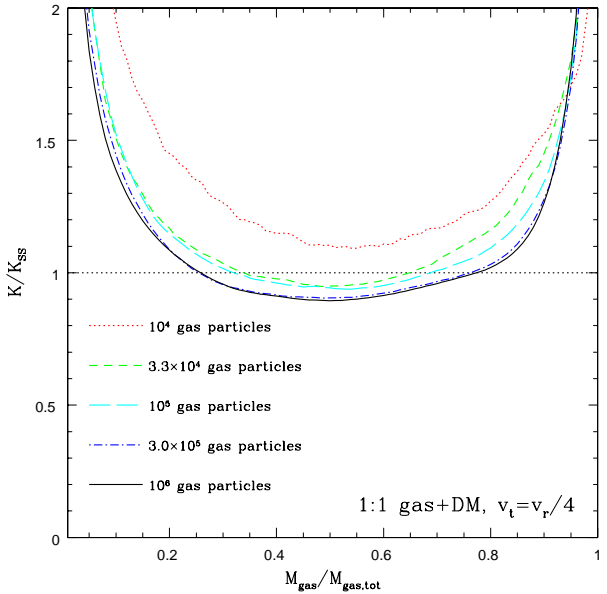}
\caption{Comparison of final entropy profile for the 1:1 gas+DM
small impact parameter merger simulated with various mass
resolutions. }
\end{figure}

\bsp

\label{lastpage}

\end{document}